\documentclass[preprint,12pt,3p]{elsarticle}

\usepackage{graphicx}

\usepackage{amssymb}

\usepackage{lineno}
\usepackage[normalem]{ulem}
\usepackage{colortbl}
\usepackage{color}
\usepackage{float}
\usepackage{hyperref}
\usepackage{booktabs}
\usepackage{multirow}
\usepackage{longtable}
\usepackage[utf8]{inputenc}
\usepackage{fontawesome}
\usepackage{comment} 

\usepackage{framed} 
\definecolor{shadecolor}{rgb}{0.8,0.8,0.8}

\definecolor{darkgreen}{cmyk}{1,0,1,0}
\definecolor{darkorange}{rgb}{0.91, 0.41, 0.17}
\definecolor{MyGray}{rgb}{0.85,0.85,0.85}
\definecolor{purple}{rgb}{0.5, 0.0, 0.5}
\definecolor{pink}{rgb}{1.0, 0.0, 0.5}
\definecolor{ballblue}{rgb}{0.13, 0.67, 0.8}
\definecolor{orange}{rgb}{1.0, 0.49, 0.0}

\newcommand{\fabiano}[1]{{{\color{black} #1}}}
\newcommand{\daniel}[1]{{\color{black} #1}}

\newcommand{\ie}{\emph{i.e.}\ }
\newcommand{\eg}{\emph{e.g.}\ }

\begin{document}
\begin{frontmatter}

\journal{Journal of Systems and Software}

\title{Initiatives and Challenges of Using Gamification in Software Engineering: A Systematic Mapping}




\author[ufscar]{Daniel de Paula Porto\corref{cor1}}
\ead{daniel.porto@ufscar.br}

\author[ufscar]{Gabriela Martins de Jesus}
\ead{gabriela.jesus@ufscar.br}

\author[ufscar]{Fabiano Cutigi Ferrari}
\ead{fcferrari@ufscar.br}

\author[ufscar]{Sandra Camargo Pinto Ferraz Fabbri}
\ead{sfabbri@ufscar.br}

\address[ufscar]{Computing Department -- Federal University of S\~ao Carlos -- Brazil}

\cortext[cor1]{Corresponding author.}


\begin{abstract}
\emph{Context:} 
Gamification is an emerging subject that has been applied in different areas, bringing contributions to different types of activities.
\emph{Objective:}
This paper aims to characterize how gamification has been adopted \fabiano{in non-educational contexts of software engineering~(SE) activities.}
\emph{Method:}
We performed a Systematic Mapping of the literature obtained from relevant databases of the area. The searches retrieved 2640 studies (published up to January 2020), of which 548 were duplicates, 82 were selected after applying the inclusion and exclusion criteria, and 21 were included via the backward snowballing technique, thus reaching a total of 103 studies to be analyzed.
\emph{Results:}
Gamification provided benefits to activities like requirements specification, development, testing, project management, and support process. 
There is evidence of gamified support to some CMMI 2.0 Practice Areas. The most commonly used gamification elements are points and \daniel{leaderboards}. The main benefit achieved is the increased engagement and motivation to perform tasks.
\emph{Conclusion:}
The number of publications and new research initiatives have increased over the years and, from the original authors' reports, many positive results were achieved in SE activities. Despite this, gamification can still be explored for many SE tasks; for the addressed ones, empirical evidence is very limited.

\end{abstract}


\end{frontmatter}



\section{Introduction} \label{Sec:Introduction}

Over the years, several advances in the Software Engineering (SE) area have led to the production of new technologies, models and techniques that, in turn, would help to mitigate the challenge of developing high quality software. 
Despite the progress made to date, the challenge still remains. 
According to \citet{ChaosReport2018}, only 36\% of projects conform to the planned time, budget and scope. This represents the same situation observed almost 20 years ago.
This occurs not only due to the lack of research in the subject, but also because of human factors present throughout the development. Consequently, motivation and discipline have become even more crucial elements for good software development \cite{Dubois2013}. One manner to introduce and keep these two key elements in the considered activities -- sometimes tedious and not challenging -- is the use of \emph{gamification}. 

Gamification is understood as the use of game elements in non-game contexts~\cite{Deterding2011}. 
It~has emerged as a phenomenon and is increasingly present in people’s everyday lives~\cite{Deterding2011}. 
Gamification makes use of game elements and mechanics to stimulate behavior, improve people’s motivation and engagement in their tasks \cite{Deterding2011, Garcia2017}.

The earliest applications of gamification were in digital marketing strategies to increase customer engagement~\cite{Deterding2011}. 
Due to its effectiveness, gamification has been spread to other domains, such as education, health, and sales. 
There are several initiatives of application of gamification in the educational context~\cite{Kapp2012, Monterrat2015, Latulipe2015, Barata2014, Muntean2011}. 
\citet{Muntean2011}, for example, observed that the use of 
gamification motivates students to study, makes them more interested, and stimulates their learning process.

Recently, the use of gamification has been extended to work environment in order to engage people in their tasks~\cite{Deterding2011b}. 
\citet{Hamari2014} mentioned the use of gamification in the following contexts (in addition to the educational one): 
market, health/fitness, internal system of organizations, sharing, sustainable consumption, work, innovations, and data collection.

Whereas gamification has been applied to several contexts, it has also been explored in SE, thus bringing joy aspects to this context. \citet{Beecham2008} stated that one of the biggest challenges for software development companies is to keep teams motivated.
\citet{Dubois2013} argued that the use of gamification in software development has several advantages due to factors like reward mechanisms. Thus, unpleasant tasks for the development team, such as writing unit testing and easy-to-do maintenance, are stimulated with rewards and joy obtained with gamification.

In particular, the application of gamification in SE activities seems to be promising, since over the years several pieces of research have emerged with this purpose~\cite{Garcia2017, Dubois2013, Munoz2017}. According to \citet{Garcia2017}, besides improving team motivation and engagement in their activities, it is expected that the use of gamification in SE tasks improves the achieved results, both in terms of product quality and project performance. Moreover, the use of gamification in SE context goes beyond motivation and involvement. There are recent pieces of work that enumerated several benefits of using gamification in the software development environment, such as: encouraging good programming practice~\cite{Singer2012}, identification and fault removal~\cite{Fraser2017}, and improvement at performing processes~\cite{Dorling2012}.

In the SE field, therefore, researchers and practitioners are aware of the potential benefits of gamification in the workplace. Gamification enables organizations to reward their developers for every aspect of their activities, each completed task, and each written unit test. The mechanics of gamification not only represent a way to reward the team members, but also make the work funnier~\cite{Garcia2017}.

As a consequence of those promising applications in the industry, researchers have begun to investigate gamification from some points of view. Even though the gamification subject is quite recent and has shown preliminary results, there are already several primary studies regarding that. 
Moreover, there are some secondary studies that attempt to group these primary studies based either on specific \cite{Munoz2017, Hernandez2017, Hernandez2016, Machuca-Villegas2018, Machuca-Villegas2018b, GomezAlvarez2017, UnkelosShpigel2018, Mannov2018, Cursino2018, Maentylae2016, Jesus2018, Alhammad2018} or on general SE perspectives \cite{Olgun2017, Pedreira2015, GarciaMireles2019}.

In order to
provide the reader with 
an up-to-date, comprehensive overview of findings
from the use of \daniel{gamification in non-educational contexts of SE activities}, this article reports on the results of a systematic literature mapping. 
The growing number of publications on this topic may signal awareness of the area about the contribution of gamification to the success of software development projects. 
More specifically, this article presents and discusses:

\begin{itemize}

    \item \daniel{a mapping study that focuses on gamification in non-educational contexts of software engineering;}

    \item the benefits that have already been achieved with gamification;
        
    \item which gamification elements have been used, and in which contexts and activities;

    \item which CMMI 2.0 Practice Areas have been, directly or indirectly, impacted by gamification;
        
    \item which tools have been used to date; and
    
    \item the challenges and difficulties to implement gamification in the SE context.
    
\end{itemize}

It is worth mentioning that in this work we used CMMI as a reference model because it is largely used by the software development industry. According to the CMMI Adoption Trends Report - 2018 Year-End Update \cite{CMMI2018b}, 
approximately 57\% of professionals
who train in CMMI are from the software development area. 
In this same report, it is possible to follow the annual growth in the number of CMMI certifications. 
Other models were not considered in our study, but they can be mapped to CMMI; examples of such models are ISO 9000 \cite{CMMI-ISO}; ISO 12207 \cite{CMMI-ISO}; ISO 15288 \cite{CMMI-ISO}; ISO 15504 \cite{Ehsan2010}. 
CMMI can also work in conjunction with Scrum \cite{CMMI2016}.

Besides this introductory section, the remainder of this article is structured as follows: 
Section \ref{Sec:MappingProcess} reports on the systematic literature mapping carried out, focusing on gamification elements, activities supported by gamification, achieved benefits, CMMI 2.0 Practice Areas impacted by gamification, used tools, and challenges and difficulties to implement gamification. 
The systematic mapping results are summarized and discussed in Section \ref{Sec:Results}.
A summary of the main findings and implications of this study is presented in Section~\ref{Sec:Implications}. Section \ref{Sec:Threats} addresses threats to validity, and, 
Section \ref{Sec:RelatedWork} summarizes the main secondary studies related to this. 
Finally, Section \ref{Sec:Conclusion} brings the final considerations and planned future work.

\section{Mapping Process} \label{Sec:MappingProcess}

Systematic mappings (SM) provide an overview of the area to be studied and help on the identification of research opportunities. In this section, we describe the SM reported on this article.\footnote{The complete protocol of this SM can be accessed in \url{https://goo.gl/pQHYV7}} 
The process adopted was the one proposed by \citet{Petersen2008, Petersen2015}, which is composed of the following activities: definition of the research questions, conduction of the searches and screening of the studies, classification scheme and data extraction, and results mapping. Each of these activities is described in the sequence.

\subsection{Research Questions}\label{subsec:ResearchQuestions}

As previously mentioned, this SM is motivated by the need for an up-to-date view of the current uses of gamification in the software engineering context. 
Our main goal is \emph{to understand how gamification has been applied to software engineering}. 
We aim at presenting an overview of where and how gamification elements have been used in a software development life cycle. This overview includes characterizing which and how gamification elements have been mostly used, in which contexts they have been applied, which benefits have been achieved, which CMMI 2.0 Practice Areas have been impacted by gamification, which tools have been used, and which are the challenges and difficulties to implement gamification.

The goal established for this study has two interest dimensions. The first one concerns the gamification elements themselves, with the aim of identifying the ones that have been mostly used and their purpose. 
The second dimension refers to the activities performed in software engineering, with the aim of identifying and characterizing in which activities gamification has been used. To better guide the study, the main goal of this SLR was mapped into four research questions of more specific contexts, as follows:

\textbf{RQ1. How is gamification inserted into software engineering activities?} This research question aims to identify which gamification elements have been mostly used, and in which software engineering activities they have shown to be useful.

\textbf{RQ2. How do software engineering activities benefit from gamification?} This research question aims to list all direct and indirect benefits achieved through gamification, and identify in which software engineering activities these benefits occur. 
To better answer this research question, the following sub-question was created: 
\textbf{RQ2.1. Which CMMI 2.0 Practice Areas have been impacted by gamification?} This sub-question aims to identify which Practice Areas from CMMI 2.0 process model have been, direct or indirectely, impacted by gamification. Here, we tried to relate all the gamification implementation initiatives with the Practice Areas of CMMI 2.0.

\textbf{RQ3. Which software has supported the gamification implementation and in which contexts it has been used?} This research question aims to identify the software systems that have been used to implement gamification and in what contexts they have been used.

\textbf{RQ4. What are the challenges and difficulties of deploying gamification in software engineering?} Finally, this research question aims to identify which are the difficulties and challenges of implementing gamification in software engineering, as reported in the literature.

\subsection{Conducting Search and Screening of Papers} \label{subsec:ConductingSearch}

\subsubsection{Search String}\label{subsubsec:SearchString}
\begin{figure}[!ht]
     \centering
     \includegraphics[width=1\linewidth]{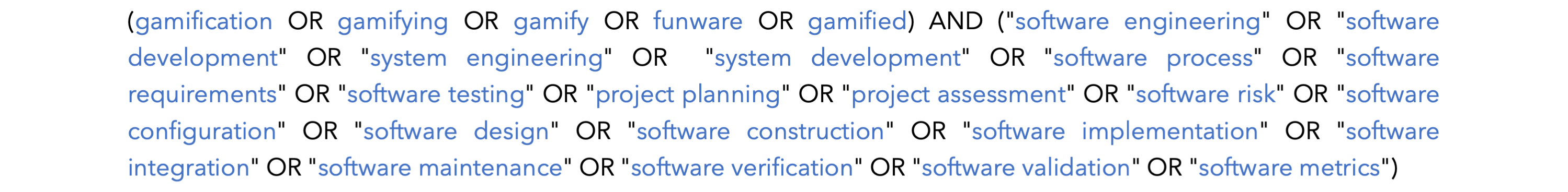}
     \vspace{-.8cm}
     \caption{Used search string.}
     \label{fig:SearchString}
\end{figure}

In order to construct the search string, we selected the main terms and synonyms found in previously known studies, following the recommendation of \citet{Kitchenham2007}. At this point, the relevant keywords found in known secondary studies were also used.
After some initial tests, we decided to complement the string with other synonyms, so that its range was as large as possible. The final search string can be seen in Figure~\ref{fig:SearchString}.

We did not define a time frame to narrow search results, given that as far as the use of gamification elements in software engineering context has began recently, any restriction in this way could disregard relevant publications to our research. To check the search string consistency, a control group was created containing some relevant and previously known articles. 
Results that were retrieved with the search string included all studies from the control group.

\subsubsection{Inclusion and Exclusion Criteria}\label{subsubsec:IECriterias}

To avoid subjectivity in study selection and focus only on gamification in the software engineering context, we defined the inclusion \textbf{(i)} and exclusion criteria \textbf{(e)} listed in the sequence. Note that a study was selected if it passed \textbf{i1} and did not pass in any of the exclusion criteria.

\begin{description}
\item [i1.] Addresses the use of gamification in the software engineering context;
\item [e1.] Considers gamification in the educational or training context; 
\item [e2.] Does not consider gamification in software engineering;
\item [e3.] Is not an end software engineering activity;
\item [e4.] Is an index or preface of another publication;
\item [e5.] Is not written in English;
\item [e6.] Addresses real games or serious games;
\item [e7.] Is a secondary study.
\item [e8.] Is not available online.
\end{description}

Regarding the exclusion criteria, concerning criteria e1, in our research, we classified as \emph{educational or training context} any study in which the applicability of the proposal occurs exclusively in these contexts. 
Even though training is an activity present in software development, it is only a secondary activity in the development process (excluded by criterion e3). Besides that, if training-related studies were selected, all studies in the educational context should also be selected as these could be used as training in a real company. To select all educational studies, the search string should be different and would bring a large number of other studies.
Regarding the e4 criterion, it is important to mention that many search engines return, in addition to studies, conference indexes, or prefaces. Therefore, e4 was created to exclude these results from the search.

\subsubsection{Search process}\label{subsubsec:SearchProcess}

After the definition of the search string and the inclusion and exclusion criteria, the next step was to identify the primary studies. Two searches were carried out. The first (Search 1) was performed on July, 2018, in five databases (details next, in this section) that index studies published in proceedings of scientific events, in journals, or books, all related to the subject. 
A second search (Search 2) was run in January, 2020, with the intent of making our results up-to-date; for this, we used the same search string and the same databases.

\begin{figure}[!ht]
     \centering
     \includegraphics[width=.7\linewidth]{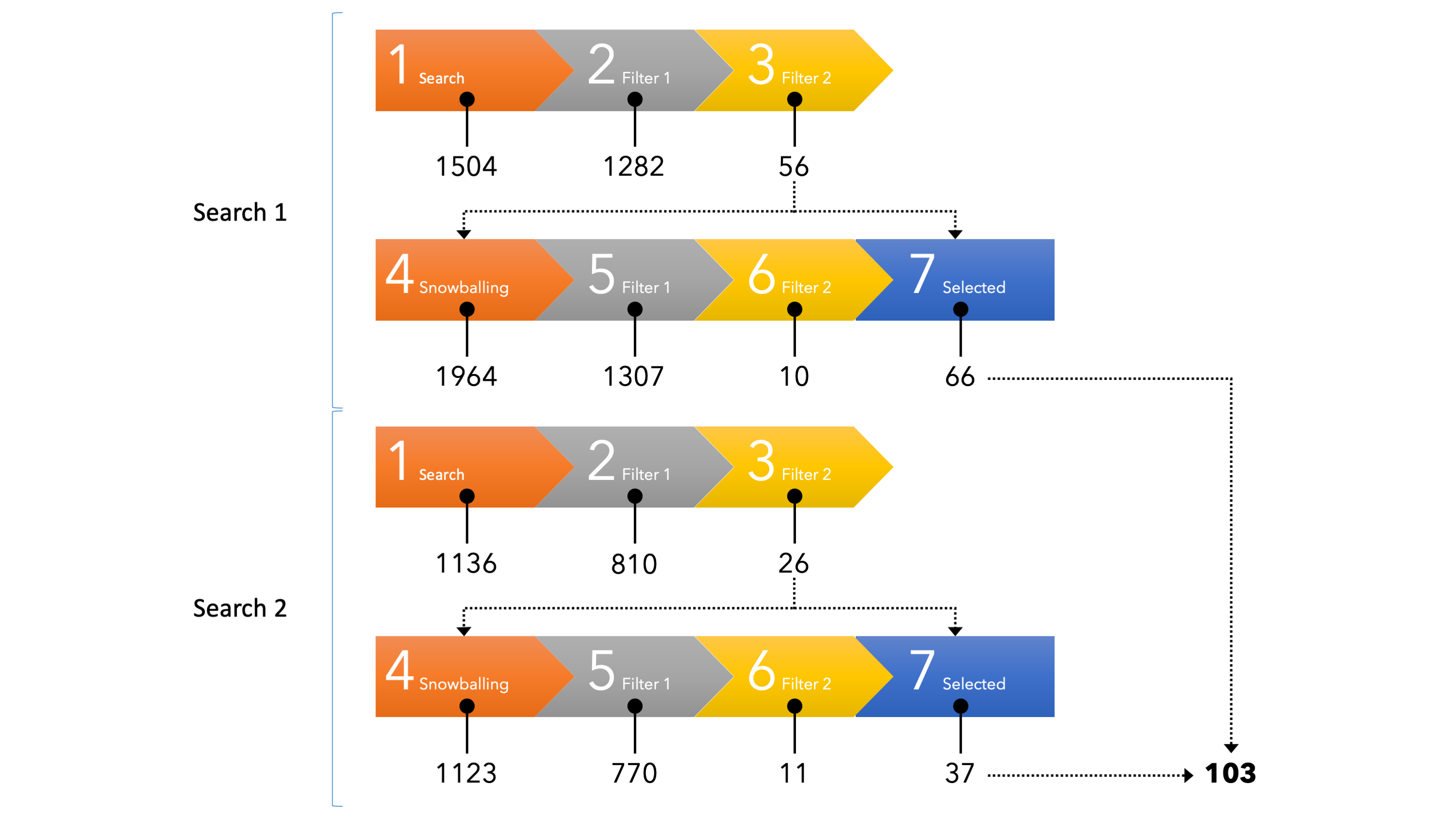}
     \caption{Search process.}
     \label{fig:SearchProcess}
\end{figure}

Searches have targeted title, abstract, and keywords of studies. Figure~\ref{fig:SearchProcess} illustrates the performed process.
For Search~1, the Figure~\ref{fig:SearchProcess} shows that 1504 studies were retrieved as a result of the application of the search string in all databases. From these studies, the duplicates were removed (Filter 1), reducing the list to 1282 studies. The application of inclusion and exclusion criteria (Filter 2) reduced the list to 56 studies. 
Based on these 56 studies and also in the studies discarded by criteria e7 (secondary studies), we applied an iteration of the backward snowballing technique~\cite{wohlin2014}.
The purpose of backward snowballing is to use the reference list of selected studies to identify new studies to be included.
The first step is to list all possible new primary studies from the list of references of the selected papers.
Afterwards, such studies go through the same selection steps carried out for the studies retrieved with automatic search (\ie Filters 1 and 2 are applied).
In total, 1964 additional items were retrieved. 
From these, 657 duplicates were removed (Filter 1), decreasing the number of entries to 1307 studies. Then, inclusion and exclusion criteria were applied (Filter 2), which reduced the list to 10 studies. The final list from Search~1 was composed by 56 studies found early and the 10 recovered by snowballing.

For the second search (Search~2), we performed exactly the same steps as performed for Search 1, and ended up with a final list of 37 studies.
The final list of selected studies includes the 66 selected studies from Search~1, as well as the 37 selected studies on Search~2, thus composing a set of 103 studies. Note that, hereafter, we refer to the primary studies selected in this SM by using the naming convention S$<$NUMBER$>$, where $<$NUMBER$>$ is a sequential number. The complete list of selected studies can be found in Appendix~B.

The surveyed search engines were:
(i) ACM Digital Library (DL);\footnote{\url{http://dl.acm.org} -- accessed on 13-September-2020} 
(ii) IEEE Xplore Digital Library;\footnote{\url{http://www.ieeexplore.ieee.org} -- accessed on 13-September-2020} 
(iii) Elsevier ScienceDirect;\footnote{\url{http://www.sciencedirect.com} -- accessed on 13-September-2020} 
(iv) Elsevier Scopus;\footnote{\url{http://www.scopus.com} -- accessed on 13-September-2020} and 
(v) Springer Nature SpringerLink.\footnote{\url{https://link.springer.com} -- accessed on 13-September-2020} 
Table \ref{table:SearchResults} shows the number of studies at the end of each phase, by type of search. The search engine that retrieved more studies was SpringerLink, finding more than three times studies than Scopus. However, after the application of filters 1 and 2, we verified that Scopus was the most effective, contributing with 65 selected studies, while SpringerLink contributed with only six. 
It is noteworthy the importance of the snowballing technique, which led to the inclusion of the second largest number of studies (21, in total).
Furthermore, due to the ``hybrid'' nature of some search engines (\eg Scopus' and ACM's search results overlap results from other queried databases), in Table~\ref{table:Publishers} we provide more precise information regarding the number of studies selected per publisher. 
Table~\ref{table:Publishers} shows that the publisher with the largest number of selected studies is IEEE with 30 studies, followed by Springer with 23 and ACM with 17. 
All other publishers contributed at most with 4 studies each.

\begin{table}[!htpb]
\caption{Studies by search type at the end of each phase.}
\label{table:SearchResults}
\centering

\fontseries{m}
\fontshape{n}
\fontsize{7.75}{8.5}
\selectfont

\begin{tabular}{lccc}
\toprule
Search
&  \multicolumn{1}{c}{Found} 
&  \multicolumn{1}{c}{Studies after} 
&  \multicolumn{1}{c}{Studies after} \\

&  \multicolumn{1}{c}{studies} 
&  \multicolumn{1}{c}{filter 1} 
&  \multicolumn{1}{c}{filter 2} \\ \midrule

String - ACM           & 228                                                      & 160                                                               & 3                                                                 \\
String - IEEE          & 174                                                      & 72                                                                & 7                                                                 \\
String - ScienceDirect & 32                                                       & 1                                                                 & 1                                                                 \\
String - Scopus        & 541                                                      & 470                                                               & 65                                                                \\
String - Springer      & 1665                                                     & 1390                                                              & 6                                                                 \\
Snowballing            & 3087                                                     & 2077                                                              & 21                                                                \\* \midrule
Total                  & 5727                                                     & 4170                                                              & 103                                                               \\* \bottomrule
\end{tabular}
\end{table}

\vspace{-.6cm}

\begin{table}[!htpb]
\caption{Studies by publisher.}
\label{table:Publishers}
\centering

\fontseries{m}
\fontshape{n}
\fontsize{7.75}{8.5}
\selectfont

\begin{tabular}{ll}
\toprule
Publisher & \# \\ \midrule
IEEE      & 30 \\
Springer  & 23 \\
ACM       & 17 \\
Others    & 33 \\ \bottomrule
\end{tabular}
\end{table}

Table \ref{table:SearchResults} shows that Filter 1 restricted the number of studies in more than 25\% of the total found (1557 of 5727). 
It should be noted that filter 1 was performed automatically using the StArt tool \cite{Fabbri2016}, which greatly reduced working time and chances of errors (in comparison with manual work).
Table~\ref{table:SearchResults} also shows that  Filter 2 (inclusion and exclusion criteria) was responsible for restricting more than 97\% of the non-duplicates (4067 out of 4170), reducing the final number of studies selected to almost 2\% of the total originally retrieved (103 out of 5727). 
Table \ref{table:Filter2Exclusion} shows the reasons that led the studies to be discarded by Filter 2. 
Note that exclusion criteria e1 and e2 were responsible for discarding almost 90\% studies (3626 out of 4067).

\begin{table*}[!htpb]
\centering
\caption{Number of studies discarded by exclusion criteria.}
\label{table:Filter2Exclusion}

\fontseries{m}
\fontshape{n}
\fontsize{7.75}{8.5}
\selectfont

\begin{tabular}{lc}
\toprule
Exclusion criteria                                 & Discarded studies \\ \midrule
e1. Considers gamification in the educational or training context & 816               \\
e2. Does not consider gamification in software engineering        & 2810              \\
e3. It is not an end software engineering activity                & 58                \\
e4. Is an index or preface of another publication                 & 127               \\
e5. Is not written in English                                     & 68                \\
e6. Addresses real games or serious games                         & 87                \\
e7. Is a secondary study                                          & 100               \\
e8. Is not available online                                       & 1                 \\* \midrule
Total                                                             & 4067              \\* \bottomrule
\end{tabular}
\end{table*}

The selection of primary studies was performed  by  one author, according to the inclusion and exclusion criteria.
All conflicts in the study selection were resolved with support of additional authors in order to reduce potential threats. 
After applying filters, and based on the questions to be answered, the data to be collected from the studies was established to support the systematic mapping synthesis. This is presented next.

\subsection{Classification Scheme and Data Extraction}\label{subsec:ClassificationScheme}

We established seven data extraction steps (and associated data fields): 
(i) retrieve study metadata; 
(ii) analyze which gamification elements have been used; 
(iii) identify the software engineering activities benefited from gamification; 
(iv) analyze gamification benefits presented in the studies;   
(v) analyze the relationship between gamified activities and CMMI 2.0 Practice Areas;
(vi) identify used tools; and
(vii) identify challenges and difficulties for implementing gamification.
In order to answer the research questions listed in Section \ref{subsec:ResearchQuestions}, the following data fields and classification scheme were adopted:

\emph{Studies metadata:} 
Publication venue, 
publication type, 
year of publication, 
type of study, 
research method, 
and 
country of author affiliation. 
Regarding the types of publications, the studies were categorized as 
Conferences Papers, 
Articles, 
Books, 
Book Chapters, 
Master’s Dissertation, 
and 
PhD Thesis.
Concerning the type of study, in relation to the addressed content, we classified the selected studies in the six categories proposed by \citet{Wieringa2005}: 
Validation Research, 
Evaluation Research, 
Proposal of Solution, 
Philosophical Papers, 
Opinion Papers, 
and 
Personal experience paper. 
Regarding the research method, we used the classification suggested by \citet{Petersen2015}: 
Case Study, 
Experiment, 
and 
Survey. 
The option \emph{No experimental study has been carried out} was added to this list.

\emph{Gamification elements (RQ1):} 
The taxonomy for gamification elements was based on \daniel{a book of \citet{werbach2012};}
The elements taken from the study are:
avatar, 
\daniel{social graphs}, 
betting\footnotemark, 
\daniel{leaderboards}, 
voting\footnotemark[\value{footnote}], 
challenges, 
levels, 
badges, 
points, 
and 
rewards.

\footnotetext{\daniel{The \emph{betting} and \emph{voting} elements do not appear in the \citet{werbach2012}'s taxonomy. Therefore, they were added to the list due to the fact that they  explicitly appear in the selected studies.}}

\emph{Software engineering activities (RQ1, RQ2):}
The software engineering activities were 
also extracted from a secondary study by other authors~\cite{Pedreira2015}.
The selected software engineering activities (derived from ISO/IEC 12207 standard) are: 
Project Management, 
Requirements, 
Development, 
Testing, 
and 
Support Processes.

\emph{Benefits achieved with gamification (RQ2):}
The studies were analyzed according to direct and indirect benefits achieved with gamification.

\emph{CMMI 2.0 Practice Areas (RQ2.1):} 
The studies were analyzed in relation to activities supported by gamification to compare them with the activities mentioned in the software maturity models. For this relationship, we used the Practice Areas of the CMMI 2.0 model as a reference. The Practice Areas in the model are: 
Requirements development and management (RDM), 
Process quality assurance (PQA), 
Verification and validation (VV), 
Peer reviews (PR), 
Technical solution (TS), 
Product integration (PI), 
Supplier agreement management (SAM), 
Estimating (EST), 
Planning (PLAN), 
Monitor and control (MC), 
Risk and opportunity management (RSK), 
Organizational training (OT), 
Causal analysis and resolution (CAR),
Decision analysis and resolution (DAR), 
Configuration management (CM), 
Governance (GOV), 
Implementation infrastructure (II), 
Process management (PCM), 
Process asset development (PAD), and 
Managing performance and measurement (MPM).

\emph{Used tools (RQ3):}
The studies were analyzed in relation to supporting tools, distinguishing whether they described any tool that supports gamification or not. In addition, we identified which tools were \daniel{presented, in which contexts they were used, and the kind of support. The kind of support can be gamified tools (tools with gamification ready to use) or tools enabling gamification (tools that make it possible to create a gamified environment).}

\emph{Challenges and difficulties regarding gamification (RQ4):} 
The studies were analyzed according to the difficulties and challenges brought about by the gamification deployment.

We used a data extraction form designed to gather the aforementioned data. 
As in the step of study selection, one author has read each study, applied the classification scheme, and performed the data extraction. When the decision for including a study was not clear, other authors were involved in the decision making process.
The data extraction step was also performed using the StArt tool~\cite{Fabbri2016}, which supports the MS process and reduces its execution time.
The extracted data was analyzed in order to answer all questions presented in section~\ref{subsec:ResearchQuestions}. A synthesis of the achieved results is presented in the next section.

\section{Results} \label{Sec:Results}

This section presents the main results of this systematic mapping. A discussion is presented in order to answer the research questions set out in Section \ref{subsec:ResearchQuestions}.

\subsection{Analysis of Study Metadata}\label{subsec:ResultsMetadata}

\begin{figure*}[!ht]
     \centering
     \includegraphics[width=1\textwidth]{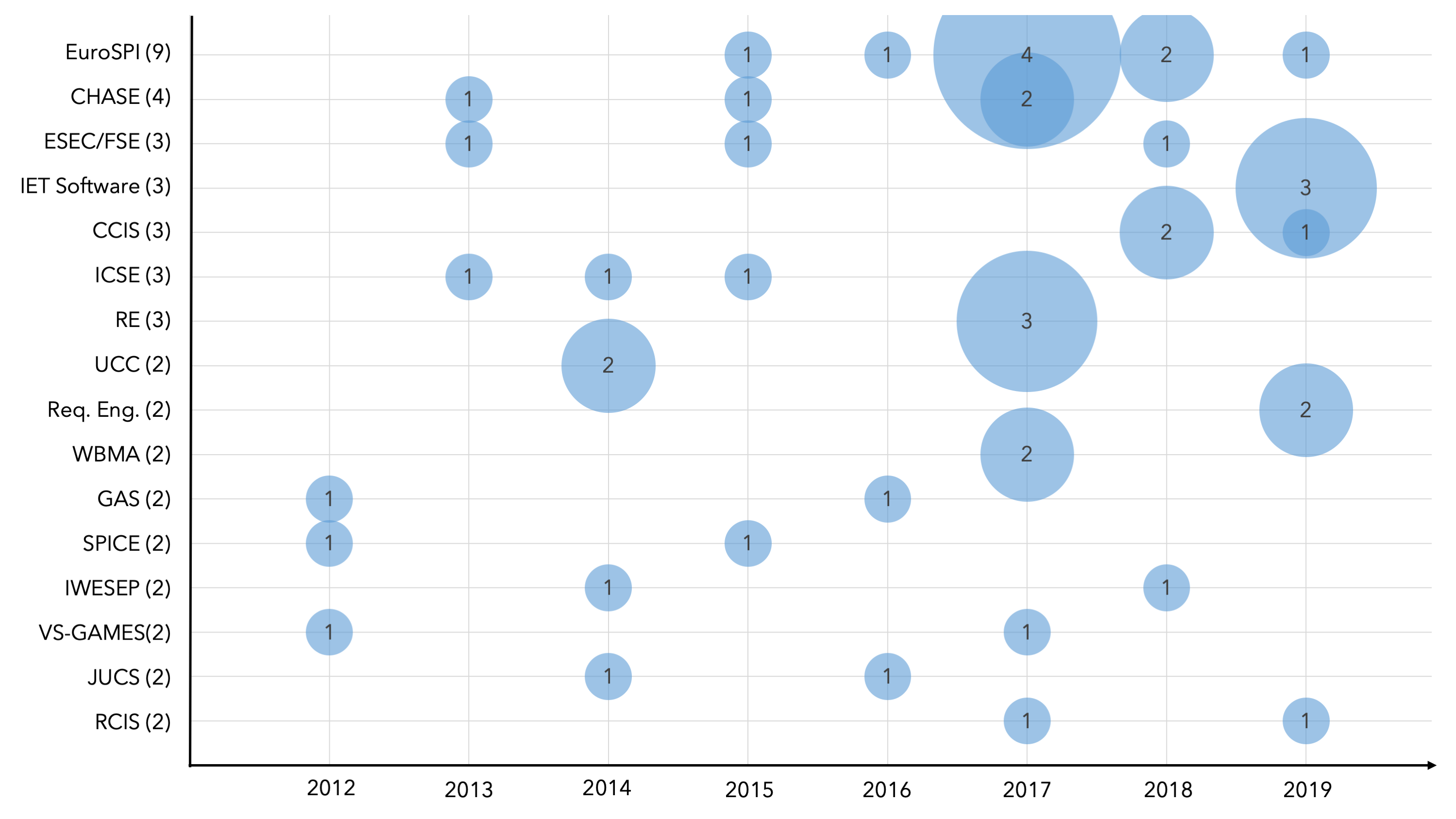}
     \caption{Main publication venues by year.}
     \label{fig:PublicationVenuesYear}
\end{figure*}

We identified 65 different publication venues.\footnote{All publication venues can be accessed in \url{https://bit.ly/353LWge}}
Figure \ref{fig:PublicationVenuesYear} highlights the 16 venues with {two or more} selected studies. 
This group represents almost 40\% (41 of 103) of the selected studies. 
The venue with the largest number of selected studies (9, in total) was the EuroSPI conference (European Conference on Software Process Improvement).

Regarding the types of publications, Table \ref{table:PublicationTypes} shows that the vast majority of studies (93) are papers or articles. In total, 76 studies were published in conferences proceedings, and 17 studies were published in journals. 
In addition to the papers and articles, we also selected 7 Master's dissertation, 1 book, 1 book chapter, and 1 PhD Thesis.

\begin{table}[!htpb]
\centering
\caption{Publication types.}
\label{table:PublicationTypes}

\fontseries{m}
\fontshape{n}
\fontsize{7.75}{8.5}
\selectfont

\begin{tabular}{lc}
\toprule
Publication type      & Number \\ \midrule
Conference Paper  & 76 \\
Article           & 17 \\
MSc. Dissertation & 7  \\
PhD. Thesis       & 1  \\
Book              & 1  \\
Book Chapter      & 1           \\ \bottomrule
\end{tabular}
\end{table}

\begin{figure}[!ht]
     \centering
     \includegraphics[width=1\linewidth]{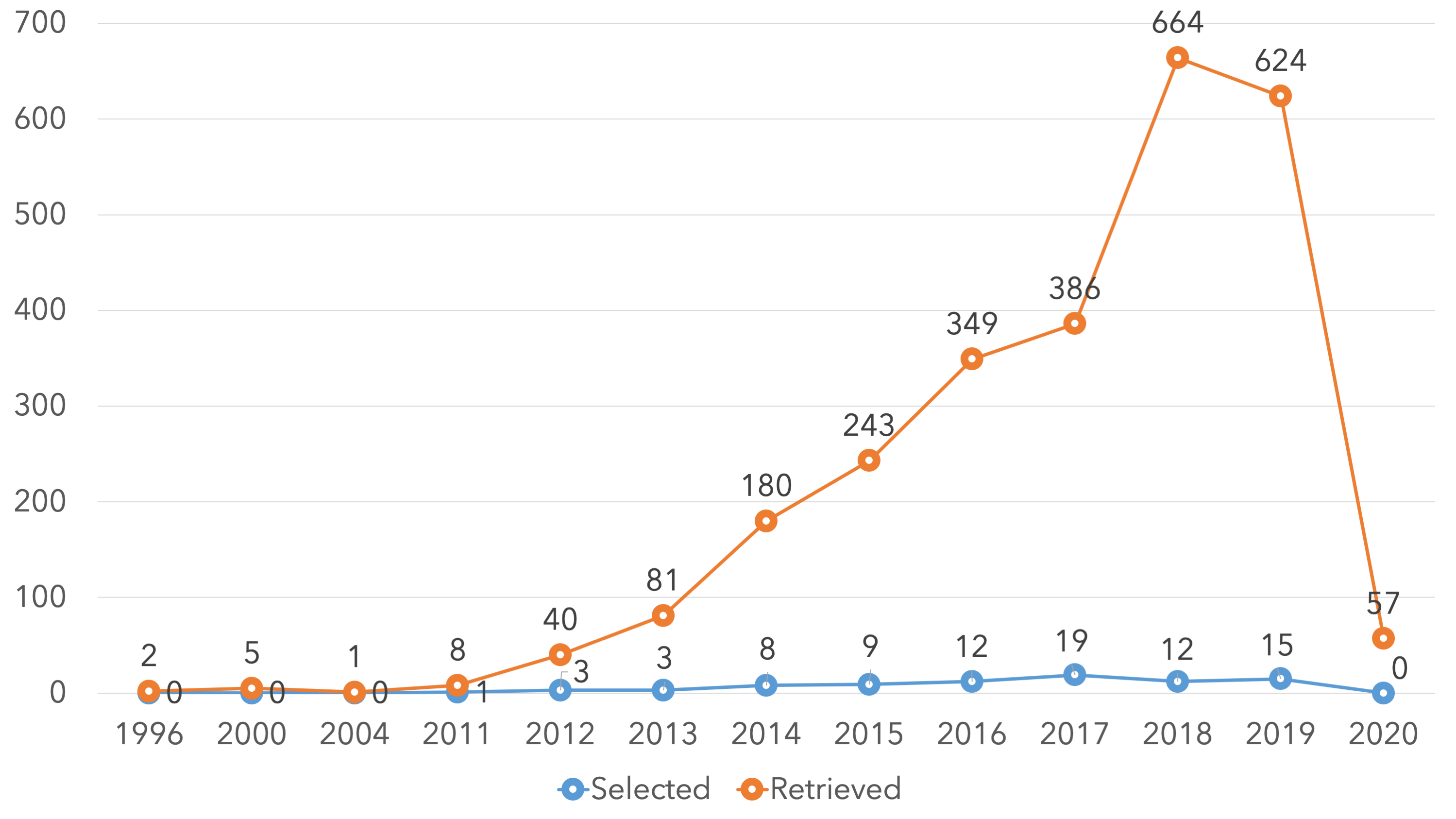}
     \caption{Number of publications per year.
     }
     \label{fig:PublicationPerYear}
\end{figure}

The distribution of selected studies along the years reveals the evolution of gamification adoption. This is depicted in Figure \ref{fig:PublicationPerYear}. 
In \citeyear{Deterding2011} \citet{Deterding2011} stated that gamification was an emerging issue.
Figure \ref{fig:PublicationPerYear} shows a gradual increase in the number of studies retrieved up to 2018. In 2019 there is a small drop (note that numbers for 2020 are only partial). 
Perhaps the subject of gamification in software engineering peaked in 2018 and its popularity has
reached a plateau ever since, though we may need some additional years to draw definite conclusions on it.
It is important to remember that the retrieved studies were filtered (according to the search string).
Given the large number of gamification studies related to other contexts, it is likely that Figure \ref{fig:PublicationPerYear} would be different if the search string was directed to gamification in general.
Another possibility is that \citet{Deterding2011}'s claim is no longer valid almost a decade later.

Information about the type of study can be seen in 
Table \ref{table:StudiesType}.
By focusing on the numbers of studies classified as Proposal of Solution and Validation Research in 
Table \ref{table:StudiesType}, we notice that both encompass 83\% of the selected studies.
These two categories represent studies at an earlier stage, for which there is no robust statistical analysis of the experiment carried out.
This may reflect the short time of use of gamification and the immaturity of several studies.

\begin{table*}[!htpb]
\caption{Selected studies type.}
\label{table:StudiesType}
\centering

\fontseries{m}
\fontshape{n}
\fontsize{7.75}{8.5}
\selectfont

\begin{tabular}{lcp{11.5cm}}
\toprule
Studies type         & Number & Studies                                                                                                                                                                                                                                                                                                                                                                                                                                                                                                              \\ \midrule
Evaluation Research  & 14     & {[}S2{]}, {[}S10{]}, {[}S14{]}, {[}S16{]}, {[}S21{]}, {[}S26{]}, {[}S46{]}, {[}S61{]}, {[}S69{]}, {[}S70{]}, {[}S75{]}, {[}S78{]}, {[}S81{]}, {[}S90{]}                                                                                                                                                                                                                                                                                                                                                              \\ \midrule
Proposal of Solution & 40     & {[}S7{]}, {[}S8{]}, {[}S9{]}, {[}S12{]}, {[}S13{]}, {[}S20{]}, {[}S23{]}, {[}S24{]}, {[}S25{]}, {[}S27{]}, {[}S28{]}, {[}S29{]}, {[}S33{]}, {[}S34{]}, {[}S38{]}, {[}S44{]}, {[}S45{]}, {[}S50{]}, {[}S52{]}, {[}S57{]}, {[}S58{]}, {[}S59{]}, {[}S64{]}, {[}S65{]}, {[}S72{]}, {[}S74{]}, {[}S76{]}, {[}S79{]}, {[}S85{]}, {[}S87{]}, {[}S89{]}, {[}S91{]}, {[}S92{]}, {[}S93{]}, {[}S94{]}, {[}S96{]}, {[}S98{]}, {[}S100{]}, {[}S102{]}, {[}S103{]}                                                               \\ \midrule
Validation Research  & 46     & {[}S1{]}, {[}S3{]}, {[}S4{]}, {[}S5{]}, {[}S6{]}, {[}S11{]}, {[}S15{]}, {[}S17{]}, {[}S18{]}, {[}S19{]}, {[}S22{]}, {[}S30{]}, {[}S31{]}, {[}S32{]}, {[}S35{]}, {[}S36{]}, {[}S37{]}, {[}S39{]}, {[}S40{]}, {[}S41{]}, {[}S42{]}, {[}S43{]}, {[}S47{]}, {[}S48{]}, {[}S49{]}, {[}S51{]}, {[}S53{]}, {[}S54{]}, {[}S56{]}, {[}S60{]}, {[}S62{]}, {[}S63{]}, {[}S66{]}, {[}S67{]}, {[}S68{]}, {[}S71{]}, {[}S73{]}, {[}S77{]}, {[}S80{]}, {[}S82{]}, {[}S83{]}, {[}S86{]}, {[}S88{]}, {[}S95{]}, {[}S99{]}, {[}S101{]} \\ \midrule
Philosophical Papers & 1      & {[}S55{]}                                                                                                                                                                                                                                                                                                                                                                                                                                                                                                            \\ \midrule
Opinion Papers       & 2      & {[}S84{]}, {[}S97{]}                                                                                                                                                                                                                                                                                                                                                                                                                                                                                                 \\ \bottomrule
\end{tabular}
\end{table*}

Regarding the research method, Table \ref{table:ResearchMethod} shows that many studies (32, in total) did not present any experimental evaluation.
This reinforces the large number of studies in the early stages shown in 
Table \ref{table:StudiesType}. 
From the studies that presented experimental evaluation, 30~presented a case study, whereas 22~presented controlled experiments, and only 19~presented surveys.

\begin{table*}[!htpb]
\caption{Research Methods.}
\label{table:ResearchMethod}
\centering

\fontseries{m}
\fontshape{n}
\fontsize{7.75}{8.5}
\selectfont

\begin{tabular}{p{3cm}cp{11cm}}
\toprule
Research Method                            & Number & Studies                                                                                                                                                                                                                                                                                                                                                       \\ \midrule
Case Study                                 & 30     & {[}S2{]}, {[}S5{]}, {[}S10{]}, {[}S15{]}, {[}S22{]}, {[}S26{]}, {[}S27{]}, {[}S32{]}, {[}S36{]}, {[}S37{]}, {[}S39{]}, {[}S47{]}, {[}S48{]}, {[}S54{]}, {[}S60{]}, {[}S61{]}, {[}S68{]}, {[}S69{]}, {[}S71{]}, {[}S73{]}, {[}S75{]}, {[}S81{]}, {[}S82{]}, {[}S83{]}, {[}S86{]}, {[}S90{]}, {[}S95{]}, {[}S97{]}, {[}S98{]}, {[}S103{]}                       \\ \midrule
Experiment                                 & 22     & {[}S1{]}, {[}S6{]}, {[}S11{]}, {[}S16{]}, {[}S17{]}, {[}S18{]}, {[}S21{]}, {[}S30{]}, {[}S31{]}, {[}S42{]}, {[}S43{]}, {[}S51{]}, {[}S53{]}, {[}S62{]}, {[}S63{]}, {[}S66{]}, {[}S67{]}, {[}S74{]}, {[}S78{]}, {[}S80{]}, {[}S88{]}, {[}S99{]}                                                                                                                \\ \midrule
Survey                                     & 19     & {[}S3{]}, {[}S9{]}, {[}S14{]}, {[}S28{]}, {[}S35{]}, {[}S40{]}, {[}S41{]}, {[}S45{]}, {[}S46{]}, {[}S49{]}, {[}S50{]}, {[}S52{]}, {[}S55{]}, {[}S59{]}, {[}S70{]}, {[}S76{]}, {[}S79{]}, {[}S92{]}, {[}S100{]}                                                                                                                                                \\ \midrule
No experimental study has been carried out & 32     & {[}S4{]}, {[}S7{]}, {[}S8{]}, {[}S12{]}, {[}S13{]}, {[}S19{]}, {[}S20{]}, {[}S23{]}, {[}S24{]}, {[}S25{]}, {[}S29{]}, {[}S33{]}, {[}S34{]}, {[}S38{]}, {[}S44{]}, {[}S56{]}, {[}S57{]}, {[}S58{]}, {[}S64{]}, {[}S65{]}, {[}S72{]}, {[}S77{]}, {[}S84{]}, {[}S85{]}, {[}S87{]}, {[}S89{]}, {[}S91{]}, {[}S93{]}, {[}S94{]}, {[}S96{]}, {[}S101{]}, {[}S102{]} \\ \bottomrule
\end{tabular}
\end{table*}

Regarding the coutries of authors' affiliation, Table~\ref{table:AuthorsAffiliationCountries} shows that, among the selected studies, Spain and Brazil are the countries with the largest number of published studies (10), followed by United States (9), and Switzerland, India and Italy (8).

\begin{table}[!htpb]
\centering
\caption{Author’s affiliation countries.}
\label{table:AuthorsAffiliationCountries}

\fontseries{m}
\fontshape{n}
\fontsize{7.75}{8.5}
\selectfont

\begin{tabular}{lclc}
\toprule
Country        & Number & Country      & Number \\ \midrule
Spain          & 10     & Sweden       & 4      \\
Brazil         & 10     & Japan        & 4      \\
United States  & 9      & Ireland      & 4      \\
Switzerland    & 8      & Turkey       & 4      \\
India          & 8      & France       & 3      \\
Italy          & 8      & Austria      & 2      \\
Germany        & 7      & Colombia     & 2      \\
United Kingdom & 7      & Australia    & 1      \\
Portugal       & 6      & Belgium      & 1      \\
Netherlands    & 6      & Saudi Arabia & 1      \\
Mexico         & 6      & Chile        & 1      \\
Canada         & 6      & Lithuania    & 1      \\
Norway         & 5      & China        & 1      \\
Israel         & 5      & Malaysia     & 1      \\* \bottomrule
\end{tabular}
\end{table}

We next analyze the results with respect to the research questions established for this study.

\subsection{(RQ1)  How is gamification inserted into software  engineering  activities?}\label{subsec:ResultsSQ1}

In order to answer this research question, it is necessary to analyze the selected studies from two perspectives: 
the use of gamification elements, 
and the activities supported by them. 
Regarding the first perspective, Table \ref{table:Elements} shows the occurrence of each gamification element and the respective selected studies that quote them. It is worth remembering that the occurrences are counted from explicit quotations within the studies.

\begin{table*}[!ht]
\centering
\caption{Occurrence of gamification elements in selected studies.
}
\label{table:Elements}

\fontseries{m}
\fontshape{n}
\fontsize{7.75}{8.5}
\selectfont

\begin{tabular}{l c p{11.3cm}}
\toprule
Gamification element & Number & Studies                                                                                                                                                                                                                                                                                                                                                                                                                                                                                                                                                                                                                                                                                                                                                                                                                                                                                                                                                                                  \\ \midrule
Points               & 87     & {[}S1{]}, {[}S2{]}, {[}S3{]}, {[}S4{]}, {[}S6{]}, {[}S7{]}, {[}S8{]}, {[}S10{]}, {[}S11{]}, {[}S13{]}, {[}S14{]}, {[}S16{]}, {[}S18{]}, {[}S19{]}, {[}S21{]}, {[}S22{]}, {[}S23{]}, {[}S24{]}, {[}S25{]}, {[}S26{]}, {[}S27{]}, {[}S28{]}, {[}S29{]}, {[}S30{]}, {[}S31{]}, {[}S32{]}, {[}S34{]}, {[}S35{]}, {[}S36{]}, {[}S37{]}, {[}S38{]}, {[}S40{]}, {[}S41{]}, {[}S42{]}, {[}S43{]}, {[}S44{]}, {[}S45{]}, {[}S46{]}, {[}S47{]}, {[}S48{]}, {[}S49{]}, {[}S51{]}, {[}S52{]}, {[}S53{]}, {[}S54{]}, {[}S55{]}, {[}S56{]}, {[}S57{]}, {[}S58{]}, {[}S59{]}, {[}S60{]}, {[}S61{]}, {[}S62{]}, {[}S63{]}, {[}S64{]}, {[}S65{]}, {[}S66{]}, {[}S67{]}, {[}S68{]}, {[}S69{]}, {[}S70{]}, {[}S71{]}, {[}S72{]}, {[}S73{]}, {[}S74{]}, {[}S75{]}, {[}S76{]}, {[}S77{]}, {[}S78{]}, {[}S79{]}, {[}S80{]}, {[}S81{]}, {[}S82{]}, {[}S83{]}, {[}S86{]}, {[}S87{]}, {[}S88{]}, {[}S89{]}, {[}S90{]}, {[}S93{]}, {[}S95{]}, {[}S97{]}, {[}S98{]}, {[}S100{]}, {[}S101{]}, {[}S102{]}, {[}S103{]} \\ \midrule
\daniel{Leaderboards}               & 62     & {[}S1{]}, {[}S2{]}, {[}S3{]}, {[}S4{]}, {[}S7{]}, {[}S8{]}, {[}S11{]}, {[}S13{]}, {[}S14{]}, {[}S16{]}, {[}S18{]}, {[}S19{]}, {[}S22{]}, {[}S23{]}, {[}S24{]}, {[}S26{]}, {[}S28{]}, {[}S30{]}, {[}S31{]}, {[}S32{]}, {[}S34{]}, {[}S35{]}, {[}S37{]}, {[}S40{]}, {[}S42{]}, {[}S44{]}, {[}S45{]}, {[}S46{]}, {[}S47{]}, {[}S48{]}, {[}S49{]}, {[}S51{]}, {[}S52{]}, {[}S54{]}, {[}S57{]}, {[}S60{]}, {[}S62{]}, {[}S63{]}, {[}S65{]}, {[}S66{]}, {[}S67{]}, {[}S68{]}, {[}S70{]}, {[}S71{]}, {[}S72{]}, {[}S73{]}, {[}S74{]}, {[}S75{]}, {[}S76{]}, {[}S78{]}, {[}S80{]}, {[}S81{]}, {[}S86{]}, {[}S87{]}, {[}S88{]}, {[}S89{]}, {[}S90{]}, {[}S93{]}, {[}S97{]}, {[}S100{]}, {[}S101{]}, {[}S102{]}                                                                                                                                                                                                                                                                                    \\ \midrule
Badges               & 49     & {[}S2{]}, {[}S7{]}, {[}S8{]}, {[}S11{]}, {[}S14{]}, {[}S15{]}, {[}S19{]}, {[}S22{]}, {[}S23{]}, {[}S25{]}, {[}S28{]}, {[}S30{]}, {[}S31{]}, {[}S34{]}, {[}S37{]}, {[}S38{]}, {[}S39{]}, {[}S40{]}, {[}S45{]}, {[}S46{]}, {[}S47{]}, {[}S49{]}, {[}S50{]}, {[}S51{]}, {[}S52{]}, {[}S53{]}, {[}S55{]}, {[}S57{]}, {[}S59{]}, {[}S61{]}, {[}S65{]}, {[}S66{]}, {[}S67{]}, {[}S68{]}, {[}S69{]}, {[}S70{]}, {[}S71{]}, {[}S77{]}, {[}S82{]}, {[}S83{]}, {[}S87{]}, {[}S89{]}, {[}S90{]}, {[}S93{]}, {[}S97{]}, {[}S98{]}, {[}S100{]}, {[}S101{]}, {[}S103{]}                                                                                                                                                                                                                                                                                                                                                                                                                                \\ \midrule
Levels               & 34     & {[}S2{]}, {[}S3{]}, {[}S7{]}, {[}S10{]}, {[}S11{]}, {[}S14{]}, {[}S19{]}, {[}S21{]}, {[}S22{]}, {[}S31{]}, {[}S35{]}, {[}S37{]}, {[}S38{]}, {[}S40{]}, {[}S47{]}, {[}S51{]}, {[}S53{]}, {[}S56{]}, {[}S57{]}, {[}S67{]}, {[}S68{]}, {[}S69{]}, {[}S71{]}, {[}S77{]}, {[}S78{]}, {[}S83{]}, {[}S86{]}, {[}S87{]}, {[}S89{]}, {[}S93{]}, {[}S95{]}, {[}S98{]}, {[}S100{]}, {[}S103{]}                                                                                                                                                                                                                                                                                                                                                                                                                                                                                                                                                                                                      \\ \midrule
Rewards              & 33     & {[}S1{]}, {[}S9{]}, {[}S10{]}, {[}S11{]}, {[}S14{]}, {[}S19{]}, {[}S22{]}, {[}S30{]}, {[}S31{]}, {[}S34{]}, {[}S40{]}, {[}S44{]}, {[}S45{]}, {[}S46{]}, {[}S47{]}, {[}S48{]}, {[}S52{]}, {[}S57{]}, {[}S59{]}, {[}S61{]}, {[}S64{]}, {[}S65{]}, {[}S68{]}, {[}S69{]}, {[}S83{]}, {[}S89{]}, {[}S90{]}, {[}S98{]}, {[}S99{]}, {[}S100{]}, {[}S101{]}, {[}S102{]}, {[}S103{]}                                                                                                                                                                                                                                                                                                                                                                                                                                                                                                                                                                                                              \\ \midrule
Challenges               & 21     & {[}S2{]}, {[}S7{]}, {[}S9{]}, {[}S11{]}, {[}S22{]}, {[}S27{]}, {[}S29{]}, {[}S30{]}, {[}S31{]}, {[}S36{]}, {[}S41{]}, {[}S49{]}, {[}S51{]}, {[}S61{]}, {[}S68{]}, {[}S78{]}, {[}S82{]}, {[}S87{]}, {[}S89{]}, {[}S98{]}, {[}S100{]}                                                                                                                                                                                                                                                                                                                                                                                                                                                                                                                                                                                                                                                                                                                                                      \\ \midrule
\daniel{Social Graphs}    & 14     & {[}S2{]}, {[}S4{]}, {[}S11{]}, {[}S13{]}, {[}S19{]}, {[}S22{]}, {[}S34{]}, {[}S51{]}, {[}S60{]}, {[}S68{]}, {[}S72{]}, {[}S73{]}, {[}S93{]}, {[}S101{]}                                                                                                                                                                                                                                                                                                                                                                                                                                                                                                                                                                                                                                                                                                                                                                                                                                  \\ \midrule
Avatar               & 14     & {[}S2{]}, {[}S11{]}, {[}S12{]}, {[}S15{]}, {[}S22{]}, {[}S31{]}, {[}S37{]}, {[}S45{]}, {[}S49{]}, {[}S50{]}, {[}S52{]}, {[}S68{]}, {[}S77{]}, {[}S78{]}                                                                                                                                                                                                                                                                                                                                                                                                                                                                                                                                                                                                                                                                                                                                                                                                                                  \\ \midrule
Voting               & 11     & {[}S5{]}, {[}S22{]}, {[}S25{]}, {[}S27{]}, {[}S29{]}, {[}S36{]}, {[}S53{]}, {[}S71{]}, {[}S79{]}, {[}S81{]}, {[}S88{]}                                                                                                                                                                                                                                                                                                                                                                                                                                                                                                                                                                                                                                                                                                                                                                                                                                                                   \\ \midrule
Betting              & 2      & {[}S17{]}, {[}S22{]}                                                                                                                                                                                                                                                                                                                                                                                                                                                                                                                                                                                                                                                                                                                                                                                                                                                                                                                                                                     \\ \bottomrule
\end{tabular}
\end{table*}

By looking at Table \ref{table:Elements}, it is notorious that points and \daniel{leaderboards} were the most used elements. 
As an example, we refer to the study by Prause and Jarke [S42], in which points were given according to code adherence to conventions set forth. In this way, a \daniel{leaderboard} shows the developers who produce code more according to the convention. Another example is the study of Snipes et al. [S35], in which points and \daniel{leaderboards} were used to motivate developers to access certain functionality on IDE, such as viewing method call hierarchies, navigating to variable definitions, and opening class diagrams.
On the other hand, voting and betting mechanisms were less explored. These last two mechanisms appeared mainly in requirements and agile process activities, respectively.

On the second perspective of this research question, Table~\ref{table:SupportedActivities} presents the list of studies that aided each software development activity. It is noted that, nowadays, development activities are the ones that have the most support of gamification. Among several development activities, code review was the most supported by gamification (10 studies). The others activities supported by gamification were: Project Management, Requirements, Testing,  and  Support Processes. Note that, in some cases, the benefits obtained with  gamification were observed by more than one activity and, therefore, the study was mapped to more than one activity.

\begin{table*}[!htpb]
\caption{Activities supported by gamification.}
\label{table:SupportedActivities}
\centering

\fontseries{m}
\fontshape{n}
\fontsize{7.75}{8.5}
\selectfont

\begin{tabular}{l c p{11.3cm}}
\toprule
Activity           & Number & Studies                                                                                                                                                                                                                                                                                                                                                                                      \\ \midrule
Project Management & 15     & {[}S2{]}, {[}S5{]}, {[}S7{]}, {[}S9{]}, {[}S17{]}, {[}S33{]}, {[}S34{]}, {[}S41{]}, {[}S44{]}, {[}S46{]}, {[}S68{]}, {[}S97{]}, {[}S98{]}, {[}S100{]}, {[}S103{]}                                                                                                                                                                                                                            \\ \midrule
Requirements       & 22     & {[}S2{]}, {[}S14{]}, {[}S20{]}, {[}S21{]}, {[}S25{]}, {[}S27{]}, {[}S29{]}, {[}S31{]}, {[}S32{]}, {[}S36{]}, {[}S47{]}, {[}S51{]}, {[}S53{]}, {[}S54{]}, {[}S55{]}, {[}S61{]}, {[}S72{]}, {[}S75{]}, {[}S81{]}, {[}S82{]}, {[}S87{]}, {[}S90{]}                                                                                                                                              \\ \midrule
Development        & 35     & {[}S1{]}, {[}S3{]}, {[}S6{]}, {[}S8{]}, {[}S11{]}, {[}S16{]}, {[}S18{]}, {[}S19{]}, {[}S24{]}, {[}S28{]}, {[}S34{]}, {[}S35{]}, {[}S37{]}, {[}S38{]}, {[}S41{]}, {[}S42{]}, {[}S49{]}, {[}S56{]}, {[}S58{]}, {[}S59{]}, {[}S62{]}, {[}S63{]}, {[}S65{]}, {[}S66{]}, {[}S70{]}, {[}S74{]}, {[}S76{]}, {[}S77{]}, {[}S78{]}, {[}S79{]}, {[}S85{]}, {[}S94{]}, {[}S96{]}, {[}S99{]}, {[}S102{]} \\ \midrule
Testing            & 15     & {[}S2{]}, {[}S12{]}, {[}S15{]}, {[}S26{]}, {[}S37{]}, {[}S50{]}, {[}S65{]}, {[}S77{]}, {[}S78{]}, {[}S80{]}, {[}S84{]}, {[}S86{]}, {[}S91{]}, {[}S94{]}, {[}S95{]}                                                                                                                                                                                                                           \\ \midrule
Support Processes  & 27     & {[}S4{]}, {[}S10{]}, {[}S13{]}, {[}S22{]}, {[}S23{]}, {[}S26{]}, {[}S30{]}, {[}S39{]}, {[}S40{]}, {[}S43{]}, {[}S45{]}, {[}S48{]}, {[}S52{]}, {[}S57{]}, {[}S60{]}, {[}S64{]}, {[}S67{]}, {[}S69{]}, {[}S71{]}, {[}S73{]}, {[}S83{]}, {[}S89{]}, {[}S91{]}, {[}S92{]}, {[}S93{]}, {[}S101{]}, {[}S103{]}                                                                                     \\ \bottomrule
\end{tabular}
\end{table*}

With regard to Project Management activities, the most cited contribution among the studies was the improvement in the people's engagement and motivation to execute the activities (6 studies).
Regarding Requirements-related activities, the most explored with gamification was the increase of stakeholder involvement during elicitation of requirements~(11~studies). 
Defect logging and improvement in the people's engagement and motivation to execute the activities, both reported in 5 studies, were the main Testing activities supported by gamification. Finally, for Support Processes, support to agile process was cited in 14 studies as the main benefit obtained through gamification.

\begin{figure*}[!ht]
     \centering
     \includegraphics[width=1\textwidth]{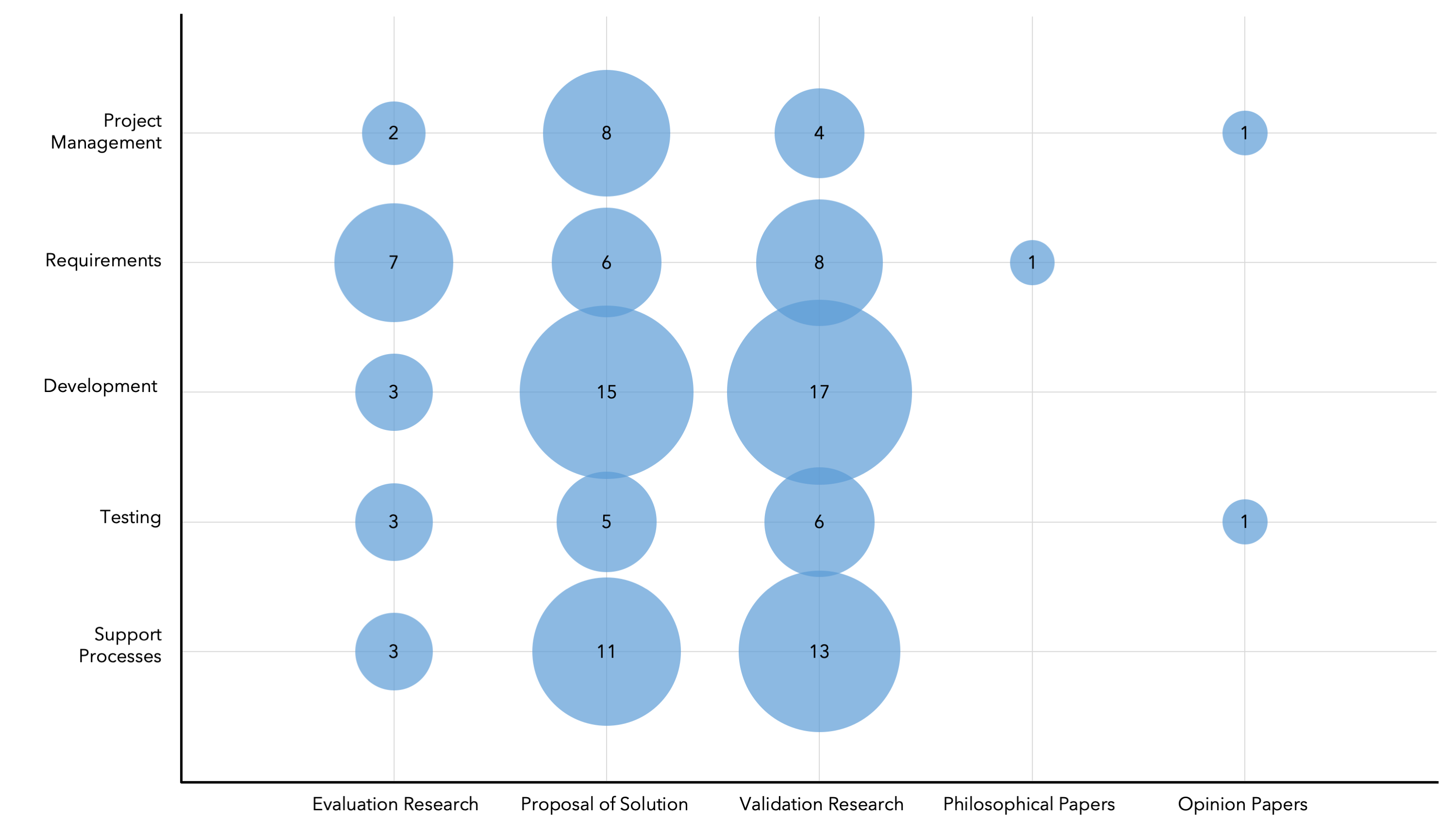}
     \caption{Relationship between study types and software engineering activities.}
     \label{fig:StudyTypeXActivity}
\end{figure*}

Another important piece of information extracted from the studies is the relationship between software development activities and the types of studies. 
This is depicted in  Figure~\ref{fig:StudyTypeXActivity}.
The studies are concentrated in the Proposal of Solution, and Validation Research types. 
It is also possible to observe a concentration of studies in the development activity. Once again, note that a given study can be related to several activities simultaneously.

\begin{figure*}[!ht]
     \centering
     \includegraphics[width=1\textwidth]{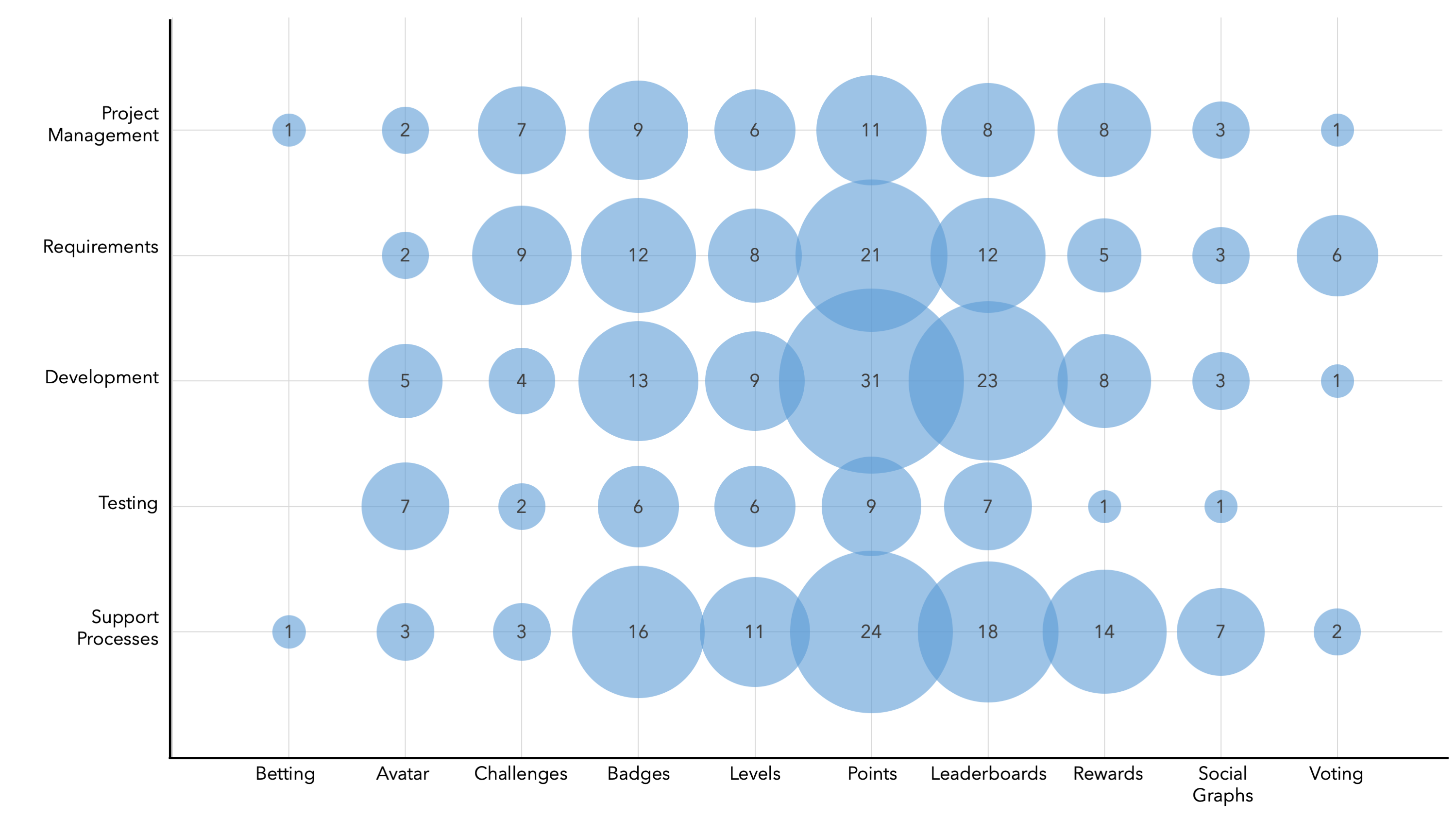}
     \caption{Number of studies that relate gamification elements to software engineering activities.}
     \label{fig:ElementsXActivity}
\end{figure*}

Figure \ref{fig:ElementsXActivity} presents a chart with the relationship between gamification elements and software engineering activities. It shows how many studies mentioned a particular element as a support for a given activity. It should be noted that the same element may have been reported in more than one study, and one study may have mentioned more than one element.

Figure \ref{fig:ElementsXActivity} provides evidence regarding the wide use of points as a gamification element, and as a great support for development activities.  
On the other hand, it is noted that the betting element has not been much explored, appearing in only two studies to support project management and testing activities.
The voting element, on the other hand, has been mainly explored in the Requirements activity, so that people involved could vote on what requirements they want for the system.
Therefore, the Figure \ref{fig:ElementsXActivity} brings a broad view of the activities in which gamification elements have been widely or little explored.

\begin{shaded}
\noindent
Thus, revisiting our research question RQ1, it becomes clear that gamification is mainly inserted in SE activities by adopting the gamification elements mentioned in Figure~\ref{fig:ElementsXActivity}. At the same time, the use of gamification in the activities of Project Management, Requirements, Development, Testing, and Support Processes is notorious. It is also noteworthy the little depth of the studies, since they focus on the Proposal of Solution and Validation Research types (Figure~\ref{fig:StudyTypeXActivity}).
\end{shaded}

\subsection{(RQ2) How do software engineering activities benefit from gamification?}\label{subsec:ResultsSQ2}

In order to answer RQ2, we identified direct and indirect benefits achieved with the use of gamification. This is summarized in Table \ref{table:Benefits}, separated by activities, the benefits obtained with gamification, and the studies in which the benefits were reported.

For the Project Management activity, the main identified benefits  were the increase in the engagement and motivation for the team to execute the activities, and also in the improvement of the quality of work performed.

\begin{table*}[!ht]
\centering
\caption{Benefits achieved with gamification.}
\label{table:Benefits}

\fontseries{m}
\fontshape{n}
\fontsize{7.75}{8.5}
\selectfont

\begin{tabular}{p{2cm}lcp{6.1cm}}
\toprule
Activity                            & Benefit                                   & Number & Studies                                                                                                                                                   \\ \midrule
\multirow{4}{2cm}{Project Management} & Engage and motivate to perform activities & 6      & {[}S5{]}, {[}S9{]}, {[}S41{]}, {[}S44{]}, {[}S98{]}, {[}S103{]}                                                                                           \\
                                    & Improve the quality of work performed     & 4      & {[}S2{]}, {[}S68{]}, {[}S97{]}, {[}S100{]}                                                                                                                \\
                                    & Facilitate activities distribution        & 3      & {[}S17{]}, {[}S34{]}, {[}S46{]}                                                                                                                           \\
                                    & Increase team integration                 & 2      & {[}S7{]}, {[}S33{]}                                                                                                                                       \\ \midrule
\multirow{5}{2cm}{Requirements}       & Improve the engagement of stakeholders    & 11     & {[}S20{]}, {[}S31{]}, {[}S32{]}, {[}S47{]}, {[}S51{]}, {[}S54{]}, {[}S61{]}, {[}S72{]}, {[}S75{]}, {[}S81{]}, {[}S87{]}                                   \\
                                    & Facilitate requirements prioritization    & 6      & {[}S21{]}, {[}S27{]}, {[}S29{]}, {[}S36{]}, {[}S53{]}, {[}S90{]}                                                                                          \\
                                    & Engage and motivate to perform activities & 3      & {[}S14{]}, {[}S55{]}, {[}S82{]}                                                                                                                           \\
                                    & Improve the quality of work performed     & 2      & {[}S2{]}, {[}S25{]}                                                                                                                                       \\ \midrule
\multirow{13}{2cm}{Development}       & Encourage code review                     & 10     & {[}S6{]}, {[}S19{]}, {[}S28{]}, {[}S56{]}, {[}S58{]}, {[}S59{]}, {[}S63{]}, {[}S65{]}, {[}S66{]}, {[}S76{]}                                               \\
                                    & Engage and motivate to perform activities & 9      & {[}S8{]}, {[}S24{]}, {[}S34{]}, {[}S35{]}, {[}S38{]}, {[}S41{]}, {[}S85{]}, {[}S99{]}, {[}S102{]}                                                         \\
                                    & Encourage good programming practices      & 3      & {[}S37{]}, {[}S49{]}, {[}S96{]}                                                                                                                           \\
                                    & Update traceability matrix                & 2      & {[}S78{]}, {[}S94{]}                                                                                                                                      \\
                                    & Improve software documentation            & 2      & {[}S1{]}, {[}S79{]}                                                                                                                                       \\
                                    & Improve the quality of work performed     & 2      & {[}S11{]}, {[}S77{]}                                                                                                                                      \\
                                    & Shorten the coding time                   & 2      & {[}S18{]}, {[}S62{]}                                                                                                                                      \\
                                    & Encourage bug removal                     & 2      & {[}S16{]}, {[}S70{]}                                                                                                                                      \\
                                    & Stimulate code convention adherence       & 1      & {[}S42{]}                                                                                                                                                 \\
                                    & Encourage frequent commit                 & 1      & {[}S74{]}                                                                                                                                                 \\
                                    & Encourage code refactoring                & 1      & {[}S3{]}                                                                                                                                                  \\ \midrule
\multirow{5}{2cm}{Testing}            & Stimulate defect log                      & 5      & {[}S65{]}, {[}S77{]}, {[}S80{]}, {[}S91{]}, {[}S95{]}                                                                                                     \\
                                    & Engage and motivate to perform activities & 5      & {[}S12{]}, {[}S15{]}, {[}S37{]}, {[}S84{]}, {[}S86{]}                                                                                                     \\
                                    & Update traceability matrix                & 2      & {[}S78{]}, {[}S94{]}                                                                                                                                      \\
                                    & Improve the quality of work performed     & 2      & {[}S2{]}, {[}S26{]}                                                                                                                                       \\
                                    & Obtain user feedback                      & 1      & {[}S50{]}                                                                                                                                                 \\ \midrule
\multirow{5}{2cm}{Support Processes}  & Support in the execution of agile process & 14     & {[}S4{]}, {[}S10{]}, {[}S13{]}, {[}S22{]}, {[}S30{]}, {[}S45{]}, {[}S48{]}, {[}S52{]}, {[}S57{]}, {[}S64{]}, {[}S69{]}, {[}S83{]}, {[}S101{]}, {[}S103{]} \\
                                    & Improve the process                       & 12     & {[}S23{]}, {[}S26{]}, {[}S39{]}, {[}S40{]}, {[}S43{]}, {[}S60{]}, {[}S71{]}, {[}S73{]}, {[}S89{]}, {[}S91{]}, {[}S92{]}, {[}S93{]}                        \\
                                    & Engage and motivate to perform activities & 1      & {[}S67{]}                                                                                                                                                 \\ \bottomrule 
\end{tabular}

\end{table*}

Regarding the Requirements-related activities, the improvement of stakeholder engagement during requirements eliciting activities was the most cited benefit. 
Studies such as the ones published by Lombriser et al. [S51] and by Ribeiro et al. [S54] advocated that the use of gamification is a positive influence on the requirements gathering process, making those involved more participatory and consequently generating requirements with higher quality.

As an activity that is sometimes ignored and considered to be less motivating, code review was the most supported activity among Development activities. As examples, we can mention the studies published by Arai et al. [S6] and by Unkelos-Shpigel and Hadar [S58], in which gamification encouraged developers to revise the codes.

On the Testing activities, gamification acted more as a way to stimulate the registration of defects, and also as an engaging and motivating factor to perform activities.
For the first case, we can mention the study  by Lotufo et al. [S95], 
which showed that gamification encouraged team members to increase the quantity and quality of defect records.
For the second case, we can mention the study published by Kohl [S84] which discussed the use of gamification as a tool to increase engagement, creativity, productivity, and fun in testing activities.

Last but not least, regarding Support Processes, gamification supported the execution of agile processes and also the process improvement. In the first case, taking the study of  Yilmaz and OConnor [S10] as an example, gamification supported the execution and also in the improvement of agile processes. 
Note that the number of studies relating gamification with agile processes is interesting; possibly due to the flexibility and adaptivity of agile process, they better accommodate the elements included by gamification. In the second case, as in studies published by Uskarci and Demir{\"o}rs [S23], Herranz et al. [S26] and Ruiz et al. [S39], gamification supported the execution of activities indicated in the process maturity models, thus supporting the improvement of the process itself.

In addition to all unique benefits of each activity, two generic benefits have been reported in more than one activity: engagement and motivation to perform activities, and improvement of the quality of the performed work. 
In addition to all occurrences in all activities, the engagement and motivation to perform activities was the most common benefit; it was cited in 23 studies. 
On the other hand, 8 studies cited improvement of the quality of work performed as a benefit achieved by gamification.

\subsubsection{(RQ2.1) Which CMMI 2.0 Practice Areas have been impacted by gamification?}

\begin{figure*}[!ht]
     \centering
     \includegraphics[width=1\textwidth]{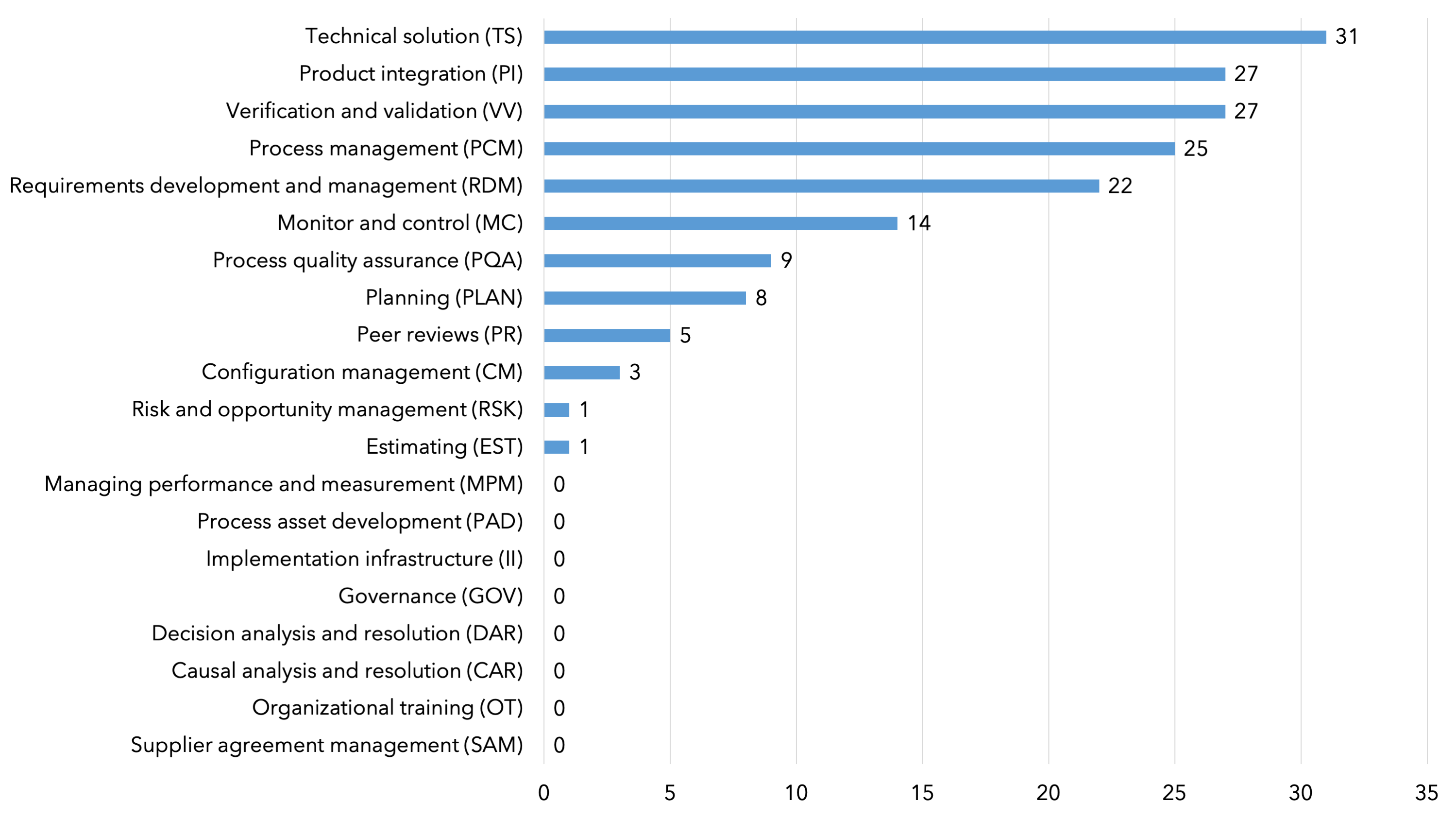}
     \caption{CMMI Practice Areas impacted by gamification}
     \label{fig:StudiesXCMMI}
\end{figure*}

The purpose of this sub-question was to identify what activities were supported by gamification and how they are related to process improvement activities. For this, the international model CMMI \cite{CMMI2018}, in its version 2.0 launched in the first semester of 2018, was used as a reference. 
The objective was to map the activities supported by gamification to the Practice Areas (PA) of the guide, which can be seen in the first column of Table~\ref{table:CMMIPAs}. 
Once again, note that a study may be related to more than one PA. The number of studies per Practice Area is also shown in Figure \ref{fig:StudiesXCMMI} in order to make it easier to compare the distribution of studies per Practice Area.

\begin{table*}[!ht]
\centering
\caption{CMMI Practice Areas impacted by gamification.}
\label{table:CMMIPAs}

\fontseries{m}
\fontshape{n}
\fontsize{7.75}{8.5}
\selectfont

\begin{tabular}{lcp{7.1cm}}
\toprule
Practice Area - PA                            & Number & Studies                                                                                                                                                                                                                                                                                                                                           \\ \midrule
Requirements development and management (RDM) & 22     & {[}S2{]}, {[}S14{]}, {[}S20{]}, {[}S21{]}, {[}S25{]}, {[}S27{]}, {[}S29{]}, {[}S31{]}, {[}S32{]}, {[}S36{]}, {[}S47{]}, {[}S51{]}, {[}S53{]}, {[}S54{]}, {[}S55{]}, {[}S61{]}, {[}S72{]}, {[}S75{]}, {[}S81{]}, {[}S82{]}, {[}S87{]}, {[}S90{]}                                                                                                   \\ \midrule
Process quality assurance (PQA)               & 9      & {[}S23{]}, {[}S26{]}, {[}S39{]}, {[}S40{]}, {[}S60{]}, {[}S89{]}, {[}S91{]}, {[}S92{]}, {[}S93{]}                                                                                                                                                                                                                                                 \\ \midrule
Verification and validation (VV)              & 27     & {[}S2{]}, {[}S6{]}, {[}S11{]}, {[}S12{]}, {[}S15{]}, {[}S16{]}, {[}S19{]}, {[}S26{]}, {[}S28{]}, {[}S37{]}, {[}S50{]}, {[}S56{]}, {[}S58{]}, {[}S59{]}, {[}S63{]}, {[}S65{]}, {[}S66{]}, {[}S70{]}, {[}S76{]}, {[}S77{]}, {[}S78{]}, {[}S80{]}, {[}S84{]}, {[}S86{]}, {[}S91{]}, {[}S94{]}, {[}S95{]}                                             \\ \midrule
Peer reviews (PR)                             & 5      & {[}S19{]}, {[}S56{]}, {[}S58{]}, {[}S66{]}, {[}S76{]}                                                                                                                                                                                                                                                                                             \\ \midrule
Technical solution (TS)                       & 31     & {[}S1{]}, {[}S3{]}, {[}S8{]}, {[}S11{]}, {[}S18{]}, {[}S24{]}, {[}S28{]}, {[}S34{]}, {[}S35{]}, {[}S37{]}, {[}S38{]}, {[}S41{]}, {[}S42{]}, {[}S49{]}, {[}S56{]}, {[}S58{]}, {[}S62{]}, {[}S63{]}, {[}S65{]}, {[}S66{]}, {[}S74{]}, {[}S76{]}, {[}S77{]}, {[}S78{]}, {[}S79{]}, {[}S85{]}, {[}S88{]}, {[}S94{]}, {[}S96{]}, {[}S99{]}, {[}S102{]} \\ \midrule
Product integration (PI)                      & 27     & {[}S1{]}, {[}S3{]}, {[}S8{]}, {[}S11{]}, {[}S24{]}, {[}S28{]}, {[}S34{]}, {[}S35{]}, {[}S37{]}, {[}S38{]}, {[}S41{]}, {[}S42{]}, {[}S56{]}, {[}S58{]}, {[}S65{]}, {[}S66{]}, {[}S74{]}, {[}S76{]}, {[}S77{]}, {[}S78{]}, {[}S79{]}, {[}S85{]}, {[}S88{]}, {[}S94{]}, {[}S96{]}, {[}S99{]}, {[}S102{]}                                             \\ \midrule
Supplier agreement management (SAM)           & 0      &                                                                                                                                                                                                                                                                                                                                                   \\ \midrule
Estimating (EST)                              & 1      & {[}S68{]}                                                                                                                                                                                                                                                                                                                                         \\ \midrule
Planning (PLAN)                               & 8      & {[}S2{]}, {[}S7{]}, {[}S33{]}, {[}S34{]}, {[}S68{]}, {[}S98{]}, {[}S100{]}, {[}S103{]}                                                                                                                                                                                                                                                            \\ \midrule
Monitor and control (MC)                      & 14     & {[}S2{]}, {[}S7{]}, {[}S9{]}, {[}S17{]}, {[}S33{]}, {[}S34{]}, {[}S41{]}, {[}S44{]}, {[}S46{]}, {[}S68{]}, {[}S97{]}, {[}S98{]}, {[}S100{]}, {[}S103{]}                                                                                                                                                                                           \\ \midrule
Risk and opportunity management (RSK)         & 1      & {[}S5{]}                                                                                                                                                                                                                                                                                                                                          \\ \midrule
Organizational training (OT)                  & 0      &                                                                                                                                                                                                                                                                                                                                                   \\ \midrule
Causal analysis and resolution (CAR)          & 0      &                                                                                                                                                                                                                                                                                                                                                   \\ \midrule
Decision analysis and resolution (DAR)        & 0      &                                                                                                                                                                                                                                                                                                                                                   \\ \midrule
Configuration management (CM)                 & 3      & {[}S67{]}, {[}S74{]}, {[}S102{]}                                                                                                                                                                                                                                                                                                                  \\ \midrule
Governance (GOV)                              & 0      &                                                                                                                                                                                                                                                                                                                                                   \\ \midrule
Implementation infrastructure (II)            & 0      &                                                                                                                                                                                                                                                                                                                                                   \\ \midrule
Process management (PCM)                      & 25     & {[}S4{]}, {[}S10{]}, {[}S13{]}, {[}S22{]}, {[}S23{]}, {[}S26{]}, {[}S30{]}, {[}S39{]}, {[}S43{]}, {[}S45{]}, {[}S48{]}, {[}S52{]}, {[}S57{]}, {[}S60{]}, {[}S64{]}, {[}S69{]}, {[}S71{]}, {[}S73{]}, {[}S83{]}, {[}S89{]}, {[}S91{]}, {[}S92{]}, {[}S93{]}, {[}S101{]}, {[}S103{]}                                                                \\ \midrule
Process asset development (PAD)               & 0      &                                                                                                                                                                                                                                                                                                                                                   \\ \midrule
Managing performance and measurement (MPM)    & 0      &                                                                                                                                                                                                                                                                                                                                                   \\ \bottomrule
\end{tabular}

\end{table*}

From Table \ref{table:CMMIPAs} and Figure \ref{fig:StudiesXCMMI}, the contribution of gamification to designing and coding activities is substantial. 
This is clear from the number of studies associated with the Technical solution (TS) and Product integration (PI) PAs. 
Within this group, for example, there are studies that dealt with code improvement and documentation. There are 3 studies associated with the Configuration management (CM) PA, which addressed the activities in the version control systems. 
There are also some studies that dealt with code review; those are associated with Verification and Validation (VV), and Peer reviews (PR).

In addition to the code review studies, several studies supporting test activities were also mapped in the Verification and validation (VV) PA. 
Most of them dealt with improved defect registration.

The gamification support for the requirements-related activities is represented by the  Requirements development and management (RDM) PA. 
It emphasizes the prioritization of requirements and, mainly, the improvement of stakeholder involvement.

Project management activities were mapped into the Planning (PLAN), Monitor and control (MC), Estimating (EST), and Risk and opportunity management (RSK) PAs. In general, these studies focused on improving the engagement and motivation of teams. There are studies that dealt with the distribution of activities, and also with risk assessment and mapping.

The Process management (PCM) and Process quality assurance (PQA) PAs mapped the studies related to process management. These studies dealt with improvements in processes through the execution of activities defined in maturity models. It is worth mentioning the large number of studies that explored gamification as a mechanism to support and improve agile processes.

Eight PAs (namely, 
Supplier agreement management (SAM), 
Organizational training (OT), 
Causal analysis and resolution (CAR), 
Decision analysis and resolution (DAR), 
Governance (GOV), 
Implementation infrastructure (II), 
Process asset development (PAD), and 
Managing performance and measurement (MPM)) 
could not be mapped to any study. 
Naturally, studies that would fit into Organizational training (OT) were excluded by the \textbf{e1} exclusion criterion (namely, \emph{Considers gamification in the educational or training context}), while the other PAs were excluded because they are not software engineering end activities (exclusion criteria \textbf{e3}).
Despite this, there is no real impediment to the application of gamification to support activities related do those PAs.

\begin{shaded}
\noindent
Revisiting our research question RQ2, the benefits of gamification in software development activities can be clearly seen in Table \ref{table:Benefits}. 
Through Table \ref{table:CMMIPAs} it is also clear that, so far, gamification has covered a considerable amount of Practice Areas defined in CMMI 2.0. 
In this research, disregarding the studies discarded by the exclusion criteria, it can be seen that  existing research on gamification to support software engineering tasks can be mapped to 60\% (12 of 20) of CMMI Practice Areas.
\end{shaded}

\subsection{(RQ3) Which software has supported the gamification implementation and in which contexts it has been used?}\label{sec:resultadosSQ3}

To answer RQ3, we identified the tools (software) that were used to support the application of gamification in software engineering. 
From the 103 selected studies, 57 explicitly mentioned this type of support. 
Table \ref{table:Tools2} shows all identified tools grouped by activity and benefit.

In Table \ref{table:Tools2} it is possible to identify the type of the tool. 
Gamified tools are marked with the \faGamepad~symbol,
while tools enabling gamification 
are identified with the \faCogs~symbol.
In the table, it is possible to identify that the vast majority of tools are gamified.
Another interesting fact is the small number of tools that have declared themselves as open-source. It is possible that more tools are open-source, but only six made that clear.
A remarkable feature of the identified tools is that the vast majority are available on the web platform (symbol \faCloud).
Only two tools are available on the mobile platform (symbol \faMobile).
Six tools are desktop (symbol \faDesktop) and all of them are related to development activities since most of these tools are integrated with the IDEs used by programmers.
Five tools do not make it clear on which platform they are available and were marked with the \faQuestionCircle~symbol. 
From all tools, only seven indicated a web address\footnote{RE-PROVO: \url{http://egov-requirements.org}; GamifiedSD : \url{https://github.com/skbly7/gamifiedSD}; GithubCC: \url{https://github.com/tzachz/github-comment-counter}; Stack overflow: \url{www.stackoverflow.com}; G-Unit: \url{https://github.com/davidarnarsson/Gunit}; Mozilla Open Badges: \url{https://openbadges.org}, and Habitica: \url{https://habitica.com} -- accessed on 13-September-2020.}  where they can be found.

\begin{table}[!htpb]
\centering
\caption{Tools that support gamification.}
\label{table:Tools2}

\fontseries{m}
\fontshape{n}
\fontsize{7.75}{8.5}
\selectfont
\setlength{\tabcolsep}{3pt}

\begin{tabular}{m{1.7cm}m{4cm}m{3.2cm}lccc}
\toprule
Activity                            & Benefit                                                                                & Tool                         & Studies                                    & Type          & \begin{tabular}[c]{c}Open \\ Source\end{tabular} & \begin{tabular}[c]{c}Plat \\ form\end{tabular}    \\ \midrule
\multirow{9}{*}{Requirements}       & \multirow{2}{4cm}{Improve the engagement of stakeholders}                                & Agon                         & {[}S14{]}, {[}S47{]}                       & \faGamepad~\faCogs &             & \faCloud        \\
                                    &                                                                                        & MAF                          & {[}S47{]}                                  & \faCogs        &             & \faQuestionCircle            \\ \cmidrule(l){2-7} 
                                    & \multirow{3}{4cm}{Facilitate requirements prioritization}                                & DMGame                       & {[}S27{]}, {[}S29{]}, {[}S36{]}, {[}S53{]} & \faGamepad        &             & \faCloud        \\
                                    &                                                                                        & Garuso                       & {[}S21{]}, {[}S55{]}, {[}S61{]}, {[}S87{]} & \faGamepad        &             & \faCloud        \\
                                    &                                                                                        & GRP                          & {[}S90{]}                                  & \faGamepad~\faCogs &             & \faCloud        \\ \cmidrule(l){2-7} 
                                    & \multirow{4}{4cm}{Improve the participation in a collaborative requirements elicitation} & iThink                       & {[}S54{]}, {[}S55{]}, {[}S75{]}            & \faGamepad        &             & \faCloud        \\
                                    &                                                                                        & REfine                       & {[}S32{]}, {[}S55{]}, {[}S81{]}            & \faGamepad        &             & \faCloud        \\
                                    &                                                                                        & RE-PROVO                     & {[}S82{]}                                  & \faGamepad        &             & \faCloud        \\
                                    &                                                                                        & REVISE                       & {[}S72{]}                                  & \faGamepad        &             & \faCloud        \\ \midrule
\multirow{19}{*}{Development}       & \multirow{6}{*}{Encourage code review}                                                 & CRA                          & {[}S28{]}                                  & \faGamepad        & Y           & \faCloud        \\
                                    &                                                                                        & CARE                         & {[}S56{]}                                  & \faGamepad        &             & \faDesktop       \\
                                    &                                                                                        & CodeBrag                     & {[}S66{]}                                  & \faGamepad        &             & \faCloud        \\
                                    &                                                                                        & GamifiedSD                   & {[}S66{]}                                  & \faGamepad        & Y           & \faCloud        \\
                                    &                                                                                        & GithubCC                     & {[}S66{]}                                  & \faGamepad        & Y           & \faCloud        \\
                                    &                                                                                        & SCRUT                        & {[}S58{]}                                  & \faGamepad        &             & \faQuestionCircle            \\ \cmidrule(l){2-7} 
                                    & \multirow{5}{4cm}{Encourage good programming practices}                                  & Themis                       & {[}S37{]}, {[}S49{]}                       & \faGamepad        &             & \faCloud        \\
                                    &                                                                                        & Blaze                        & {[}S8{]}, {[}S35{]}                        & \faGamepad        &             & \faDesktop       \\
                                    &                                                                                        & Teamfeed                     & {[}S8{]}, {[}S74{]}                        & \faGamepad        &             & \faCloud        \\
                                    &                                                                                        & OO Practices                 & {[}S8{]}                                   & \faGamepad        &             & \faQuestionCircle            \\
                                    &                                                                                        & Beehive                      & {[}S8{]}                                   & \faGamepad        &             & \faQuestionCircle            \\ \cmidrule(l){2-7} 
                                    & \multirow{2}{*}{Encourage code refactoring}                                            & CodeArena                    & {[}S3{]}                                   & \faGamepad        &             & \faCloud~\faDesktop \\
                                    &                                                                                        & GBC & {[}S6{]}                                   & \faGamepad        &             & \faDesktop       \\ \cmidrule(l){2-7} 
                                    & \multirow{3}{4cm}{Improve software documentation}                                        & AKB                          & {[}S88{]}                                  & \faGamepad        &             & \faCloud        \\
                                    &                                                                                        & CollabReview                 & {[}S1{]}                                   & \faGamepad        &             & \faCloud        \\
                                    &                                                                                        & QuoDocs                      & {[}S79{]}                                  & \faGamepad        &             & \faCloud        \\ \cmidrule(l){2-7} 
                                    & \multirow{2}{*}{Update traceability matrix}                                            & Eclipse Capra                & {[}S67{]}                                  & \faGamepad        &             & \faDesktop       \\
                                    &                                                                                        & GamiTracify                  & {[}S78{]}                                  & \faGamepad        &             & \faDesktop       \\ \cmidrule(l){2-7} 
                                    & Encourages knowledge exchange between developers                                       & stack overflow               & {[}S8{]}, {[}S95{]}                        & \faGamepad        &             & \faCloud        \\ \midrule
\multirow{2}{*}{Testing}            & \multirow{2}{4cm}{Engage and motivate to perform test activities}                        & G-Unit                       & {[}S70{]}                                  & \faGamepad        & Y           & \faCloud        \\
                                    &                                                                                        & Rank-Me                      & {[}S80{]}                                  & \faGamepad        &             & \faCloud        \\ \midrule
\multirow{5}{1.5cm}{Project Management} & \multirow{5}{4cm}{Facilitate activities distribution and control}                        & Agile Workbench              & {[}S4{]}, {[}S13{]}                        & \faGamepad        &             & \faCloud        \\
                                    &                                                                                        & RUPGY                        & {[}S22{]}, {[}S103{]}                      & \faGamepad        &             & \faCloud        \\
                                    &                                                                                        & DevRPG                       & {[}S98{]}                                  & \faGamepad        &             & \faQuestionCircle            \\
                                    &                                                                                        & Scraim                       & {[}S41{]}                                  & \faGamepad        &             & \faCloud        \\
                                    &                                                                                        & Trogon                       & {[}S46{]}                                  & \faGamepad        &             & \faCloud        \\ \midrule
\multirow{5}{1.5cm}{Support Processes}  & \multirow{2}{4cm}{Support in software process improvement}                               & GamiSPI                      & {[}S71{]}                                  & \faGamepad        &             & \faCloud        \\
                                    &                                                                                        & SysDyn                       & {[}S39{]}                                  & \faGamepad        &             & \faMobile        \\ \cmidrule(l){2-7} 
                                    & \multirow{3}{4cm}{Support in the execution of agile process}                             & SD project gamification      & {[}S44{]}                                  & \faGamepad        &             & \faCloud        \\
                                    &                                                                                        & Gaming Scrum                 & {[}S101{]}                                 & \faGamepad        &             & \faCloud        \\
                                    &                                                                                        & XGamify                      & {[}S57{]}                                  & \faGamepad        &             & \faCloud        \\ \midrule
\multirow{6}{1.5cm}{General Activities} & \multirow{6}{4cm}{Generic support for gamification insertion}                            & Gamiware                     & {[}S26{]}, {[}S40{]}, {[}S60{]}, {[}S73{]} & \faGamepad~\faCogs &             & \faCloud        \\
                                    &                                                                                        & Mozilla Open Badges          & {[}S97{]}                                  & \faCogs        & Y           & \faCloud        \\
                                    &                                                                                        & GOAL                         & {[}S2{]}                                   & \faGamepad~\faCogs &             & \faCloud        \\
                                    &                                                                                        & Habitica                     & {[}S8{]}, {[}S68{]}                        & \faGamepad        & Y           & \faCloud~\faMobile  \\
                                    &                                                                                        & OpenBadgesUCA                & {[}S39{]}                                  & \faGamepad        &             & \faCloud        \\
                                    &                                                                                        & GamAnalyze                   & {[}S39{]}                                  & \faGamepad        &             & \faCloud        \\ \bottomrule
\end{tabular}

\end{table}

\begin{shaded}
\noindent
According to Table \ref{table:Tools2}, most of the identified tools support software development activities. 
Despite this, among the identified tools we highlight DMGame, Garuso and Gamiware that were either presented or used in more studies (4 each, in total).
DMGame and Garuso are gamified tools while Gamiware is at same time a gamified tool and a tool enabling gamification.

\end{shaded}

\subsection{RQ4. What are the challenges and difficulties of deploying gamification in software engineering?} \label{sec:analysisRQ4}

As already shown in 
Table \ref{table:StudiesType}, a substantial number of studies were classified as proposals of solutions. 
Table \ref{table:ResearchMethod} shows a high number of studies with no experimental evaluation carried out.
Indeed, the majority of the selected studies presented research in very early stage and did not address any type of difficulties and challenges on the implementation of gamification. 
More precisely, 65 studies addressed specific problems in which gamification can help, without clearly addressing the challenges and difficulties of implementing the gamification itself. On the other hand, 44 studies already pointed out difficulties and challenges for implementing gamification. The items identified in these 44 studies are listed in Table~\ref{table:Difficulties}.

\begin{table}[!ht]
\centering
\caption{Challenges and difficulties of implementing gamification.
}
\label{table:Difficulties}

\fontseries{m}
\fontshape{n}
\fontsize{7.75}{8.5}
\selectfont

\begin{tabular}{p{9cm}cp{5cm}}
\toprule
Challenges and difficulties                                                                                    & Number & Studies                                                                                          \\ \midrule
Fair assignment of points or reward and, at the same time, enjoyable to the players                            & 9      & {[}S1{]}, {[}S11{]}, {[}S28{]}, {[}S37{]}, {[}S69{]}, {[}S70{]}, {[}S74{]}, {[}S80{]}, {[}S97{]} \\ \midrule
Conducting empirical studies                                                                                   & 9      & {[}S2{]}, {[}S14{]}, {[}S22{]}, {[}S37{]}, {[}S43{]}, {[}S49{]}, {[}S73{]}, {[}S81{]}, {[}S88{]} \\ \midrule
Implementation of the tool or gamified environment                                                             & 7      & {[}S2{]}, {[}S14{]}, {[}S19{]}, {[}S65{]}, {[}S68{]}, {[}S70{]}, {[}S75{]}                       \\ \midrule
Cheats                                                                                                         & 6      & {[}S3{]}, {[}S16{]}, {[}S25{]}, {[}S26{]}, {[}S99{]}, {[}S102{]}                                 \\ \midrule
Find elements of gamification that motivate the whole team                                                     & 4      & {[}S32{]}, {[}S41{]}, {[}S77{]}, {[}S78{]}                                                       \\ \midrule
Changing the focus of the activity: Having a better score is more important than having the activity performed & 4      & {[}S12{]}, {[}S16{]}, {[}S32{]}, {[}S102{]}                                                      \\ \midrule
People demotivation                                                                                            & 4      & {[}S1{]}, {[}S32{]}, {[}S39{]}, {[}S48{]}                                                        \\ \midrule
Commitment of top-managers                                                                                     & 3      & {[}S26{]}, {[}S71{]}, {[}S97{]}                                                                  \\ \midrule
Find elements that motivate the long term                                                                      & 3      & {[}S46{]}, {[}S81{]}, {[}S97{]}                                                                  \\ \midrule
People stressed with the competitiveness generated by gamification                                             & 2      & {[}S6{]}, {[}S60{]}                                                                              \\ \midrule
Fear with data privacy                                                                                         & 2      & {[}S12{]}, {[}S37{]}                                                                             \\ \midrule
Decrease people's creativity                                                                                   & 2      & {[}S25{]}, {[}S31{]}                                                                             \\ \midrule
Integrate the tool with the company's existing tools                                                           & 2      & {[}S26{]}, {[}S47{]}                                                                             \\ \midrule
Atmosphere to be impersonal: people stop interacting in person                                                 & 2      & {[}S51{]}, {[}S90{]}                                                                             \\ \midrule
Decreases autonomy                                                                                             & 1      & {[}S35{]}                                                                                        \\ \midrule
Find professionals with experience in gamification                                                             & 1      & {[}S40{]}                                                                                        \\ \midrule
Motivating women                                                                                               & 1      & {[}S8{]}                                                                                         \\ \midrule
Segregate people into groups                                                                                   & 1      & {[}S25{]}                                                                                        \\ \bottomrule
\end{tabular}
\end{table}

As shown in the Table~\ref{table:Difficulties}, one of the main challenges reported by the studies was finding a fair assignment of points or reward. 
This was not simple for 9 studies. At the same time, since people and teams have different personalities, finding the ideal elements that motivate everyone was also cited as a difficult factor in 4 studies. The question of the motivation period was quoted in three studies, which report that, over time, the gamification elements loose their motivating effect.

Another major difficulty was the execution of experimental studies. Finding a company ready to conduct experimental studies seems to be a very important factor since we found a low number of evaluation studies (\emph{cf.} 
Table \ref{table:StudiesType}
). Next to this, it was also mentioned the lack of commitment of top-managers, turning it difficult to implement gamification in the real environment.

Other difficulty pointed out was the implementation of a tool or a gamified environment. Automation was always a pursued goal, however the construction of tools is far from being trivial, especially considering the integration between the gamified environment and the tools already existing in the company.

Regarding the level of motivation, even though one can have a good choice for  gamification elements, and consequently succeed in motivating everyone, ``excessive motivation'' should be carefully considered. 
In some cases, for example, it was related to situations where people cheated to improve their score. 
In other cases, there was an inversion of values: having a good score became more important than doing the final activity. In other cases, yet, the environment became impersonal and this led to some people to stop interacting in person and beginning to do so only through the tools.

Another point also cited as hard is the excess of competitiveness generated by gamification. In some cases, the competition made some people feeling stressed and unmotivated. 
In other cases, this excess was enough to segregate people into groups, which in turn became competitors, thus reducing the collaboration between them. Yet in other cases, people's creativity was reduced as they became more involved and focused on the elements of games.

Finally, the difficulty in finding professionals with experience in gamification, and the greater difficulty in motivating women, were also remembered among the challenge factors and difficulties to introduce gamification.

\begin{shaded}
\noindent
Thus, revisiting our research question RQ4, based on the results summarized in Table \ref{table:Difficulties}, we conclude that there are still a considerable number of challenges and difficulties inherent in the gamification implementation process. Any initiative in the direction of insertion of gamification should take into consideration the items presented as results for this research question.
\end{shaded}

\subsection{Evaluating the mapping process} \label{sec:EvaluateMappingProcess}

This section presents the application of the evaluation rubric proposed by \citet{Petersen2015}. The evaluation rubric contains 26 actions to be taken when a systematic mapping is performed.
Table \ref{table:PetersenRubrics} shows all the actions suggested in the rubric. The actions taken are marked with the \faCheckCircle~symbol and represent more than 53\% of the total suggested actions.
Tables \ref{table:PetersenRubric1}-\ref{table:PetersenRubric5} presents the scoring rubrics. The scores identified in this mapping study are highlighted with an *.

\begin{table}[!htpb]
\centering
\caption{Activities conducted in this research.}
\label{table:PetersenRubrics}

\fontseries{m}
\fontshape{n}
\fontsize{7.75}{8.5}
\selectfont

\begin{tabular}{llc}
\toprule
Phase                 & Actions                                                                 & Applied \\ \midrule
Need for map          & Motivate the need and relevance                                         & \faCheckCircle      \\
                      & Define objectives and questions                                         & \faCheckCircle      \\
                      & Consult with target audience to define questions                        & -       \\ \midrule
Study ident.          & Choosing search strategy                                                &         \\
                      & \ \ \ Snowballing                                                             & \faCheckCircle      \\
                      & \ \ \ Manual                                                                  & -       \\
                      & \ \ \ Conduct database search                                                 & \faCheckCircle      \\
                      & Develop the search                                                      &         \\
                      & \ \ \ PICO                                                                    & -       \\
                      & \ \ \ Consult librarians or experts                                           & -       \\
                      & \ \ \ Iteratively try finding more relevant papers                            & -       \\
                      & \ \ \ Keywords from known papers                                              & \faCheckCircle      \\
                      & \ \ \ Use standards, encyclopedias, and thesaurus                             & -       \\
                      & Evaluate the search                                                     &         \\
                      & \ \ \ Test-set of known papers                                                & \faCheckCircle      \\
                      & \ \ \ Expert evaluates result                                                 & -       \\
                      & \ \ \ Search web-pages of key authors                                         & -       \\
                      & \ \ \ Test–retest                                                             & -       \\
                      & Inclusion and Exclusion                                                 &         \\
                      & \ \ \ Identify objective criteria for decision                                & \faCheckCircle      \\
                      & \ \ \ Add additional reviewer, resolve disagreements between them when needed & \faCheckCircle      \\
                      & \ \ \ Decision rules                                                          & \faCheckCircle      \\ \midrule
Data extr. and class. & Extraction process                                                      &         \\
                      & \ \ \ Identify objective criteria for decision                                & -       \\
                      & \ \ \ Obscuring information that could bias                                   & -       \\
                      & \ \ \ Add additional reviewer, resolve disagreements between them when needed & \faCheckCircle      \\
                      & \ \ \ Test–retest                                                             & -       \\
                      & Classification scheme                                                   &         \\
                      & \ \ \ Research type                                                           & \faCheckCircle      \\
                      & \ \ \ Research method                                                         & \faCheckCircle      \\
                      & \ \ \ Venue type                                                              & \faCheckCircle      \\ \midrule
Validity discussion   & Validity discussion/limitations provided                                & \faCheckCircle      \\ \bottomrule
\end{tabular}
\end{table}

\begin{table}[!htpb]
\centering
\caption{Rubric: need for review.}
\label{table:PetersenRubric1}

\fontseries{m}
\fontshape{n}
\fontsize{7.75}{8.5}
\selectfont

\begin{tabular}{lp{11cm}l}
\toprule
Evaluation         & Description                                                                                          & Score \\ \midrule
No description     & The study is not motivated and the goal is not stated                                                & 0     \\
Partial evaluation & Motivations and questions are provided                                                               & 1*    \\
Full evaluation    & Motivations and questions are provided, and have been defined in correspondence with target audience & 2     \\ \bottomrule
\end{tabular}
\end{table}
\begin{table}[!htpb]
\centering
\caption{Rubric: choosing the search strategy. }
\label{table:PetersenRubric2}

\fontseries{m}
\fontshape{n}
\fontsize{7.75}{8.5}
\selectfont

\begin{tabular}{lll}
\toprule
Evaluation         & Description                                & Score \\ \midrule
No description     & Only one type of search has been conducted & 0     \\
Minimal evaluation & Two search strategies have been used       & 1*    \\
Full evaluation    & All three search strategies have been used & 2     \\ \bottomrule
\end{tabular}
\end{table}
\begin{table}[!htpb]
\centering
\caption{Rubric: evaluation of the search.}
\label{table:PetersenRubric3}

\fontseries{m}
\fontshape{n}
\fontsize{7.75}{8.5}
\selectfont

\begin{tabular}{lp{11cm}l}
\toprule
Evaluation         & Description                                                                                                                & Score \\ \midrule
No description     & No actions have been reported to improve the reliability of the search and inclusion/exclusion                             & 0     \\
Minimal evaluation & At least one action has been taken to improve the reliability of the search xor the reliability of the inclusion/exclusion & 1     \\
Partial evaluation & At least one action has been taken to improve the reliability of the search and the inclusion/ exclusion                   & 2*     \\
Full evaluation    & All actions identified have been taken                                                                                     & 3    \\ \bottomrule
\end{tabular}
\end{table}
\begin{table}[!htpb]
\centering
\caption{Rubric: extraction and classification.}
\label{table:PetersenRubric4}

\fontseries{m}
\fontshape{n}
\fontsize{7.75}{8.5}
\selectfont

\begin{tabular}{lp{11cm}l}
\toprule
Evaluation         & Description                                                                                                                                            & Score \\ \midrule
No description     & No actions have been reported to improve on the extraction process or enable comparability between studies through the use of existing classifications & 0     \\
Minimal evaluation & At least one action has been taken to increase the reliability of the extraction process                                                               & 1     \\
Partial evaluation & At least one action has been taken to increase the reliability of the extraction process, and research type and method have been classified            & 2*    \\
Full evaluation    & All actions identified have been taken                                                                                                                 & 3     \\ \bottomrule
\end{tabular}
\end{table}
\begin{table}[!htpb]
\centering
\caption{Rubric: study validity.}
\label{table:PetersenRubric5}

\fontseries{m}
\fontshape{n}
\fontsize{7.75}{8.5}
\selectfont

\begin{tabular}{lll}
\toprule
Evaluation      & Description                             & Score \\ \midrule
No description  & No threats or limitations are described & 0     \\
Full evaluation & Threats and limitations are described   & 1*    \\ \bottomrule
\end{tabular}
\end{table}

It is important to note that in this systematic mapping a quality assessment was not carried out on the selected studies. As \citet{Petersen2015} argue, the quality assessment is more essential in systematic reviews to determine the rigor and relevance of the primary studies. 
In systematic maps no quality assessment needs to be performed.
If we consider the \citet{Wieringa2005}'s research types classification, the category of solution proposals would contain papers with no empirical evidence. 
Even though such studies would not be included in a systematic review, they are important to spot trends of topics under investigation in systematic maps \citep{Petersen2015}.

\section{Research Implications} \label{Sec:Implications}

This section presents a compilation concerning the use of gamification in the software engineering context. 
In addition, it presents how gamification has been used to achieve the proposed goals.
For each presented topic in this section, we discuss the potential contexts not yet explored with the application of gamification.

\subsection{Requirements}

Regarding the \emph{requirements} topic, most of the selected studies ([S2], [S14], [S20], [S25], [S31], [S32], [S47], [S51], [S54], [S55], [S61], [S72], [S75], [S81], [S82], [S87]) explored gamification as a facilitator for the requirements elicitation process. In these studies, the gamification elements were used to encourage the participation of stakeholders, making them more committed to the elicited requirements, specially when they were geographically distributed. Consequently, more requirements were elicited, and with better quality.

The main gamification elements used to assist the requirements elicitation process were points, \daniel{leaderboards}, and levels. They were useful to motivate the stakeholders to propose new requirements, stimulate comments and discussions about the requirements aiming at improving their precision. 

In addition, gamification was explored in the requirements prioritization context. 
Some studies ([S21], [S27], [S29], [S36], [S53], [S90]) pointed to the use of gamification elements to encourage stakeholders to agree with each other, and to speed up the creation of prioritized lists of requirements.

\vspace{.3cm}
\noindent
\textsf{\emph{Research opportunities:}}
There are areas already explored within the requirements elicitation process. 
However, we noticed a lack of studies that applied gamification to motivate and improve requirements writing and documentation, and studies that deal with modeling the requirements in diagrams. 
In addition, although the management of changes and the maintenance of traceability between artifacts have been used in other activities in the software development life cycle, these activities were not explored in the studies related to requirements.

\subsection{Development}

Activities of \emph{development and coding} represent other topic related to gamification. There are several ways in which the development process can benefit from gamification. All of them aim at making the process easier, and at delivering a higher quality source code. The main gamification elements used in this scenario were points, \daniel{leaderboards}, and badges.

Most studies related to development have the direct objective of improving the quality of the developed code. Some studies ([S16], [S65], [S70]) explored the use of gamification for the creation and execution of unit tests. In studies [S49] and [S42], gamification was used to encourage the removal of technical debits and adherence to code conventions, respectively.

Inspection and code review are other major software development activities addressed by gamification. These activities, sometimes considered boring, time consuming, unattractive, but directly related to code quality, were the target of several studies ([S19], [S28], [S56], [S58], [S59], [S66], [S76]).

In addition to code review, refactoring is another aspect directly related to code quality.  Some studies ([S3], [S6]) described the use of gamification to incentive and motivate a constant code refactoring aiming at making the code lesser complex, updated, and with higher quality. These activities were often encouraged by inserting gamification into the IDEs. Besides this, some studies ([S3], [S8], [S96]) speculated for more interesting results when inserting gamification elements as close as possible to the developers' environment~(IDEs).

Some studies ([S8], [S18], [S24], [S38], [S62], [S63], [S77], [S99]) used the gamification elements to keep developers engaged and motivated while performing their programming tasks. Other ways of using gamification is to ensure a coding standardization or to adopt good programming practices. Specifically, some studies ([S35], [S37], [S41], [S74], [S96], [S102]) applied gamification to encourage desirable good practices such as an adoption of new tools and also frequent commit in version control tools. In this case, in addition to good practices, gamification also supported the configuration management of the artifacts.

Documentation, as an often ignored activity by many programmers, was also a target of gamification. A subset of the selected studies ([S1], [S78], [S79], [S94]) pointed to the use of gamification elements as a way of documenting artifacts, mostly as code comments.
Maintaining a traceability matrix between code and test cases was also reported and fits into this category.

\vspace{.3cm}
\noindent
\textsf{\emph{Research opportunities:}}
Despite all this range of presented possibilities, we noticed gamification has not been explored to support for other activities related to the development phase.
While coding activities are well supported by gamification, design activities remain disregarded. Activities related to software modeling and architecture, project-level documentation and the use of design patterns were not directly explored in the selected studies. Another relevant point regards software maintenance. While refactoring has been explored in many studies, reengineering has also been disregarded.

\subsection{Testing}

Regarding testing-related activities, the main gamification elements (points and \daniel{leaderboards}) were used as a stimulus for performing activities. Some studies ([S12], [S15], [S37], [S84], [S86]) used gamification elements to motivate and commit the team to test case generation as well as test execution.

Another important part of the testing process is defect records. Some studies ([S65], [S80], [S95]) took advantage of gamification precisely to improve and stimulate the team in bug reporting. Obtaining user feedback is another target of gamification. 
In study [S50], avatar, badges, and points were used to encourage the user to provide feedback regarding the software. In that study, one of the aspects evaluated by the feedback may be the usability of the software. 

Traceability supported by gamification was also explored in testing context. As examples of this, two studies ([S78], [S94]) explored gamification as a tool for maintaining defect tracking and also keeping the traceability matrix between code and defect updated. 

\vspace{.3cm}
\noindent
\textsf{\emph{Research opportunities:}}
We believe there is a big gap to be investigated when it comes to software testing techniques and criteria. We did not identified any study that explicitly aimed to explore the traditional techniques such as functional, structural and fault-based testing, as well as their associated test selection criteria. 
Even studies that supported test case generation and test execution did not mention any of these techniques. 
Another identified research opportunity is regarding test types. 
The studies did not directly mention the types of tests being performed. With this, we can indicate as research opportunities, but not limited to this, the following types of tests: acceptance, installation, alpha and beta, usability, reliability and evaluation, regression, performance, safety, stress, and recovery.

\subsection{Project Management}

A few studies ([S2], [S41], [S68], [S97]) showed that the management of software development projects can benefit from the use of gamification, particularly in activities related to overall project monitoring and control. 
In those studies, the project manager used gamification elements such as points, \daniel{leaderboards}, levels and challenges as a way to keep the team motivated while performing their activities. Thus the team was encouraged to use the process and tools as desired by the project manager. Since the process and tools were used as planned, metrics could provide a better overview of what is going on with the project, making decision making easier ([S2], [S68]).

In addition to these studies, several others ([S5], [S34], [S41], [S44], [S46], [S97], [S98], [S103]) pointed out the use of gamification in project task management. These studies, which dealt with project scope and time control, basically used points, \daniel{leaderboards}, levels, and badges to control the distribution and execution of project activities. These studies largely advocated the use of gamification in task tracking and logging tools.

Another important aspect of project management is project resource control and communication. In the first case, some studies ([S7], [S33]) used gamification to perform team profile mapping, and thus the project's human resources could be more appropriately managed. In the second case, other studies ([S9], [S100]) tried to solve with gamification a serious problem in projects: the collaboration and communication between team members.

\vspace{.3cm}
\noindent
\textsf{\emph{Research opportunities:}}
Despite the support to aspects already mentioned regarding project management activities, gamification could also be explored for controlling costs, acquisitions, as well as stakeholder control. The last one only appears in stakeholder control when the requirements are raised.
It is also important to remember that many studies that addressed project management to be applied in the context of software development may have been left out of this research because they did apply their research directly to other types of projects but software development-related ones.

\subsection{Support Processes}

Studies categorized as support process used gamification to support software development processes. These studies supported two aspects of software development. The majority of studies of this category advocated for support of gamification in the context of agile processes. 
Several studies ([S10], [S22], [S30], [S45], [S48], [S52], [S57], [S64], [S69], [S83], [S101], [S103]) pointed to the use of points, \daniel{leaderboards}, avatars, and badges as ways to motivate and encourage the team to perform software development activities in an agile context. 
Other studies ([S4], [S13]) used points to motivate teams to promote fast sprint delivery. In these studies, teams received points corresponding with the delivery speed of the sprint to finish the sprint before the scheduled time. 

Another issue that was also widely addressed in support activities ([S23], [S26], [S39], [S40], [S43], [S60], [S71], [S73], [S89], [S92], [S93]) was the support tool for monitoring and process improvement. In these cases, simple gamification elements such as points, \daniel{leaderboards}, and badges were used to promote an improvement in the execution of software development activities. These studies then argued that improving the execution of each activity can improve the process as a whole.

\vspace{.3cm}
\noindent
\textsf{\emph{Research opportunities:}}
Many studies made it clear and reinforced that the application of gamification was in an agile processes. 
However, there is still a lack of research regarding the use of gamification in non-agile processes.

\subsection{Other General Observations and Current Research Limitations}

One way to evaluate the current research landscape and identify new opportunities is to evaluate the future research possibilities presented by the authors of the analyzed studies.
Table \ref{table:nextResearches} summarizes such information. 
The main themes presented in this table can be grouped into two distinct groups: 
\emph{need for further empirical studies} (lines 1, 3, 5 and 6); and 
\emph{need for tool support and exploration of a wider range of elements} (lines 2 and 7).

\begin{table}[!htpb]
\centering
\caption{Upcoming research presented in the studies.}
\label{table:nextResearches}

\fontseries{m}
\fontshape{n}
\fontsize{7.75}{8.5}
\selectfont

\begin{tabular}{lc}
\toprule
Upcoming research theme                                    & Number of studies \\ \midrule
Perform most significant statistical tests                 & 43                \\
Create / Improve tools                                     & 24                \\
Implement the proposal / Analyze the proposal feasibility  & 16                \\
Do not cite upcoming studies                               & 11                \\
Conduct experiments outside the academic environment       & 10                \\
Probe the research / Mature the proposal                   & 10                \\
Explore more deep the gamification use, its elements, etc. & 8                 \\ \bottomrule
\end{tabular}
\end{table}

The first group presents a discomfort of the authors with the simplicity of their research. This is a point that was evident during the analysis of the selected studies. Most of them did not present data that allows for further analysis of the study. 
This is downside of current research because besides having low credibility, it does not allow replication and comparison of the achieved results.

It is also alarming the fact that many studies did not show any result, whilst others just presented preliminary results.
Another point that caught our attention is that many of the studies that showed some results were conducted with students in an academic environment. When something was done in the industry, a considerable number of studies were just a survey, which was always done with a very small number of participants. 
Unfortunately, this is a barrier for definite conclusions on the subject, and reinforces the idea that the community still needs to deepen research on gamification in software development.

The second group focused their concerns on the study of gamification itself, on the computer-based support for gamification application, and on the best use of the elements. 
The use of tools for the implementation of gamification was a largely addressed issue in the studies. Although the number of studies that addressed the construction and use of tools is large, practically all studies indicated difficulties for the construction of these tools.
One of the difficulties pointed out in the studies was the integration of the gamification tool with the tools already used by the teams. This is because most studies attempted to use the elements of gamification more automatically. For example, researchers did not want to keep counting participants' points manually. This really makes building tools far from trivial.

As already shown in Figure \ref{fig:ElementsXActivity}, most studies supported gamification with basic elements such as \daniel{leaderboards}, levels, badges, and (mainly) points.
In general, we can notice a simple and direct application of these elements. Therefore, it is clear that there is still a long way to go for maturing gamification in the context of software development. This makes us think of the following questions:

\begin{itemize}

    \item Can the other elements be better explored?
    
    \item Are there other elements (other than those shown in Figure \ref{fig:ElementsXActivity}) that can contribute to software development?
    
    \item Are the other gamification elements not used because they are more difficult to implement?
    
\end{itemize}

\begin{figure*}[!htp]
     \centering
     \includegraphics[width=0.9\textwidth]{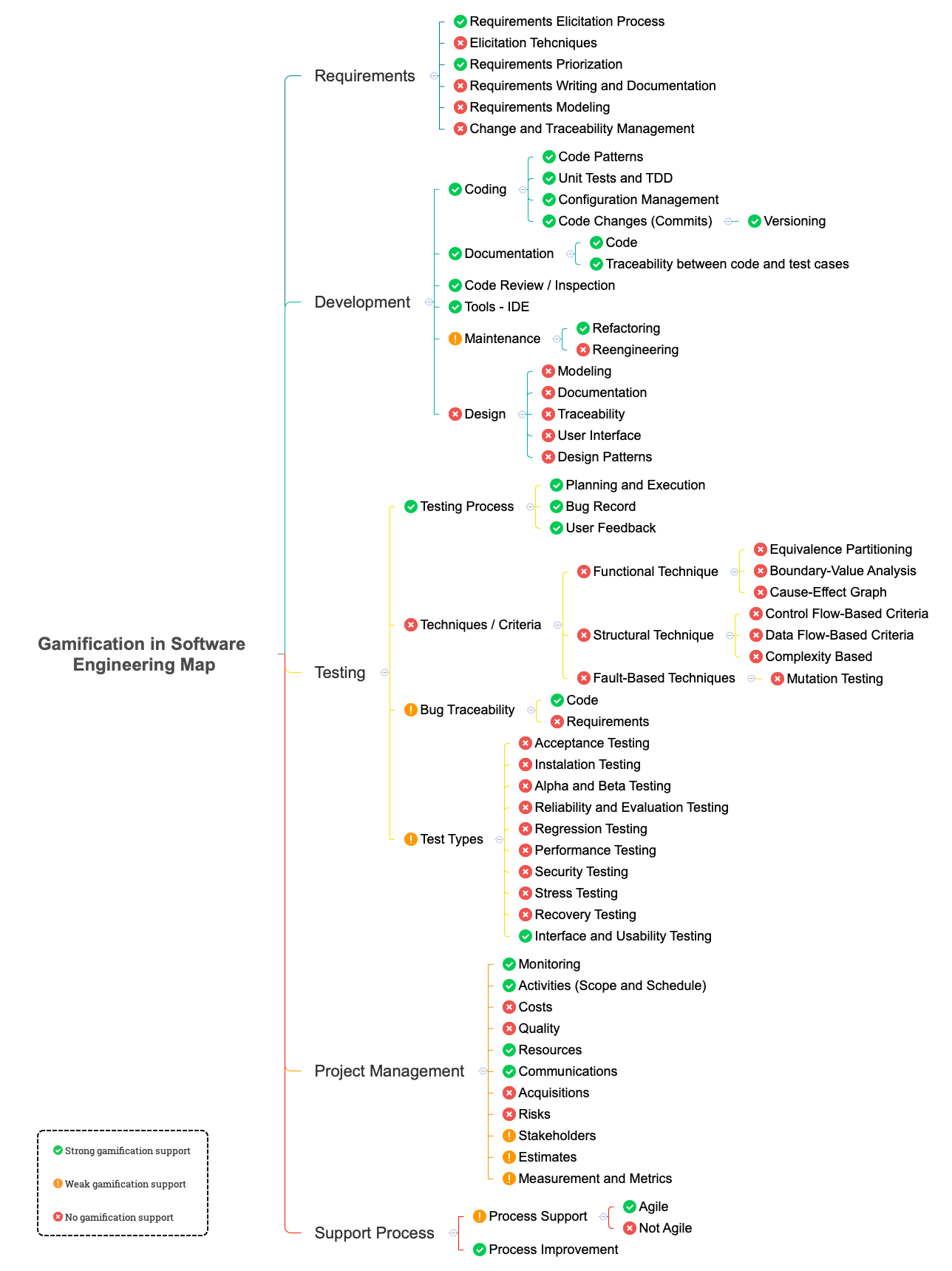}
     \caption{Map of the use of gamification in software development.}
     \label{fig:GamificationMap}
\end{figure*}

These are still open questions. They show that there is still a lot of research opportunity in this area.
In general, we can summarize this whole section in Figure~\ref{fig:GamificationMap}. 
It presents a mind map that describes which topics have already been studied with gamification, and what remains without investigation.

\vspace{.3cm}
\section{Threats to Validity} \label{Sec:Threats}

This section presents the possible threats associated to this SM based on the main threats defined by \citet{Zhou2016}.

\vspace{.25cm}\noindent\textbf{\emph{Inappropriate research question}} and \textbf{\emph{Incomprehensible venues or database:}}
In this study, the research questions may not address all aspects of gamification in software engineering. To minimize this threat, we developed a set of research questions that explore different perspectives on the use of gamification in software engineering. Regarding the search engines and their associated databases, they are well-known sources that return studies from relevant scientific events and journals on the subject under investigation. In addition, as mentioned earlier, we tried to mitigate the possibility of a study not to be indexed in the search engines by performing a round of backward snowballing \cite{wohlin2014} in the selected studies.

\vspace{.25cm}\noindent\textbf{\emph{Primary study duplication:}}
We used the Start tool \cite{Fabbri2016} to facilitate and decrease the chances of errors when dealing with duplicate papers. With the tool, duplicate papers are automatically removed from the list of papers.

\vspace{.25cm}\noindent\textbf{\emph{Incorrect search method}} and \textbf{\emph{Inappropriate or incomplete search terms in automatic search:}}
Regarding the search string, when compared with search strings used in related studies, 
our string includes a larger variety of synonyms retrieved from the already known studies. To validate the string, we checked if the string was able to return all previously known studies.

\vspace{.25cm}\noindent\textbf{\emph{Bias in study selection}} and \textbf{\emph{Identification error of primary studies in the searching process:}}
We must consider the subjective decisions that may have occurred during the selection of primary studies. Consequently, relevant studies might not have been selected. To minimize this threat, a rigorous plan was followed, which was guided by the well-defined inclusion and exclusion criteria that were carefully applied to the selected studies. This step was performed by an author and, when there was doubt about the criteria application, the decision was made with the support of additional authors. 
As a way of standardizing the application of the criteria, a round of cross-validation was carried out in which all authors received five studies and applied the criteria in the same way.
In addition, to reduce fatigue and, consequently, human error, each review session lasted, at most, four hours.

\vspace{.25cm}\noindent\textbf{\emph{Bias in data extraction}}, \textbf{\emph{Misclassification of primary studies}}, and \textbf{\emph{Subjective interpretation about the extracted data:}}
We must consider the subjective decisions that may have occurred during the extraction of data.
In this study, during the data extraction process, in which we established the relationships between the use of gamification and software engineering activities, a second author was consulted to mitigate doubts. When no consensus was found, a group discussion was performed until the conflicting ideas were sorted out.
Regarding the mapping of the CMMI-2 areas, cross-validation was carried out. The primary studies were analyzed trying to find areas of CMMI-2 in which the studies fit, and the model was also analyzed with the aim of finding, in the primary studies, the areas described in the model.
This classification was carefully done to reduce the likelihood of misclassifications.
It is also worth mentioning that the conclusions drawn in this SM were made based on the reports contained in each study. Thus, only the characteristics explicitly mentioned in the studies were mapped. Another point is that many of the studies are still incipient (\ie research in early stage). 
Consequently, perhaps not all benefits, difficulties, and characteristics have been clearly stated in the studies. This fact may compromise the conclusions reported in this SM.
At this point, we decided not to apply any quality criteria to the studies. As recommended by \citet{Petersen2015}, if we discarded studies of possibly low quality, we could erroneously conclude that the studies are more mature than they are. Besides, important information such as tools used and challenges and difficulties encountered could be overlooked by refusing some studies.

\section{Related Work} \label{Sec:RelatedWork}

This section presents other secondary studies that relate gamification and SE. 
As gamification is a new trend in SE, we believe it is important to monitor its evolution. 
Several secondary studies have been found. 
Some studies addressed gamification in SE in a general context, and most of the others addressed the use of gamification in some specific context of SE.

Starting with the specific studies, some of them addressed the use of gamification in teamwork. 
In this context, it is possible to mention the study by \citet{Munoz2017}, who performed an informal literature review to understand how gamification influences collaborative work in software development teams. 
\citeauthor{Munoz2017} found 31 primary studies and, differently from our study, they reported on the use of gamification focused on facilitating teamwork, both in the educational and in the enterprise context. They reported that, in the first case, gamification works by improving the students' skills and knowledge. In the second case, gamification improves the social interaction of the teams. In both cases, gamification attempts to improve the motivation and commitment of team members.
An interesting point raised by \citeauthor{Munoz2017} is that most of the gamification tools are web-based. 
Similarly to one of the analysis we present in our article, \citeauthor{Munoz2017} reported on  major difficulties and benefits achieved by the use of gamification. However, the key difference is that they presented such information from the teamwork viewpoint.

In the same line, we refer to \citeauthor{Hernandez2016}'s studies~\cite{Hernandez2016, Hernandez2017}. Both studies reported on the same literature review, which focused on the use of gamification as a motivating factor in software development teams. 
In total, \citeauthor{Hernandez2016} found 31~primary studies. In general, the selected studies addressed gamification as a way to create and support teamwork, with the main objective of inducing the accomplishment of the activities. Different gamification elements were found, and one of the main contributions of the work is the identification of the main factors to consider when choosing the most suitable gamification elements, namely: the environment in which they will be implemented; 
the way they are applied; and 
the target audience.

\citeauthor{Hernandez2017}, as well as \citeauthor{Munoz2017} (and unlike this work), took into account studies of gamification in the educational context. In this context, studies on the application of gamification to support teamwork in SE courses were selected. Regarding the application in the software industry, \citeauthor{Hernandez2016} identified the use of gamification as a way to improve team skills, both individually and at the team level.

In 2018, \citeauthor{Machuca-Villegas2018} published a Systematic Mapping \cite{Machuca-Villegas2018} and a Systematic Literature Review \cite{Machuca-Villegas2018b}. Both investigated the use of gamification specifically for software project management initiatives.
In the Systematic Mapping~\cite{Machuca-Villegas2018}, the authors selected 55 studies published between 2011 and 2017.
Similarly to us, \citeauthor{Machuca-Villegas2018} recognized that the gamification area is under development, and that the number of studies has increased over the years. The lack of experimental studies was also noted. 
However, unlike our study, the authors kept the focus on the project management part, and considered studies in both academic and industrial contexts.

In the Systematic Literature Review~\cite{Machuca-Villegas2018b}, for the same period, 49 studies were selected. The achieved results indicate a predominance of studies in project management areas related to integration, resources, and scoping. As other secondary studies, \citeauthor{Machuca-Villegas2018} found studies in a very preliminary stage. This indicates a need to evolve the software project management area. Both reviews by \citeauthor{Machuca-Villegas2018} took into account serious games as a manifestation of gamification. 
Studies with focus on serious games were not selected in our study.

Regarding software process improvement, \citet{GomezAlvarez2017} performed a Systematic Mapping 
to investigate the use of gamification in process improvement approaches, identifying and categorizing existing proposals. 
The low number of selected studies (13, in total) reflects how recent is the use of gamification in that context. 
Another fact that reinforces this lack of maturity in the use of gamification is the scarcity of experimental studies. As we emphasized along our article, there is a substantial number of studies that lack experimental evaluation.

There are three studies addressing the application of gamification on requirements engineering activities. 
Two of them are informal reviews~\cite{UnkelosShpigel2018, Mannov2018}, whereas the other is a Systematic Literature Review~\cite{Cursino2018}.
In one informal review,
\citet{UnkelosShpigel2018} analyzed 62 primary studies retrieved with Google Scholar. 
The study established a relationship between the elements of gamification and their use in improving the performance of the participants, as well as their participation and engagement in the requirements collection process. All of these relationships have been found and are described in more detail in our study.
In the other informal review, 
\citet{Mannov2018}
provided a compilation of gamification usage presented within the IEEE Requirements Engineering (RE) Conference from 2007 to 2017. 
Altogether, 8 studies were found. 
The author identified many serious games that used to assist in requirements engineering. 
All of these serious games were also found in our study, but were discarded because the focus of our research is on gamification only.
Another interesting point discussed by \citeauthor{Mannov2018} is that shifting a gamified part to a mobile application increases the access to data and facilitates the involvement of stakeholders.

In addition to the two aforementioned informal reviews, \citet{Cursino2018} conducted a Systematic Literature Review on the application of gamification in requirements engineering. 
As previous secondary studies, a low number of primary studies was found (8 studies, in total).
The concentration of the use of game elements on points, badges, and \daniel{leaderboards} calls the author's attention. The major consequence of applying gamification was to increase of stakeholders engagement on requirements engineering activities. Improving the cooperation and communication between teams and stakeholders, or increasing the quality of requirements, are other consequences achieved by applying game elements in the activity.

Other studies were performed by \citet{Maentylae2016} and \citet{Jesus2018}. 
In those studies, the authors addressed gamification initiatives in software testing.
\citeauthor{Maentylae2016} selected 20 studies and showed that gamification has been used for a more technical context (such as unit testing) and end-user testing (such as beta-testing and exploratory testing). 
\citeauthor{Jesus2018}, on the other hand,
selected 15 studies and showed that gamification has been used with the aim of increasing engagement and motivation, and improving skills, but without any clear focus on particular testing technique, level or process phase.
Like our study, \citeauthor{Maentylae2016}, and \citeauthor{Jesus2018}, listed several used gamification elements. Among them, the most used was points. 
Another important detail raised by \citet{Maentylae2016} regards the challenges of implementing gamification in the context of software testing.

As discussed in this article, another focus of interest in the studies is the association of gamification with agile processes. In this case, \citeauthor{Alhammad2018} conducted a Systematic Mapping that retrieved studies from 2011 to 2017~\cite{Alhammad2018}. In total, 6 studies were found. 
Just like the other secondary studies, \citeauthor{Alhammad2018} revealed that current research in the field is at the very early stages. There are very few studies and most of them reported on very preliminary results and did not provide empirical evidence of the impact of gamification on the agile process.

In addition to the secondary studies presented above, there is a group of three studies that address the use of gamification more broadly in SE and are a little closer to our study~\cite{Olgun2017, Pedreira2015, GarciaMireles2019}. 
The first is a Systematic Literature Review by \citet{Olgun2017} 
that encompassed 10 studies published from 2010 to 2017 on the application of gamification in the context of software development. 
According to the authors, one of the main benefits of gamification in the context of software development is the increase in user motivation, engagement, and collaboration. 
Moreover, they reported that gamification also helps to increase software quality and performance, and to resolve the obstacles related to human factors.
In spite of these observations, one of the main raised points is that the decision for adopting  gamification in real projects requires explicit evidence produced through empirical studies.

The second study, which is the closest to ours, is the Systematic Mapping reported by \citet{Pedreira2015}. 
The authors analyzed studies published up to June 2014 with the aim of 
characterizing the state of the art of gamification in the context of software engineering. 

\begin{figure}[!ht]
     \centering
     \includegraphics[width=1\linewidth]{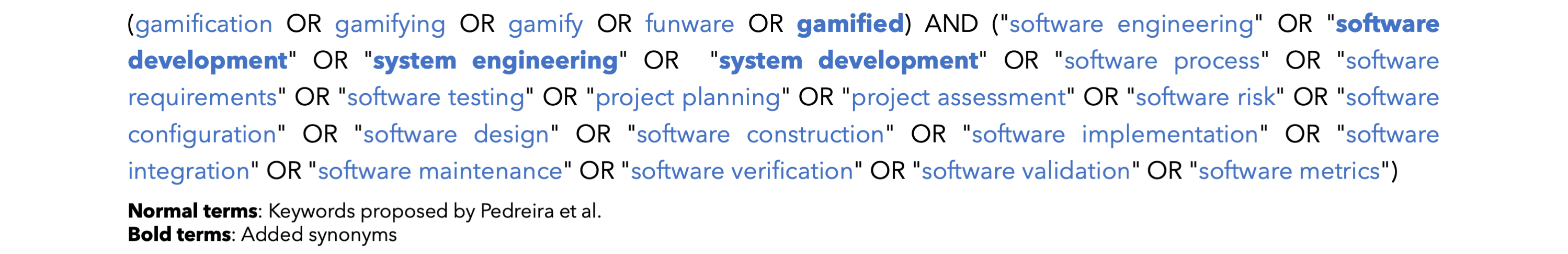}
     \vspace{-.9cm}
     \caption{Used search string.}
     \label{fig:SearchStringPedreira}
\end{figure}

The search string used in this article is based on the \citet{Pedreira2015}'s one.
For comparison purposes, in Figure~\ref{fig:SearchStringPedreira} we highlight the differences between our string the string used by them.
We added keywords --- extracted from some other previously known studies ---
with the aim of enlarging the set of retrieved studies.

\citet{Pedreira2015} analyzed primary studies published up to June 2014; that is, such study is outdated. Moreover, at that time, evidence concerning gamification applied in the SE field was mostly preliminary. As a way to confirm the need for a new secondary study on the topic of gamification in SE, we applied to \citet{Pedreira2015}' study the framework originally proposed by \citet{Garner2016} and evaluated in the context of SE by \citet{Mendes2020}. 
Specifically, \citeauthor{Mendes2020} applied the framework in the context of systematic literature reviews; despite this, we understand that the questions are generic enough to be applied also in the context of SMs. 
The results in our case signaled positively for a new secondary study (details can be checked in Appendix~A).

It is important to notice that there are substantial differences between \citeauthor{Pedreira2015}'s study and this one.
In their study, the authors aimed to identify which software processes are covered by gamification; 
for this, the ISO/IEC 12207 standard (with two additional process areas) was used. 
In our study, CMMI 2.0 practice areas were taken into account. Another point of difference between both studies is that only our study has specific research questions regarding the tools and the challenges and difficulties of implementing gamification in software engineering. 
In total, \citeauthor{Pedreira2015} analyzed 29~studies, and this set includes both academic non-academic studies --- the latter ones retrieved with the ordinary Google search engine.\footnote{https://www.google.com - accessed on 13-September-2020}
\citeauthor{Pedreira2015}'s study set and our study set have only 14 primary studies in common due to two main reasons: we did not use ordinary Google search engine, and we applied more rigorous exclusion criteria.
As results, the authors showed how recent the implementation of gamification in SE was until that moment, as well as that most of the studies focused on development activities. 
They also noticed the use of a few gamification elements (such as points and badges), as well as the lack of empirical evidence of the impact of gamification.
We highlight that many other studies in the topic have emerged since \citeauthor{Pedreira2015}'s study was published. 
Ever since, as we report in this study, the diversity of gamification elements and areas where gamification is applied have increased. 
Nevertheless, there is still little empirical evidence and a lack of reports on how to integrate gamification tools with existing tools in software companies.

The third study closest to ours is a tertiary study by \citet{GarciaMireles2019} that aimed to analyze the application of gamification in software engineering. The study retrieved 12 secondary studies published between 2015 and 2018. The majority of studies reported on the usage of points, badges, and \daniel{leaderboards} as game elements in software engineering process, software engineering methods and tools, and software engineering management. 
From the studies analyzed by \citet{GarciaMireles2019}, 7 out of 12 are described in this section: 
\cite{Machuca-Villegas2018b, Cursino2018, Jesus2018, GomezAlvarez2017, Hernandez2017, Olgun2017, Pedreira2015}.
It is worth noting that not all studies recovered by \citet{GarciaMireles2019} are described in this section as some of them addressed the application of gamification in an educational context.

For general comparison purposes, all studies described in this section are listed in Table~\ref{table:SecondaryStudies}. 
Specifically, the comparison is made between the research questions of this work (see Section~\ref{subsec:ResearchQuestions} for more details) with those of the other studies. 
The symbol \faCircle~is used if a research question is fully answered in the related study. 
If a research question is not fully answered in the related study (\eg the question is not a research question in the related study, but it somehow commented on throughout the text.) the symbol \faAdjust~is used. 
Finally, the symbol \faCircleO~is used if the research question is not answered in the related study. 
For example, \citeauthor{Hernandez2017}'s study~\cite{Hernandez2017}
has a research question about gamification elements that is very similar to our RQ1. Consequently, in Table~\ref{table:SecondaryStudies}, RQ1 is marked with the \faCircle \ symbol. 
Likewise, our RQ2 is not a research question defined in the study of \citet{Hernandez2017}, but is commented on throughout the text. Thus, RQ2 is marked with the \faAdjust \ symbol.  
It is also worth noting that most of the related studies are restricted to a specific area of software engineering such as teamworks~\cite{Hernandez2017, Hernandez2016, Munoz2017} and project management~\cite{Machuca-Villegas2018, Machuca-Villegas2018b},  
Therefore, even if a research question is marked with \faCircle, it does not bring the same results presented in our study.
Note that even in the more general studies~\cite{Olgun2017, Pedreira2015, GarciaMireles2019} (\ie the last three listed in Table~\ref{table:SecondaryStudies}) have low coverage of the research questions defined for this study, and analyzed much smaller sets of primary studies.

\begin{table*}[!htpb]
\setlength{\tabcolsep}{3pt} 
\centering
\caption{Secondary and tertiary studies related to our study.}
\label{table:SecondaryStudies}

\fontseries{m}
\fontshape{n}
\fontsize{7.75}{12}
\selectfont

\begin{tabular}{cp{2.7cm}p{4cm}ccp{1.8cm}ccccc}  
\toprule
Ref                   & Authors & Title                                                                                                                                    & Year & \begin{tabular}[c]{@{}c@{}}\# Analyzed \\ studies\end{tabular} & Main focus                   & RQ1 & RQ2 & RQ2.1 & RQ3 & RQ4 \\ \midrule  

\cite{Hernandez2017}         & \citeauthor{Hernandez2017} & A systematic literature review focused on the use of gamification in software engineering teamworks                                      & 2017 & 31                                                          & Teamworks                    & \faCircle                                                       & \faAdjust                                                       & \faCircleO                                                         & \faCircle                                                       & \faCircleO                                                       \\
\cite{Hernandez2016}         & \citeauthor{Hernandez2016} & Gamification in software engineering teamworks: A systematic literature review                                                           & 2016 & 31                                                          & Teamworks                    & \faCircle                                                       & \faAdjust                                                       & \faCircleO                                                         & \faCircleO                                                       & \faCircleO                                                       \\
\cite{Munoz2017}             & \citeauthor{Munoz2017} & State of the use of gamification elements in software development teams                                                                  & 2017 & 31                                                          & Teamworks                    & \faCircle                                                       & \faAdjust                                                       & \faCircleO                                                         & \faCircle                                                       & \faCircleO                                                       \\
\cite{Machuca-Villegas2018}  & \citeauthor{Machuca-Villegas2018} & Gamification for improving software project: Systematic mapping in project management                                                    & 2018 & 55                                                          & Project Management           & \faCircleO                                                       & \faAdjust                                                       & \faCircleO                                                         & \faCircleO                                                       & \faCircleO                                                       \\
\cite{Machuca-Villegas2018b} & \citeauthor{Machuca-Villegas2018b} & Gamification for improving software project management processes: A systematic literature review                                         & 2018 & 49                                                          & Project Management           & \faCircle                                                       & \faAdjust                                                       & \faCircleO                                                         & \faCircleO                                                       & \faCircleO                                                       \\
\cite{GomezAlvarez2017}     & \citeauthor{GomezAlvarez2017} & Gamification as strategy for software process improvement: A systematic mapping                                                          & 2017 & 13                                                          & Software process improvement & \faCircleO                                                       & \faCircleO                                                       & \faCircleO                                                         & \faCircleO                                                       & \faCircle                                                       \\
\cite{UnkelosShpigel2018}             & \citeauthor{UnkelosShpigel2018} & Leveraging Motivational Theories for Designing Gamification for RE                                                                       & 2018 & 62                                                          & Requirements Engineering     & \faCircle                                                       & \faAdjust                                                       & \faCircleO                                                         & \faCircleO                                                       & \faCircleO                                                       \\
\cite{Mannov2018}            & \citeauthor{Mannov2018} & Freud, Kierkegaard, and gamification in RE                                                                                               & 2018 & 8                                                           & Requirements Engineering     & \faCircleO                                                       & \faCircleO                                                       & \faCircleO                                                         & \faAdjust                                                       & \faCircleO                                                       \\
\cite{Cursino2018}           & \citeauthor{Cursino2018} & Gamification in Requirements Engineering: A Systematic Review                                                                            & 2018 & 8                                                           & Requirements Engineering     & \faCircle                                                       & \faCircle                                                       & \faCircleO                                                         & \faCircleO                                                       & \faCircle                                                       \\
\cite{Maentylae2016}         & \citeauthor{Maentylae2016} & Gamification of software testing - An MLR                                                                                                & 2016 & 20                                                          & Testing                      & \faCircle                                                       & \faCircleO                                                       & \faCircleO                                                         & \faCircle                                                       & \faCircle                                                       \\
\cite{Jesus2018}             & \citeauthor{Jesus2018} & Gamification in software testing: A characterization study                                                                               & 2018 & 15                                                          & Testing                      & \faCircle                                                       & \faCircle                                                       & \faCircleO                                                         & \faAdjust                                                       & \faCircleO                                                       \\
\cite{Alhammad2018}             & \citeauthor{Alhammad2018} & What is going on in agile gamification?                                                                                                  & 2018 & 6                                                           & Agile process                & \faCircle                                                       & \faCircle                                                       & \faCircleO                                                         & \faCircle                                                       & \faCircleO                                                       \\
\cite{Olgun2017}             & \citeauthor{Olgun2017}  & A systematic investigation into the use of game elements in the context of software business landscapes: a systematic literature review. & 2017 & 10                                                          & ES in general                & \faCircle                                                       & \faCircleO                                                       & \faCircleO                                                         & \faAdjust                                                       & \faCircleO                                                       \\
\cite{Pedreira2015}          & \citeauthor{Pedreira2015} & Gamification in software engineering - A systematic mapping                                                                              & 2015 & 29                                                          & ES in general                & \faCircle                                                       & \faAdjust                                                       & \faCircleO                                                         & \faCircleO                                                       & \faAdjust                                                       \\
\cite{GarciaMireles2019}           & \citeauthor{GarciaMireles2019} & Gamification in Software Engineering: A Tertiary Study                                                                                   & 2019 & 12                                                          & ES in general                & \faAdjust                                                       & \faAdjust                                                       & \faCircleO                                                         & \faCircleO                                                       & \faAdjust                                                       \\ \bottomrule
\end{tabular}
\end{table*}  

\daniel{Many things have changed since the secondary and tertiary studies described in this section have been published.}
If on the one hand, some information remains the same (\eg points as the most use gamification element, and the existence of immature and preliminary studies), on the other hand, much new information has emerged from this study. 
The main new findings of this study,  \daniel{particularly in comparison with to the study of \citet{Pedreira2015},} are:

\begin{itemize}
	\item Even with many immature and preliminary studies, \daniel{22 selected studies reported controlled experiments in this mapping};
	
	\item \daniel{The number of studies 
	selected in our mapping was substantially higher. In total, we found and analyzed 103 studies against 29 studies analyzed in the prior mapping, what represents 3.5 times more studies}. This means that the subject is still being addressed by the researchers. 
	\fabiano{We highlight that \daniel{17} studies published in journals were found in this mapping, whereas only 2 were found in the prior mapping \cite{Pedreira2015};}
	
	\item \daniel{This mapping is the first that explicitly presents a list of difficulties and challenges of implementing gamification in SE activities in general (that is, not focused on a narrow set of SE activities);} 
	
	\item More studies were found on project management \daniel{(15 against 5)}, requirements \daniel{(22 against 2)}, and configuration management \daniel{(3 against 2)};
	
	\item This mapping revealed the existence of \daniel{46 tools to support gamification};

	\item Some tools are already evolving, including \daniel{six} that are already integrated with the existing company's tools.
	
\end{itemize}

\section{Conclusion and Future Work} \label{Sec:Conclusion}

The purpose of this paper  was reporting on the results of a systematic literature mapping about gamification and its application in the context of software engineering. 
The scope of interest planned in this study, in order to reach the main goal, involved: the evolution of research on this topic based on the number of publications; the gamification elements that have been adopted, and in which software engineering activities; the benefits that have been achieved; the relationship between the activities supported by gamification and the process maturity, having the CMMI 2.0 model as reference; the tools that have been used to support gamification in certain activities; and, finally, the challenges and difficulties of deploying gamification in the software engineering context. The conclusions herein presented build from the analysis of 103 selected studies.

Based on those points of view, we noticed an annual increase in the number of publications regarding the use of gamification in software engineering activities. This indicates that this area is still new, since the first publications date from 2011. More importantly, we noticed that there is not strong empirical evidence, thus suggesting many gaps for investigations. These gaps were discussed in Section~\ref{Sec:Implications}.

Despite the existence of several gamification elements, we found out that the most used ones in the investigated context are points and \daniel{leaderboards}; both are present in the activities of Project Management, Requirements, Development, Testing, and Support Processes. 
Some major benefits achieved with the use of gamification were: 
engagement and motivation to perform the activities;
encouragement for code review tasks;  
engagement of stakeholders during requirements elicitation; and
improvement in agile processes, in which the more dynamic profile seems to make the use of gamification more propitious.

Regarding the software process maturity, the activities supported by gamification were mapped to the CMMI 2.0 Practice Areas. 
As a result, we identified gamification initiatives related mainly to: Technical solution (TS); Product integration (PI); Verification and validation (VV); and Requirements development and management (RDM). 
With respect to the use of gamified tools, this is an important matter in the gamification adoption, and 57 out of 103 selected studies explicitly cited this type of support.

Despite the fact that gamification is a recent topic of interest, several studies already pointed out difficulties and challenges for its implementation.
The main challenges reported by the studies are set a fair assignment of points or reward and, at the same time, enjoyable to the players; conducting empirical studies; and the difficulty for implementing a tool or a gamified environment. Other 15 issues were reported as difficulties faced while introducing gamification in the software development process.

As future work, we intend to investigate more deeply how gamification may be used to improve software processes, and what is the relationship between gamification and agile processes. We will also investigate what are the reasons for the lack of studies that address gamification applied to some CMMI Practice Areas, and how to solve the challenges and difficulties found in this review.

\section*{Acknowledgements} \label{Sec:Acknowledgements}
This study was financed in part by 
the Coordenação de
Aperfeiçoamento de Pessoal de Nível Superior - Brasil (CAPES) - Finance Code 001,  
and Conselho Nacional de Desenvolvimento Cient\'ifico e Tecnol\'ogico - Brasil (CNPq) - 
grants 306310/2016-3  
and 167513/2017-6. 



\begin{thebibliography}{45}
\expandafter\ifx\csname natexlab\endcsname\relax\def\natexlab#1{#1}\fi
\providecommand{\bibinfo}[2]{#2}
\ifx\xfnm\relax \def\xfnm[#1]{\unskip,\space#1}\fi
\bibitem[{{The Standish Group}(2018)}]{ChaosReport2018}
\bibinfo{author}{{The Standish Group}}, \bibinfo{title}{CHAOS Report: Decision
  Latency Theory: It Is All About the Interval}, \bibinfo{publisher}{The
  Standish Group}, \bibinfo{edition}{2} edition, \bibinfo{year}{2018}.
\bibitem[{Dubois and Tamburrelli(2013)}]{Dubois2013}
\bibinfo{author}{D.~J. Dubois}, \bibinfo{author}{G.~Tamburrelli},
\newblock \bibinfo{title}{Understanding gamification mechanisms for software
  development},
\newblock in: \bibinfo{booktitle}{Proceedings of the 9th Joint Meeting on
  Foundations of Software Engineering}, \bibinfo{publisher}{{ACM} Press},
  \bibinfo{year}{2013}, pp. \bibinfo{pages}{659--662}.
\bibitem[{Deterding et~al.(2011)Deterding, Dixon, Khaled, and
  Nacke}]{Deterding2011}
\bibinfo{author}{S.~Deterding}, \bibinfo{author}{D.~Dixon},
  \bibinfo{author}{R.~Khaled}, \bibinfo{author}{L.~Nacke},
\newblock \bibinfo{title}{{From Game Design Elements to Gamefulness}},
\newblock in: \bibinfo{booktitle}{Proceedings of the 15th International
  Academic MindTrek Conference: Envisioning Future Media Environments},
  \bibinfo{publisher}{{ACM} Press}, \bibinfo{year}{2011}, pp.
  \bibinfo{pages}{9--15}.
\bibitem[{Garc{\'{\i}}a et~al.(2017)Garc{\'{\i}}a, Pedreira, Piattini,
  Cerdeira-Pena, and Penabad}]{Garcia2017}
\bibinfo{author}{F.~Garc{\'{\i}}a}, \bibinfo{author}{O.~Pedreira},
  \bibinfo{author}{M.~Piattini}, \bibinfo{author}{A.~Cerdeira-Pena},
  \bibinfo{author}{M.~Penabad},
\newblock \bibinfo{title}{A framework for gamification in software
  engineering},
\newblock \bibinfo{journal}{Journal of Systems and Software}
  \bibinfo{volume}{132} (\bibinfo{year}{2017}) \bibinfo{pages}{21--40}.
\bibitem[{Kapp(2012)}]{Kapp2012}
\bibinfo{author}{K.~M. Kapp}, \bibinfo{title}{The Gamification of Learning and
  Instruction: Game-based Methods and Strategies for Training and Education},
  \bibinfo{publisher}{Pfeiffer \& Company}, \bibinfo{edition}{1st} edition,
  \bibinfo{year}{2012}.
\bibitem[{Monterrat et~al.(2015)Monterrat, Lavou{\'{e}}, and
  George}]{Monterrat2015}
\bibinfo{author}{B.~Monterrat}, \bibinfo{author}{E.~Lavou{\'{e}}},
  \bibinfo{author}{S.~George},
\newblock \bibinfo{title}{Toward an adaptive gamification system for learning
  environments},
\newblock in: \bibinfo{booktitle}{Proceedings of the 6th International
  Conference on Computer Supported Education}, \bibinfo{publisher}{Springer},
  \bibinfo{year}{2015}, pp. \bibinfo{pages}{115--129}.
\bibitem[{Latulipe et~al.(2015)Latulipe, Long, and Seminario}]{Latulipe2015}
\bibinfo{author}{C.~Latulipe}, \bibinfo{author}{N.~B. Long},
  \bibinfo{author}{C.~E. Seminario},
\newblock \bibinfo{title}{Structuring flipped classes with lightweight teams
  and gamification},
\newblock in: \bibinfo{booktitle}{Proceedings of the 46th Technical Symposium
  on Computer Science Education}, \bibinfo{publisher}{{ACM} Press},
  \bibinfo{year}{2015}, pp. \bibinfo{pages}{392--397}.
\bibitem[{Barata et~al.(2014)Barata, Gama, Jorge, and
  Gon{\c{c}}alves}]{Barata2014}
\bibinfo{author}{G.~Barata}, \bibinfo{author}{S.~Gama},
  \bibinfo{author}{J.~A.~P. Jorge}, \bibinfo{author}{D.~J.~V. Gon{\c{c}}alves},
\newblock \bibinfo{title}{{Relating Gaming Habits with Student Performance in a
  Gamified Learning Experience}},
\newblock in: \bibinfo{booktitle}{Proceedings of the 1st annual symposium on
  Computer-human interaction in play}, \bibinfo{publisher}{{ACM} Press},
  \bibinfo{year}{2014}, pp. \bibinfo{pages}{17--25}.
\bibitem[{Muntean(2011)}]{Muntean2011}
\bibinfo{author}{C.~Muntean},
\newblock \bibinfo{title}{Raising engagement in e-learning through
  gamification},
\newblock in: \bibinfo{booktitle}{Proceedings of the 6th International
  Conference on Virtual Learning}, \bibinfo{publisher}{Bucharest University
  Press}, \bibinfo{year}{2011}, pp. \bibinfo{pages}{323--329}.
\bibitem[{Deterding et~al.(2011)Deterding, Khaled, Nacke, and
  Dixon}]{Deterding2011b}
\bibinfo{author}{S.~Deterding}, \bibinfo{author}{R.~Khaled},
  \bibinfo{author}{L.~Nacke}, \bibinfo{author}{D.~Dixon},
\newblock \bibinfo{title}{{Gamification: Toward a Definition}},
\newblock in: \bibinfo{booktitle}{Proceedings of the 29th Conference on Human
  Factors in Computing Systems - Gamification workshop},
  \bibinfo{publisher}{{ACM} Press}, \bibinfo{year}{2011}, pp.
  \bibinfo{pages}{12--15}.
\bibitem[{Hamari et~al.(2014)Hamari, Koivisto, and Sarsa}]{Hamari2014}
\bibinfo{author}{J.~Hamari}, \bibinfo{author}{J.~Koivisto},
  \bibinfo{author}{H.~Sarsa},
\newblock \bibinfo{title}{Does gamification work? -- a literature review of
  empirical studies on gamification},
\newblock in: \bibinfo{booktitle}{Proceedings of the 47th Hawaii International
  Conference on System Sciences}, \bibinfo{publisher}{{IEEE}},
  \bibinfo{year}{2014}, pp. \bibinfo{pages}{3025--3034}.
\bibitem[{Beecham et~al.(2008)Beecham, Baddoo, Hall, Robinson, and
  Sharp}]{Beecham2008}
\bibinfo{author}{S.~Beecham}, \bibinfo{author}{N.~Baddoo},
  \bibinfo{author}{T.~Hall}, \bibinfo{author}{H.~Robinson},
  \bibinfo{author}{H.~Sharp},
\newblock \bibinfo{title}{{Motivation in Software Engineering: A Systematic
  Literature Review}},
\newblock \bibinfo{journal}{Information and Software Technology}
  \bibinfo{volume}{50} (\bibinfo{year}{2008}) \bibinfo{pages}{860--878}.
\bibitem[{Mu{\~{n}}oz et~al.(2017)Mu{\~{n}}oz, Hern{\'{a}}ndez, Mejia,
  Gasca-Hurtado, and G{\'{o}}mez-Alvarez}]{Munoz2017}
\bibinfo{author}{M.~Mu{\~{n}}oz}, \bibinfo{author}{L.~Hern{\'{a}}ndez},
  \bibinfo{author}{J.~Mejia}, \bibinfo{author}{G.~P. Gasca-Hurtado},
  \bibinfo{author}{M.~C. G{\'{o}}mez-Alvarez},
\newblock \bibinfo{title}{State of the use of gamification elements in software
  development teams},
\newblock in: \bibinfo{booktitle}{Proceedings of the 24th European Conference
  on Software Process Improvement}, \bibinfo{publisher}{Springer},
  \bibinfo{year}{2017}, pp. \bibinfo{pages}{249--258}.
\bibitem[{Singer and Schneider(2012)}]{Singer2012}
\bibinfo{author}{L.~Singer}, \bibinfo{author}{K.~Schneider},
\newblock \bibinfo{title}{It was a bit of a race: Gamification of version
  control},
\newblock in: \bibinfo{booktitle}{Proceedings of the 2nd International Workshop
  on Games and Software Engineering}, \bibinfo{publisher}{{IEEE}},
  \bibinfo{year}{2012}, pp. \bibinfo{pages}{5--8}.
\bibitem[{Fraser(2017)}]{Fraser2017}
\bibinfo{author}{G.~Fraser},
\newblock \bibinfo{title}{Gamification of software testing},
\newblock in: \bibinfo{booktitle}{Proceedings of the 12th International
  Workshop on Automation of Software Testing}, \bibinfo{publisher}{{IEEE}},
  \bibinfo{year}{2017}, pp. \bibinfo{pages}{2--7}.
\bibitem[{Dorling and McCaffery(2012)}]{Dorling2012}
\bibinfo{author}{A.~Dorling}, \bibinfo{author}{F.~McCaffery},
\newblock \bibinfo{title}{The gamification of {SPICE}},
\newblock in: \bibinfo{booktitle}{Proceedings of the 12th International
  Conference on Software Process Improvement and Capability},
  \bibinfo{publisher}{Springer}, \bibinfo{year}{2012}, pp.
  \bibinfo{pages}{295--301}.
\bibitem[{Hern{\'{a}}ndez et~al.(2017)Hern{\'{a}}ndez, Mu{\~{n}}oz,
  Mej{\'{\i}}a, Pe{\~{n}}a, Rangel, and Torres}]{Hernandez2017}
\bibinfo{author}{L.~Hern{\'{a}}ndez}, \bibinfo{author}{M.~Mu{\~{n}}oz},
  \bibinfo{author}{J.~Mej{\'{\i}}a}, \bibinfo{author}{A.~Pe{\~{n}}a},
  \bibinfo{author}{N.~Rangel}, \bibinfo{author}{C.~Torres},
\newblock \bibinfo{title}{Una revisi{\'{o}}n sistem{\'{a}}tica de la literatura
  enfocada en el uso de gamificaci{\'{o}}n en equipos de trabajo en la
  ingenier{\'{\i}}a de software},
\newblock \bibinfo{journal}{Revista Ib{\'{e}}rica de Sistemas e Tecnologias de
  Informa{\c{c}}{\~{a}}o}  (\bibinfo{year}{2017}) \bibinfo{pages}{33--50}.
\bibitem[{Hern{\'{a}}ndez et~al.(2016)Hern{\'{a}}ndez, Mu{\~{n}}oz,
  Mej{\'{\i}}a, and Pe{\~{n}}a}]{Hernandez2016}
\bibinfo{author}{L.~Hern{\'{a}}ndez}, \bibinfo{author}{M.~Mu{\~{n}}oz},
  \bibinfo{author}{J.~Mej{\'{\i}}a}, \bibinfo{author}{A.~Pe{\~{n}}a},
\newblock \bibinfo{title}{Gamification in software engineering teamworks: A
  systematic literature review},
\newblock in: \bibinfo{booktitle}{Proceedings of the 5th International
  Conference on Software Proceess Improvement}, \bibinfo{publisher}{{IEEE}},
  \bibinfo{year}{2016}, pp. \bibinfo{pages}{1--8}.
\bibitem[{Machuca-Villegas and
  Gasca-Hurtado(2018{\natexlab{a}})}]{Machuca-Villegas2018}
\bibinfo{author}{L.~Machuca-Villegas}, \bibinfo{author}{G.~P. Gasca-Hurtado},
\newblock \bibinfo{title}{Gamification for improving software project:
  Systematic mapping in project management},
\newblock in: \bibinfo{booktitle}{Proceedings of the 13th Iberian Conference on
  Information Systems and Technologies}, \bibinfo{publisher}{{IEEE}},
  \bibinfo{year}{2018}{\natexlab{a}}, pp. \bibinfo{pages}{1--6}.
\bibitem[{Machuca-Villegas and
  Gasca-Hurtado(2018{\natexlab{b}})}]{Machuca-Villegas2018b}
\bibinfo{author}{L.~Machuca-Villegas}, \bibinfo{author}{G.~P. Gasca-Hurtado},
\newblock \bibinfo{title}{Gamification for improving software project
  management processes: A systematic literature review},
\newblock in: \bibinfo{booktitle}{Proceedings of the 7th International
  Conference on Software Process Improvement}, \bibinfo{publisher}{Springer
  International Publishing}, \bibinfo{year}{2018}{\natexlab{b}}, pp.
  \bibinfo{pages}{41--54}.
\bibitem[{Gomez-Alvarez et~al.(2017)Gomez-Alvarez, Gasca-Hurtado, and
  Hincapie}]{GomezAlvarez2017}
\bibinfo{author}{M.~C. Gomez-Alvarez}, \bibinfo{author}{G.~P. Gasca-Hurtado},
  \bibinfo{author}{J.~A. Hincapie},
\newblock \bibinfo{title}{Gamification as strategy for software process
  improvement: A systematic mapping},
\newblock in: \bibinfo{booktitle}{Proceedings of the 12th Iberian Conference on
  Information Systems and Technologies}, \bibinfo{publisher}{{IEEE}},
  \bibinfo{year}{2017}, pp. \bibinfo{pages}{1--7}.
\bibitem[{Unkelos-Shpigel and Hadar(2018)}]{UnkelosShpigel2018}
\bibinfo{author}{N.~Unkelos-Shpigel}, \bibinfo{author}{I.~Hadar},
\newblock \bibinfo{title}{Leveraging motivational theories for designing
  gamification for {RE}},
\newblock in: \bibinfo{booktitle}{Proceedings of the 11th International
  Workshop on Cooperative and Human Aspects of Software Engineering},
  \bibinfo{publisher}{{ACM} Press}, \bibinfo{year}{2018}, pp.
  \bibinfo{pages}{69--72}.
\bibitem[{Mannov(2018)}]{Mannov2018}
\bibinfo{author}{N.~Mannov},
\newblock \bibinfo{title}{Freud, kierkegaard, and gamification in {RE}},
\newblock in: \bibinfo{booktitle}{Proceedings of the 1st International Workshop
  on Learning from other Disciplines for Requirements Engineering},
  \bibinfo{publisher}{{IEEE}}, \bibinfo{year}{2018}, pp.
  \bibinfo{pages}{29--32}.
\bibitem[{Cursino et~al.(2018)Cursino, Ferreira, Lencastre, Fagundes, and
  Pimentel}]{Cursino2018}
\bibinfo{author}{R.~Cursino}, \bibinfo{author}{D.~Ferreira},
  \bibinfo{author}{M.~Lencastre}, \bibinfo{author}{R.~Fagundes},
  \bibinfo{author}{J.~Pimentel},
\newblock \bibinfo{title}{Gamification in requirements engineering: A
  systematic review},
\newblock in: \bibinfo{booktitle}{Proceedings of the 11th International
  Conference on the Quality of Information and Communications Technology},
  \bibinfo{publisher}{{IEEE}}, \bibinfo{year}{2018}, pp.
  \bibinfo{pages}{119--125}.
\bibitem[{Mäntylä and Smolander(2016)}]{Maentylae2016}
\bibinfo{author}{M.~V. Mäntylä}, \bibinfo{author}{K.~Smolander},
\newblock \bibinfo{title}{Gamification of software testing - an {MLR}},
\newblock in: \bibinfo{booktitle}{Proceedings of the 17th International
  Conference on Product-Focused Software Process Improvement},
  \bibinfo{publisher}{Springer}, \bibinfo{year}{2016}, pp.
  \bibinfo{pages}{611--614}.
\bibitem[{Jesus et~al.(2018)Jesus, Ferrari, Porto, and Fabbri}]{Jesus2018}
\bibinfo{author}{G.~M. Jesus}, \bibinfo{author}{F.~C. Ferrari},
  \bibinfo{author}{D.~P. Porto}, \bibinfo{author}{S.~C. P.~F. Fabbri},
\newblock \bibinfo{title}{{Gamification in Software Testing: A Characterization
  Study}},
\newblock in: \bibinfo{booktitle}{Proceedings of the 3rd Annual ACM Brazilian
  Symposium on Systematic and Automated Software Testing},
  \bibinfo{publisher}{{ACM} Press}, \bibinfo{year}{2018}, pp.
  \bibinfo{pages}{39--48}.
\bibitem[{Alhammad and Moreno(2018)}]{Alhammad2018}
\bibinfo{author}{M.~M. Alhammad}, \bibinfo{author}{A.~M. Moreno},
\newblock \bibinfo{title}{What is going on in agile gamification?},
\newblock in: \bibinfo{booktitle}{Proceedings of the 19th International
  Conference on Agile Software Development Companion},
  \bibinfo{publisher}{{ACM} Press}, \bibinfo{year}{2018}, pp.
  \bibinfo{pages}{1--4}.
\bibitem[{Olgun et~al.(2017)Olgun, Yilmaz, Clarke, and O'Connor}]{Olgun2017}
\bibinfo{author}{S.~Olgun}, \bibinfo{author}{M.~Yilmaz}, \bibinfo{author}{P.~M.
  Clarke}, \bibinfo{author}{R.~V. O'Connor},
\newblock \bibinfo{title}{A systematic investigation into the use of game
  elements in the context of software business landscapes: A systematic
  literature review},
\newblock in: \bibinfo{booktitle}{Proceedings of the 17th International
  Conference Process Improvement and Capability Determination},
  \bibinfo{publisher}{Springer International Publishing}, \bibinfo{year}{2017},
  pp. \bibinfo{pages}{384--398}.
\bibitem[{Pedreira et~al.(2015)Pedreira, Garc{\'{\i}}a, Brisaboa, and
  Piattini}]{Pedreira2015}
\bibinfo{author}{O.~Pedreira}, \bibinfo{author}{F.~Garc{\'{\i}}a},
  \bibinfo{author}{N.~Brisaboa}, \bibinfo{author}{M.~Piattini},
\newblock \bibinfo{title}{Gamification in software engineering {\textendash} a
  systematic mapping},
\newblock \bibinfo{journal}{Information and Software Technology}
  \bibinfo{volume}{57} (\bibinfo{year}{2015}) \bibinfo{pages}{157--168}.
\bibitem[{Garc{\'{\i}}a-Mireles and Morales-Trujillo(2019)}]{GarciaMireles2019}
\bibinfo{author}{G.~A. Garc{\'{\i}}a-Mireles}, \bibinfo{author}{M.~E.
  Morales-Trujillo},
\newblock \bibinfo{title}{Gamification in software engineering: A tertiary
  study},
\newblock in: \bibinfo{booktitle}{Proceedings of the 8th International
  Conference on Software Process Improvement}, \bibinfo{publisher}{Springer
  International Publishing}, \bibinfo{year}{2019}, pp.
  \bibinfo{pages}{116--128}.
\bibitem[{{CMMI Institute }(2018)}]{CMMI2018b}
\bibinfo{author}{{CMMI Institute }}, \bibinfo{title}{{CMMI Adoption Trends -
  2018 Year End Update}}, \bibinfo{year}{2018}. \bibinfo{note}{Avaliable at:
  https://cmmiinstitute.com/resource-files/public/cmmi-adoption-trends-2018-year-end-update.
  Accessed on 23-April-2020.}
\bibitem[{{CMMI Institute}(2009)}]{CMMI-ISO}
\bibinfo{author}{{CMMI Institute}}, \bibinfo{title}{{Do ISO Standards And CMMI
  Work Together?}}, \bibinfo{year}{2009}. \bibinfo{note}{Avaliable at:
  https://cmmiinstitute.zendesk.com/hc/en-us/articles/115004587567-Do-ISO-standards-and-CMMI-work-together-.
  Accessed on 23-April-2020.}
\bibitem[{Ehsan et~al.(2010)Ehsan, Perwaiz, Arif, Mirza, and
  Ishaque}]{Ehsan2010}
\bibinfo{author}{N.~Ehsan}, \bibinfo{author}{A.~Perwaiz},
  \bibinfo{author}{J.~Arif}, \bibinfo{author}{E.~Mirza},
  \bibinfo{author}{A.~Ishaque},
\newblock \bibinfo{title}{{CMMI} / {SPICE} based process improvement},
\newblock in: \bibinfo{booktitle}{Proceedings of the 5th International
  Conference on Management of Innovation {\&} Technology},
  \bibinfo{publisher}{{IEEE}}, \bibinfo{year}{2010}, pp.
  \bibinfo{pages}{859--862}.
\bibitem[{{CMMI Institute}(2016)}]{CMMI2016}
\bibinfo{author}{{CMMI Institute}}, \bibinfo{title}{{A Guide to Scrum and
  CMMI®: Improving Agile Performance with CMMI}}, \bibinfo{year}{2016}.
  \bibinfo{note}{Avaliable at:
  https://cmmiinstitute.com/resource-files/public/marketing/document/a-guide-to-scrum-and-cmmi®-improving-agile-perfor.
  Accessed on 23-April-2020.}
\bibitem[{Petersen et~al.(2008)Petersen, Feldt, Mujtaba, and
  Mattsson}]{Petersen2008}
\bibinfo{author}{K.~Petersen}, \bibinfo{author}{R.~Feldt},
  \bibinfo{author}{S.~Mujtaba}, \bibinfo{author}{M.~Mattsson},
\newblock \bibinfo{title}{Systematic mapping studies in software engineering},
\newblock in: \bibinfo{booktitle}{Proceedings of the 12th international
  conference on Evaluation and Assessment in Software Engineering},
  \bibinfo{publisher}{BCS Learning \& Development Ltd.}, \bibinfo{year}{2008},
  pp. \bibinfo{pages}{68--77}.
\bibitem[{Petersen et~al.(2015)Petersen, Vakkalanka, and
  Kuzniarz}]{Petersen2015}
\bibinfo{author}{K.~Petersen}, \bibinfo{author}{S.~Vakkalanka},
  \bibinfo{author}{L.~Kuzniarz},
\newblock \bibinfo{title}{Guidelines for conducting systematic mapping studies
  in software engineering: An update},
\newblock \bibinfo{journal}{Information and Software Technology}
  \bibinfo{volume}{64} (\bibinfo{year}{2015}) \bibinfo{pages}{1--18}.
\bibitem[{Kitchenham and Charters(2007)}]{Kitchenham2007}
\bibinfo{author}{B.~Kitchenham}, \bibinfo{author}{S.~Charters},
  \bibinfo{title}{Guidelines for performing Systematic Literature Reviews in
  Software Engineering}, \bibinfo{type}{Technical Report}
  \bibinfo{number}{Technical Report EBSE 2007-001}, School of Computer Science
  and Mathematics, Keele University, \bibinfo{year}{2007}.
\bibitem[{Wohlin(2014)}]{wohlin2014}
\bibinfo{author}{C.~Wohlin},
\newblock \bibinfo{title}{Guidelines for snowballing in systematic literature
  studies and a replication in software engineering},
\newblock in: \bibinfo{booktitle}{Proceedings of the 18th International
  Conference on Evaluation and Assessment in Software Engineering},
  \bibinfo{publisher}{{ACM} Press}, \bibinfo{year}{2014}, pp.
  \bibinfo{pages}{321--330}.
\bibitem[{Fabbri et~al.(2016)Fabbri, Silva, Hernandes, Octaviano, Thommazo, and
  Belgamo}]{Fabbri2016}
\bibinfo{author}{S.~Fabbri}, \bibinfo{author}{C.~Silva},
  \bibinfo{author}{E.~Hernandes}, \bibinfo{author}{F.~Octaviano},
  \bibinfo{author}{A.~D. Thommazo}, \bibinfo{author}{A.~Belgamo},
\newblock \bibinfo{title}{Improvements in the {StArt} tool to better support
  the systematic review process},
\newblock in: \bibinfo{booktitle}{Proceedings of the 20th International
  Conference on Evaluation and Assessment in Software Engineering},
  \bibinfo{publisher}{{ACM} Press}, \bibinfo{year}{2016}, pp.
  \bibinfo{pages}{1--5}.
\bibitem[{Wieringa et~al.(2005)Wieringa, Maiden, Mead, and
  Rolland}]{Wieringa2005}
\bibinfo{author}{R.~Wieringa}, \bibinfo{author}{N.~Maiden},
  \bibinfo{author}{N.~Mead}, \bibinfo{author}{C.~Rolland},
\newblock \bibinfo{title}{Requirements engineering paper classification and
  evaluation criteria: a proposal and a discussion},
\newblock \bibinfo{journal}{Requirements Engineering} \bibinfo{volume}{11}
  (\bibinfo{year}{2005}) \bibinfo{pages}{102--107}.
\bibitem[{Werbach and Hunter(2012)}]{werbach2012}
\bibinfo{author}{K.~Werbach}, \bibinfo{author}{D.~Hunter}, \bibinfo{title}{For
  the win: How game thinking can revolutionize your business},
  \bibinfo{publisher}{Wharton Digital Press}, \bibinfo{year}{2012}.
\bibitem[{{CMMI Institute}(2018)}]{CMMI2018}
\bibinfo{author}{{CMMI Institute}}, \bibinfo{title}{Cmmi development v2.0 quick
  reference guide}, \bibinfo{year}{2018}.
\bibitem[{Zhou et~al.(2016)Zhou, Jin, Zhang, Li, and Huang}]{Zhou2016}
\bibinfo{author}{X.~Zhou}, \bibinfo{author}{Y.~Jin},
  \bibinfo{author}{H.~Zhang}, \bibinfo{author}{S.~Li},
  \bibinfo{author}{X.~Huang},
\newblock \bibinfo{title}{A map of threats to validity of systematic literature
  reviews in software engineering},
\newblock in: \bibinfo{booktitle}{Proceedings of the 23rd Asia-Pacific Software
  Engineering Conference}, \bibinfo{publisher}{{IEEE}}, \bibinfo{year}{2016},
  pp. \bibinfo{pages}{153--160}.
\bibitem[{Garner et~al.(2016)Garner, Hopewell, Chandler, MacLehose,
  Schünemann, Akl, Beyene, Chang, Churchill, Dearness, Guyatt, Lefebvre,
  Liles, Marshall, Garc{\'{\i}}a, Mavergames, Nasser, Qaseem, Sampson,
  Soares-Weiser, Takwoingi, Thabane, Trivella, Tugwell, Welsh, and
  Wilson}]{Garner2016}
\bibinfo{author}{P.~Garner}, \bibinfo{author}{S.~Hopewell},
  \bibinfo{author}{J.~Chandler}, \bibinfo{author}{H.~MacLehose},
  \bibinfo{author}{H.~J. Schünemann}, \bibinfo{author}{E.~A. Akl},
  \bibinfo{author}{J.~Beyene}, \bibinfo{author}{S.~Chang},
  \bibinfo{author}{R.~Churchill}, \bibinfo{author}{K.~Dearness},
  \bibinfo{author}{G.~Guyatt}, \bibinfo{author}{C.~Lefebvre},
  \bibinfo{author}{B.~Liles}, \bibinfo{author}{R.~Marshall},
  \bibinfo{author}{L.~M. Garc{\'{\i}}a}, \bibinfo{author}{C.~Mavergames},
  \bibinfo{author}{M.~Nasser}, \bibinfo{author}{A.~Qaseem},
  \bibinfo{author}{M.~Sampson}, \bibinfo{author}{K.~Soares-Weiser},
  \bibinfo{author}{Y.~Takwoingi}, \bibinfo{author}{L.~Thabane},
  \bibinfo{author}{M.~Trivella}, \bibinfo{author}{P.~Tugwell},
  \bibinfo{author}{E.~Welsh}, \bibinfo{author}{E.~C. Wilson},
\newblock \bibinfo{title}{When and how to update systematic reviews: consensus
  and checklist},
\newblock \bibinfo{journal}{{BMJ}}  (\bibinfo{year}{2016})
  \bibinfo{pages}{1--10}.
\bibitem[{Mendes et~al.(2020)Mendes, Wohlin, Felizardo, and
  Kalinowski}]{Mendes2020}
\bibinfo{author}{E.~Mendes}, \bibinfo{author}{C.~Wohlin},
  \bibinfo{author}{K.~Felizardo}, \bibinfo{author}{M.~Kalinowski},
\newblock \bibinfo{title}{When to update systematic literature reviews in
  software engineering},
\newblock \bibinfo{journal}{Journal of Systems and Software}
  \bibinfo{volume}{167} (\bibinfo{year}{2020}) \bibinfo{pages}{1--24}.

\end{thebibliography}

\appendix

\section*{Appendix A. Is a new secondary study needed?}

In the application of the framework~\cite{Garner2016, Mendes2020}, we used as a baseline the study of \citet{Pedreira2015} -- as it is the closest study to what we wanted -- and executed the proposed checklist. The checklist and responses are shown in the table below. 
The table reveals that the responses for the three questions in step 1 are YES, which enables us to proceed to the next step. At least one YES response in step 2 enables us to move on to the last step. In the third step, at least one YES response gives us confirmation to proceed with the study update (or, as in our case, for a new study).

\begin{center}

\vspace{.7cm}

\fontseries{m}
\fontshape{n}
\fontsize{7.75}{8.5}
\selectfont

\begin{tabular}{lc}
\toprule
Framework Step                                                                                             & Response \\ \midrule
Step 1.a - Does the published SLR still address a current question?                                        & YES      \\
Step 1.b - Has the SLR had good access or use?                                                             & YES      \\
Step 1.c - Has the SLR used valid methods and was well-conducted?                                          & YES      \\
Step 2.a - Are there any new relevant methods?                                                             & YES      \\
Step 2.b - Are there any new studies, or new information?                                                  & YES      \\
Step 3.a - Will the adoption of new methods change the findings, conclusions or credibility?               & YES      \\
Step 3.b - Will the inclusion of new studies/information/data change findings, conclusions or credibility? & YES      \\ \bottomrule
\end{tabular}

\vspace{.7cm}

\end{center}

\section*{Appendix B. List of primary studies} 

\centering

\fontseries{m}
\fontshape{n}
\fontsize{7.75}{8.5}
\selectfont

\begin{longtable}{p{0.5cm}p{4cm}p{5.3cm}p{0.5cm}p{4cm}}
\toprule
Id         & Author                                                                                                                                                                              & Title                                                                                                                                     & Year & Journal/Event                                                                                     \\ \midrule
{[}S1{]}   & C. R. Prause and J. Nonnen and M. Vinkovits                                                                                                                                         & A Field Experiment on Gamification of Code Quality in Agile Development                                                                   & 2012 & Workshop Psychology of Programming Interest Group                                                 \\
{[}S2{]}   & F. Garc{\'{\i}}a and O. Pedreira and M. Piattini and A. Cerdeira-Pena and M. Penabad                                                              & A framework for gamification in software engineering                                                                                      & 2017 & Journal of Systems and Software                                                                   \\
{[}S3{]}   & L. Elezi and S. Sali and S. Demeyer and A. Murgia and J. P{\`{e}}rez                                                                                             & A game of refactoring: Studying the Impact of Gamification in Software Refactoring                                                        & 2016 & Scientific Workshop Proceedings of XP2016                                                         \\
{[}S4{]}   & V. S. Sharma and V. Kaulgud and P. Duraisamy                                                                                                                                        & A gamification approach for distributed agile delivery                                                                                    & 2016 & International Workshop on Games and Software Engineering                                          \\
{[}S5{]}   & G.P. Gasca-Hurtado and M.C. Gómez-Alvarez and M. Muñoz and A. Peña                                                                                                                  & A Gamified Proposal for Software Risk Analysis in Agile Methodologies                                                                     & 2019 & European Conference on Software Process Improvement                                               \\
{[}S6{]}   & S. Arai and K. Sakamoto and H. Washizaki and Y. Fukazawa                                                                                                                            & A gamified tool for motivating developers to remove warnings of bug pattern tools                                                         & 2014 & Workshop on Empirical Software Engineering in Practice                                            \\
{[}S7{]}   & M. Mu{\~{n}}oz and L. Hern{\'{a}}ndez and J. Mejia and A. Pe{\~{n}}a and N. Rangel and C. Torres and G. Sauberer & A model to integrate highly effective teams for software development                                                                      & 2017 & European Conference on Software Process Improvement                                               \\
{[}S8{]}   & T. Barik and E. Murphy-Hill and T. Zimmermann                                                                                                                                       & A perspective on blending programming environments and games: Beyond points, badges, and leaderboards                                     & 2016 & Symposium on Visual Languages and Human-Centric Computing                                         \\
{[}S9{]}   & F. Steffens and S. Marczak and F. F. Filho and C. Treude and C. R. B. de Souza                                                                                                      & A preliminary evaluation of a gamification framework to jump start collaboration behavior change                                          & 2017 & International Workshop on Cooperative and Human Aspects of Software Engineering                   \\
{[}S10{]}  & M. Yilmaz and R. Oconnor                                                                                                                                                            & A scrumban integrated gamification approach to guide software process improvement: A Turkish case study                                   & 2016 & Tehnicki vjesnik - Technical Gazette                                                              \\
{[}S11{]}  & I. Chow and L. Huang                                                                                                                                                                & A software gamification model for cross-cultural software development teams                                                               & 2017 & International Conference on Management Engineering, Software Engineering and Service Sciences     \\
{[}S12{]}  & S.A. Scherr and F. Elberzhager and K. Holl                                                                                                                                          & Acceptance testing of mobile applications: Automated emotion tracking for large user groups                                               & 2018 & International Conference on Mobile Software Engineering and Systems                               \\
{[}S13{]}  & V. S. Sharma and V. Kaulgud                                                                                                                                                         & Agile workbench: Tying people, process, and tools in distributed agile delivery                                                           & 2016 & International Conference on Global Software Engineering                                           \\
{[}S14{]}  & L. Piras                                                                                                                                                                            & Agon: a Gamification-Based Framework for Acceptance Requirements                                                                          & 2018 & Università degli Studi di Trento                                                                  \\
{[}S15{]}  & J.I. Galván-Tejada and J.G. Arceo-Olague and J.M. Celaya-Padilla and R. Solis-Robles                                                                                                & An approach to make software testing for users with down syndrome a little more pleasant                                                  & 2018 & International Conference on Human-Computer Interaction                                            \\
{[}S16{]}  & M. Johansson and E. Ivarsson                                                                                                                                                        & An Experiment on the Effectiveness of Unit Testing when Introducing Gamification                                                          & 2014 & Chalmers University of Technology                                                                 \\
{[}S17{]}  & A. McClean                                                                                                                                                                          & An Exploration of the Use of Gamification in Agile Software Development                                                                   & 2015 & Technological University Dublin                                                                   \\
{[}S18{]}  & M. Tsunoda and H. Yumoto                                                                                                                                                            & Applying Gamification and Posing to Software Development                                                                                  & 2018 & Asia-Pacific Software Engineering Conference                                                      \\
{[}S19{]}  & S. K. Sripada and Y. R. Reddy and S. Khandelwal                                                                                                                                     & Architecting an extensible framework for gamifying software engineering concepts                                                          & 2016 & India Software Engineering Conference                                                             \\
{[}S20{]}  & N. U. Shpigel                                                                                                                                                                       & Be ahead of the game: Gamification for inclusive RE                                                                                       & 2018 & Workshop on Facilitating Inclusive Requirements Engineering                                       \\
{[}S21{]}  & M. Z. H. Kolpondinos and M. Glinz                                                                                                                                                   & Behind Points and Levels – The Influence of Gamification Algorithms on Requirements Prioritization                                        & 2017 & International Requirements Engineering Conference                                                 \\
{[}S22{]}  & P.S. Neto and D.B. Medeiros and I. Ibiapina and O.C. Da Costa Castro                                                                                                                & Case study of the introduction of game design techniques in software development                                                          & 2019 & IET Software                                                                                      \\
{[}S23{]}  & A. Uskarci and O. Demir{\"o}rs                                                                                                                                     & Causes of Continuity and Participation Problems in Process Improvement with Staged Maturity Models                                        & 2015 & International Conference on Software Process Improvement and Capability Determination             \\
{[}S24{]}  & T. D. LaToza and W. Ben Towne and A. van der Hoek and J. D. Herbsleb                                                                                                                & Crowd development                                                                                                                         & 2013 & International Workshop on Cooperative and Human Aspects of Software Engineering                   \\
{[}S25{]}  & R. Snijders and F. Dalpiaz and M. Hosseini and A. Shahri and R. Ali                                                                                                                 & Crowd-centric Requirements Engineering                                                                                                    & 2014 & International Conference on Utility and Cloud Computing                                           \\
{[}S26{]}  & E. Herranz and R. Colomo-Palacios and A. Al-Barakati                                                                                                                                & Deploying a gamification framework for software process improvement: Preliminary results                                                  & 2017 & European Conference on Software Process Improvement                                               \\
{[}S27{]}  & L. {Piras} and D. {Dellagiacoma} and A. {Perini} and A. {Susi} and P. {Giorgini} and J. {Mylopoulos}                                                                    & Design Thinking and Acceptance Requirements for Designing Gamified Software                                                               & 2019 & International Conference on Research Challenges in Information Science                            \\
{[}S28{]}  & W. {Fracz} and J. {Dajda}                                                                                                                                                       & Developers' game: A preliminary study concerning a tool for automated developers assessment                                               & 2018 & International Conference on Software Maintenance and Evolution                                    \\
{[}S29{]}  & F. Kifetew and D. Munante and A. Perini and A. Susi and A. Siena and P. Busetta                                                                                                     & DMGame: A Gamified Collaborative Requirements Prioritisation Tool                                                                         & 2017 & International Requirements Engineering Conference                                                 \\
{[}S30{]}  & W. Sisomboon and N. Phakdee and N. Denwattana                                                                                                                                       & Engaging and Motivating Developers by Adopting Scrum Utilizing Gamification                                                               & 2019 & International Conference on Information Technology                                                \\
{[}S31{]}  & P. Lombriser                                                                                                                                                                        & Engaging Stakeholders in Scenario-Based Requirements Engineering with Gamification                                                        & 2015 & Utrecht University                                                                                \\
{[}S32{]}  & F. Dalpiaz and R. Snijders and S. Brinkkemper and M. Hosseini and A. Shahri and R. Ali                                                                                              & Engaging the crowd of stakeholders in requirements engineering via gamification                                                           & 2016 & Gamification                                                                                      \\
{[}S33{]}  & M. Mu{\~{n}}oz and J. Mejia and A. Pe{\~{n}}a and N. Rangel                                                                         & Establishing Effective Software Development Teams: An Exploratory Model                                                                   & 2016 & European Conference on Software Process Improvement                                               \\
{[}S34{]}  & {\c{C}}. Usfekes and M. Yilmaz and E. Tuzun and P. M. Clarke and R. V. O'Connor                                                                                  & Examining reward mechanisms for effective usage of application lifecycle management tools                                                 & 2017 & European Conference on Software Process Improvement                                               \\
{[}S35{]}  & W. Snipes and A. R. Nair and E. Murphy-Hill                                                                                                                                         & Experiences gamifying developer adoption of practices and tools                                                                           & 2014 & International Conference on Software Engineering                                                  \\
{[}S36{]}  & A. {Perini} and N. {Seyff} and M. {Stade} and A. {Susi}                                                                                                                   & Exploring RE knowledge for gamification: Can RE achieve a high score?                                                                     & 2018 & International Workshop on Affective Computing for Requirements Engineering                        \\
{[}S37{]}  & M. Foucault and X. Blanc and J.-R. Falleri and M.-A. Storey                                                                                                                         & Fostering good coding practices through individual feedback and gamification: an industrial case study                                    & 2019 & Empirical Software Engineering                                                                    \\
{[}S38{]}  & R. Minelli and A. Mocci and M. Lanza                                                                                                                                                & Free Hugs - Praising Developers for Their Actions                                                                                         & 2015 & International Conference on Software Engineering                                                  \\
{[}S39{]}  & M. Ruiz and M. Trinidad and A. Calderón                                                                                                                                             & Gamification and functional prototyping to support motivation towards software process improvement                                        & 2016 & International Conference on Product-Focused Software Process Improvement                          \\
{[}S40{]}  & E. Herranz and R. C. Palacios and A. A. Seco and M. Yilmaz                                                                                                                          & Gamification as a disruptive factor in software process improvement initiatives                                                           & 2014 & Journal of Universal Computer Science                                                             \\
{[}S41{]}  & D. Silva and A. Coelho and C. Duarte and P. C. Henriques                                                                                                                            & Gamification at scraim                                                                                                                    & 2016 & International Conference on Serious Games, Interaction, and Simulation                            \\
{[}S42{]}  & C. R. Prause and M. Jarke                                                                                                                                                           & Gamification for enforcing coding conventions                                                                                             & 2015 & Joint Meeting on Foundations of Software Engineering                                              \\
{[}S43{]}  & E. {Herranz} and J. G. {Guzmán} and A. {de Amescua-Seco} and X. {Larrucea}                                                                                                  & Gamification for software process improvement: A practical approach                                                                       & 2019 & IET Software                                                                                      \\
{[}S44{]}  & V. Platonova and S. Berzisa                                                                                                                                                         & Gamification framework for software development project processes                                                                         & 2019 & International Scientific Practical Conference                                                     \\
{[}S45{]}  & M. Češka                                                                                                                                                                            & Gamification in the SCRUM Software Development Framework                                                                                  & 2015 & Masaryk University                                                                                \\
{[}S46{]}  & D. Ašeriškis and R. Damaševičius                                                                                                                                                    & Gamification of a project management system                                                                                               & 2014 & International Conference on Advances in Computer-Human Interactions                               \\
{[}S47{]}  & L. Piras and E. Paja and P. Giorgini and J. Mylopoulos and R. Cuel and D. Ponte                                                                                                     & Gamification solutions for software acceptance: A comparative study of Requirements Engineering and Organizational Behavior techniques    & 2017 & International Conference on Research Challenges in Information Science                            \\
{[}S48{]}  & I. M. Pereira and V. J.P. Amorim and M. A. Cota and G. C. Gonçalves                                                                                        & Gamification Use in Agile Project Management: An Experience Report                                                                        & 2017 & Brazilian Workshop on Agile Methods                                                               \\
{[}S49{]}  & M. Foucault and X. Blanc and M.-A. Storey and J.-R. Falleri and C. Teyton                                                                                                           & Gamification: a Game Changer for Managing Technical Debt? A Design Study                                                                  & 2018 & arXiv                                                                                             \\
{[}S50{]}  & M. Almaliki and N. Jiang and R. Ali and F. Dalpiaz                                                                                                                                  & Gamified culture-aware feedback acquisition                                                                                               & 2014 & International Conference on Utility and Cloud Computing                                           \\
{[}S51{]}  & P. Lombriser and F. Dalpiaz and G. Lucassen and S. Brinkkemper                                                                                                                      & Gamified Requirements Engineering: Model and Experimentation                                                                              & 2016 & International Working Conference on Requirements Engineering: Foundation for Software Quality     \\
{[}S52{]}  & S. Hermanto and E.R. Kaburuan and N. Legowo                                                                                                                                         & Gamified SCRUM Design in Software Development Projects                                                                                    & 2018 & International Conference on Orange Technologies                                                   \\
{[}S53{]}  & F. M. Kifetew and D. Munante and A. Perini and A. Susi and A. Siena and P. Busetta and D. Valerio                                                                                   & Gamifying Collaborative Prioritization: Does Pointsification Work?                                                                        & 2017 & International Requirements Engineering Conference                                                 \\
{[}S54{]}  & C. Ribeiro and C. Farinha and J. Pereira and M. M.  Silva                                                                                                                           & Gamifying requirement elicitation: Practical implications and outcomes in improving stakeholders collaboration                            & 2014 & Entertainment Computing                                                                           \\
{[}S55{]}  & M.E.A. Tebib                                                                                                                                                                        & Gamifying requirements engineering for better practice                                                                                    & 2019 & CEUR Workshop                                                                                     \\
{[}S56{]}  & N. Unkelos-Shpigel. and I. Hadar                                                                                                                                                    & Gamifying software development environments using cognitive principles                                                                    & 2015 & International Conference on Advanced Information Systems Engineering                              \\
{[}S57{]}  & R. Marques and G. Costa and M. M. Silva and P. Goncalves                                                                                                                            & Gamifying software development scrum projects                                                                                             & 2017 & International Conference on Virtual Worlds and Games for Serious Applications                     \\
{[}S58{]}  & N. Unkelos-Shpigel and I. Hadar                                                                                                                                                     & Gamifying software engineering tasks based on cognitive principles: The case of code review                                               & 2015 & International Workshop on Cooperative and Human Aspects of Software Engineering                   \\
{[}S59{]}  & L.N.Q. Do and E. Bodden                                                                                                                                                             & Gamifying static analysis                                                                                                                 & 2018 & European Software Engineering Conference and Symposium on the Foundations of Software Engineering \\
{[}S60{]}  & E. Herranz and R. Colomo-Palacios and A. A. Seco                                                                                                                                    & Gamiware: A gamification platform for software process improvement                                                                        & 2015 & European Conference on Software Process Improvement                                               \\
{[}S61{]}  & M. Z. Kolpondinos and M. Glinz                                                                                                                                                      & GARUSO: a gamification approach for involving stakeholders outside organizational reach in requirements engineering                       & 2019 & Requirements Engineering                                                                          \\
{[}S62{]}  & M. {Tsunoda} and T. {Hayashi} and S. {Sasaki} and K. {Yoshigami} and H. {Uwano} and K. {Matsumoto}                                                                      & How do gamification rules and personal preferences affect coding?                                                                         & 2018 & International Workshop on Empirical Software Engineering in Practice                              \\
{[}S63{]}  & K. Yoshigami and T. Hayashi and M. Tsunoda and H. Uwano and S. Sasaki and K. Matsumoto                                                                                              & How does time conscious rule of gamification affect coding and review?                                                                    & 2019 & IEICE Transactions on Information and Systems                                                     \\
{[}S64{]}  & A. Poth and M. Kottke                                                                                                                                                               & How to Assure Agile Method and Process Alignment in an Organization?                                                                      & 2018 & European Conference on Software Process Improvement                                               \\
{[}S65{]}  & T. D. Sasso and A. Mocci and M. Lanza and E. Mastrodicasa                                                                                                                           & How to gamify software engineering                                                                                                        & 2017 & International Conference on Software Analysis, Evolution and Reengineering                        \\
{[}S66{]}  & S. Khandelwal and S. K. Sripada and Y. R. Reddy                                                                                                                                     & Impact of Gamification on Code Review Process: An Experimental Study                                                                      & 2017 & Innovations in Software Engineering Conference                                                    \\
{[}S67{]}  & S. Maro and E. Sundklev and C.-O. Persson and G. Liebel and J.-P. Steghöfer                                                                                                         & Impact of Gamification on Trace Link Vetting: A Controlled Experiment                                                                     & 2019 & International Working Conference on Requirements Engineering                                      \\
{[}S68{]}  & D. Porto and F. Ferrari and S. Fabbri                                                                                                                                               & Improving project manager decision with gamification                                                                                      & 2019 & Brazilian Symposium on Software Quality                                                           \\
{[}S69{]}  & R. Marques and G. Costa and M.M. Da Silva and D. Gonçalves and P. Gonçalves                                                                                                         & Improving scrum adoption with gamification                                                                                                & 2018 & Americas Conference on Information Systems                                                        \\
{[}S70{]}  & D. Arnarsson and Í. Jóhannesson                                                                                                                                                     & Improving Unit Testing Practices With the Use of Gamification                                                                             & 2018 & Chalmers University of Technology                                                                 \\
{[}S71{]}  & G. Ivan and P. Carla and C.-M.J. Antonio                                                                                                                                            & Introducing gamification to increase staff involvement and motivation when conducting SPI initiatives in small-sized software enterprises & 2019 & IET Software                                                                                      \\
{[}S72{]}  & N. Unkelos-Shpigel and I. Hadar                                                                                                                                                     & Inviting everyone to play: Gamifying collaborative requirements engineering                                                               & 2015 & International Workshop on Empirical Requirements Engineering                                      \\
{[}S73{]}  & E. Herranz and R. Colomo-Palacios                                                                                                                                                   & Is Gamification a Way to a Softer Software Process Improvement? A Preliminary Study of Success Factors                                    & 2018 & European Conference on Software Process Improvement                                               \\
{[}S74{]}  & L. Singer and K. Schneider                                                                                                                                                          & It was a bit of a race: Gamification of version control                                                                                   & 2012 & International Workshop on Games and Software Engineering                                          \\
{[}S75{]}  & J. Fernandes and D. Duarte and C. Ribeiro and C. Farinha and J. M. Pereira and M. M. Silva                                                                                          & IThink : A game-based approach towards improving collaboration and participation in requirement elicitation                               & 2012 & International Conference on Games and Virtual Worlds for Serious Applications                     \\
{[}S76{]}  & N. Unkelos-Shpigel and I. Hadar                                                                                                                                                     & Let's make it fun: Gamifying and formalizing Code review                                                                                  & 2016 & International Conference on Evaluation of Novel Software Approaches to Software Engineering       \\
{[}S77{]}  & E. S. Mastrodicasa                                                                                                                                                                  & Ludus opus proficit - A gamification framework for software engineering                                                                   & 2014 & Università della Svizzera Italiana                                                                \\
{[}S78{]}  & R. M. Parizi                                                                                                                                                                        & On the gamification of human-centric traceability tasks in software testing and coding                                                    & 2016 & International Conference on Software Engineering Research, Management and Applications            \\
{[}S79{]}  & R. Sukale and M. Pfaff                                                                                                                                                              & QuoDocs: Improving Developer Engagement in Software Documentation Through Gamification                                                    & 2014 & International Conference on Computer Human Interaction                                            \\
{[}S80{]}  & N. K. Nagwani and S. Verma                                                                                                                                                          & Rank-Me: A Java Tool for Ranking Team Members in Software Bug Repositories                                                                & 2012 & Journal of Software Engineering and Applications                                                  \\
{[}S81{]}  & R. Snijders and F. Dalpiaz and S. Brinkkemper and M. Hosseini and R. Ali and A. Ozum                                                                                                & REfine: A gamified platform for participatory requirements engineering                                                                    & 2015 & International Workshop on Crowd-Based Requirements Engineering                                    \\
{[}S82{]}  & A. Alexandrova and L. Rapanotti                                                                                                                                                     & Requirements analysis gamification in legacy system replacement projects                                                                  & 2019 & Requirements Engineering                                                                          \\
{[}S83{]}  & J. P. Souza and A. R. Zavan and D. E. Fl{\^{o}}r                                                                                                & Scrum hero: Gamifying the scrum framework                                                                                                 & 2017 & Brazilian Workshop on Agile Methods                                                               \\
{[}S84{]}  & J. Kohl                                                                                                                                                                             & Software Testing Is a Game                                                                                                                & 2013 & Better Software                                                                                   \\
{[}S85{]}  & S. Marczak and F. F. Filho and L. Singer and C. Treude and F. Steffens and D. Redmiles and B. Al-Ani                                                                                & Studying Gamification as a Collaboration Motivator for Virtual Software Teams: Social Issues, Cultural Issues, and Research Methods       & 2015 & Conference on Computer-Supported Collaborative Work and Social Computing                          \\
{[}S86{]}  & O. Liechti and J. Pasquier and R. Reis                                                                                                                                              & Supporting agile teams with a test analytics platform: A case study                                                                       & 2017 & International Workshop on Automation of Software Testing                                          \\
{[}S87{]}  & M. Z. H. Kolpondinos and M. Glinz                                                                                                                                                   & Tailoring gamification to requirements elicitation: A stakeholder-centric motivation concept                                              & 2017 & International Workshop on Cooperative and Human Aspects of Software Engineering                   \\
{[}S88{]}  & B. Mayer and R. Weinreich                                                                                                                                                           & The effect of gamification on software architecture knowledge management: A student experiment and focus group study                      & 2019 & Symposium on Applied Computing                                                                    \\
{[}S89{]}  & A. Dorling and F. McCaffery                                                                                                                                                         & The gamification of SPICE                                                                                                                 & 2012 & International Conference on Software Process Improvement and Capability Determination             \\
{[}S90{]}  & P. Busetta and F. M. Kifetew and D. Munante and A. Perini and A. Siena and A. Susi                                                                                                  & Tool-Supported Collaborative Requirements Prioritisation                                                                                  & 2017 & Annual Computer Software and Applications Conference                                              \\
{[}S91{]}  & G. P. G. Hurtado and M. C. G{\'o}mez-Alvarez and M. Mu{\~n}oz and J. Mejia                                                                   & Toward an assessment framework for gamified environments                                                                                  & 2017 & European Conference on Software Process Improvement                                               \\
{[}S92{]}  & E. Herranz and R. C. Palacios and A. A. Seco and M. S{\'a}nchez-Gord{\'o}n                                                                        & Towards a gamification framework for software process improvement initiatives: Construction and validation                                & 2016 & Journal of Universal Computer Science                                                             \\
{[}S93{]}  & E. Herranz and R. C. Palacios and A. A. Seco                                                                                                                                        & Towards a New Approach to Supporting Top Managers in SPI Organizational Change Management                                                 & 2013 & Conference on ENTERprise Information Systems                                                      \\
{[}S94{]}  & R. M. Parizi and A. Kasem and A. Abdullah                                                                                                                                           & Towards gamification in software traceability: Between test and code artifacts                                                            & 2015 & International Joint Conference on Software Technologies                                           \\
{[}S95{]}  & R. Lotufo and L. Passos and K. Czarnecki                                                                                                                                            & Towards Improving Bug Tracking Systems with Game Mechanisms                                                                               & 2012 & Working Conference on Mining Software Repositories                                                \\
{[}S96{]}  & W. Snipes and V. Augustine and A. R. Nair and E. Murphy-Hill                                                                                                                        & Towards recognizing and rewarding efficient developer work patterns                                                                       & 2013 & International Conference on Software Engineering                                                  \\
{[}S97{]}  & E. C. Prakash and M. Rao                                                                                                                                                            & Transforming learning and it management through gamification                                                                              & 2015 & Springer International Publishing                                                                 \\
{[}S98{]}  & E. B. Passos and D. B. Medeiros and P. A. S. Neto and E. W. G. Clua                                                                                                                 & Turning real-world software development into a game                                                                                       & 2011 & Brazilian Symposium on Games and Digital Entertainment                                            \\
{[}S99{]}  & D. J. Dubois and G. Tamburrelli                                                                                                                                                     & Understanding gamification mechanisms for software development                                                                            & 2013 & Joint Meeting on Foundations of Software Engineering                                              \\
{[}S100{]} & F. Steffens and S. Marczak and F. F. Filho and C. Treude and L. Singer and D. Redmiles and B. Al-Ani                                                                                & Using gamification as a collaboration motivator for software development teams: A preliminary framework                                   & 2015 & Brazilian Symposium on Collaborative Systems                                                      \\
{[}S101{]} & D. Redmiles and B. Al-Ani                                                                                                                                                           & Using Gamification to Increase Scrum Adoption                                                                                             & 2015 & Brazilian Symposium in Collaborative Systems                                                      \\
{[}S102{]} & A. A. Melo and M. Hinz and G. Scheibel and C. D. M. Berkenbrock and I. Gasparini and F. Baldo                                                                                       & Version control system gamification: A proposal to encourage the engagement of developers to collaborate in software projects             & 2014 & International Conference on Social Computing and Social Media                                     \\
{[}S103{]} & D. B. Medeiros and P. A. S. Neto and E. B. Passos and W. S. Ara{\'{u}}jo                                                                                         & Working and Playing with Scrum                                                                                                            & 2015 & International Journal of Software Engineering and Knowledge Engineering                           \\ \bottomrule
\end{longtable}

\end{document}